\documentclass[
aip,
amsmath,amssymb,
reprint,
]{revtex4-1}

\usepackage{graphicx}
\usepackage{dcolumn}
\usepackage{bm}
\usepackage[utf8]{inputenc}
\usepackage[T1]{fontenc}
\usepackage{mathptmx}
\usepackage{etoolbox}

\usepackage{amsfonts}
\usepackage{mathtools}
\usepackage{subcaption}
\usepackage{multirow}
\usepackage{array}
\usepackage{booktabs}
\usepackage{mathrsfs}
\usepackage{numprint}
\usepackage{xcolor}
\usepackage{float}
\usepackage{tabularx}

\def\d{\mathrm{d}}

\usepackage{soul}

\makeatletter
\def\@email#1#2{
 \endgroup
 \patchcmd{\titleblock@produce}
  {\frontmatter@RRAPformat}
  {\frontmatter@RRAPformat{\produce@RRAP{*#1\href{mailto:#2}{#2}}}\frontmatter@RRAPformat}
  {}{}
}
\makeatother

\begin{document}

\preprint{AIP/123-QED}

\title[Jacobian-Free Variational Method for Constructing Connecting Orbits]{Jacobian-Free Variational Method for Constructing Connecting Orbits in Nonlinear Dynamical Systems}
\author{Omid Ashtari}
\author{Tobias M. Schneider}
\email{tobias.schneider@epfl.ch}
\affiliation{ 
Emergent Complexity in Physical Systems Laboratory (ECPS), \'Ecole Polytechnique F\'ed\'erale de Lausanne (EPFL), CH-1015 Lausanne, Switzerland
}

\begin{abstract}
In a dynamical systems description of spatiotemporally chaotic PDEs including those describing turbulence, chaos is viewed as a trajectory evolving within a network of non-chaotic, dynamically unstable, time-invariant solutions embedded in the chaotic attractor of the system.
While equilibria, periodic orbits and invariant tori can be constructed using existing methods, computations of heteroclinic and homoclinic connections mediating the evolution between the former invariant solutions remain challenging. We propose a robust matrix-free variational method for computing connecting orbits between equilibrium solutions of a dynamical system that can be applied to high-dimensional problems. 
Instead of a common shooting-based approach, we define a minimization problem in the space of smooth state space curves that connect the two equilibria with a cost function measuring the deviation of a connecting curve from an integral curve of the vector field. Minimization deforms a trial curve until, at a global minimum, a connecting orbit is obtained. The method is robust, has no limitation on the dimension of the unstable manifold at the origin equilibrium, and does not suffer from exponential error amplification associated with time-marching a chaotic system. Owing to adjoint-based minimization techniques, no Jacobian matrices need to be constructed and the memory requirement scales linearly with the size of the problem. The robustness of the method is demonstrated for the one-dimensional Kuramoto-Sivashinsky equation.
\end{abstract}

\maketitle

\begin{quotation}
    The chaotic evolution of a dynamical system can be described in terms of the non-chaotic time-invariant solutions embedded within its chaotic attractor. Heteroclinic and homoclincic connecting orbits between these invariant solutions mediate the evolution of the chaotic trajectory from the vicinity of one invariant solution to the vicinity of another one. While a complete characterization of the state space structures supporting chaos requires the identification of connecting orbits, constructing those has remained a numerical challenge. We introduce a robust and memory-efficient method for constructing connecting orbits between equilibrium solutions. Thereby, a more complete characterization of the state space structures supporting chaotic dynamics becomes feasible.
\end{quotation}

\section{\label{sec:introduction}Introduction}
Many nonlinear driven out-of-equilibrium systems including those describing fluid flows, nonlinear optics and active suspensions exhibit spatiotemporally chaotic dynamics. 
Within a dynamical systems description the spatiotemporal chaos can be viewed as the evolution of a chaotic trajectory in the state space of the governing equations. 
Embedded in the state space are non-chaotic, time-invariant solutions including equilibria, periodic orbits and higher-dimensional invariant tori. These invariant solutions are dynamically unstable so that the chaotic trajectory visits them transiently, yet recurringly. Spatiotemporal chaos can thus be viewed as a walk through a forest of invariant solutions that form the elementary building blocks of the chaotic solution \cite{Cvitanovic2013,Chandler2013,Crowley2022}. Consequently, individual invariant solutions can provide remarkable information about the spatiotemporal chaos and physical mechanisms underlying it, and collectively they promise an avenue towards quantitatively predicting statistical properties of the chaotic dynamics. 
Due to the significantly increasing computational resources and algorithmic advances, these concepts, originally developed in the context of low-dimensional chaotic dynamical systems, are now applied to very high-dimensional problems including transitional fluid turbulence where dynamical systems descriptions based on the analysis of invariant solutions have proven to be particularly useful \cite{Kerswell2005, Eckhardt2007, Gibson2008, Kawahara2012, Suri2017, Graham2021}.

While equilibria and periodic orbits form the building blocks of the dynamics, the chaotic evolution from the neighborhood of one unstable invariant solution to another is mediated by \emph{connecting orbits}. These hetero- and homoclinic connections provide dynamic pathways between different periodic orbits or equilibria within the chaotic attractor.  
Therefore, a characterization of the chaotic dynamics in terms of state-space structures requires both to identify equilibria, periodic orbits and invariant tori embedded in the chaotic attractor, and to compute connecting orbits between them. 
In the context of fluid dynamics for example, van Veen and Kawahara use connecting orbits to explain the turbulent bursting in plane Couette flow \cite{vanVeen2011}; Suri \textit{et al.} study the network of connecting orbits that underpins the transient dynamics in a quasi-two-dimensional Kolmogorov flow \cite{Suri2019}; and Reetz and Schneider characterize the time-dependent dynamics of inclined layer convection using connecting orbits between coexisting invariant solutions \cite{Reetz2020e}.

We specifically focus on connecting orbits between equilibrium solutions. Such connecting orbits have been identified as dynamically relevant in fluid systems \cite{vanVeen2011,Suri2019,Reetz2020e} and they are involved in global bifurcations, when for instance a periodic orbit bifurcates off a homoclinic orbit or a heteroclinic cycle \cite{Homburg2010,Reetz2020b}. 
Connecting orbits are located within the intersection of the unstable manifold of one equilibrium with the stable manifold of another or the same equilibrium solution if they are of heteroclinic or homoclinic type, respectively.
In the vicinity of an equilibrium solution, a trajectory approaches/departs the equilibrium along its stable/unstable manifold exponentially in time. Consequently, the time required to traverse the entire connecting orbit is not finite. This infinite passage time makes computing connecting orbits very challenging. 

One approach to handle the computational challenge of the infinite passage time is to truncate the connecting orbit and compute an approximating part of the orbit that is traversed in finite time. Under favourable conditions, the truncated orbit can be computed using shooting methods. 
Geometrically, the truncation approach attempts to construct a trajectory that starts at some point on the unstable manifold of the origin equilibrium and ends at some other point on the stable manifold of the destination equilibrium. 
Due to their curvature, parametrizations of stable and unstable manifolds are usually not accessible. Consequently, they need to be approximated locally by the corresponding tangent spaces associated to the origin and destination equilibrium. 
Practically, a connecting orbit is thus found by identifying an initial condition in the intersection of the unstable tangent space of and a hypersphere around the origin equilibrium, which after forward time integration reaches a distance below a chosen threshold from the destination equilibrium \cite{Beyn1990}.
If the hypersphere is chosen small enough, the unstable tangent space accurately approximates the unstable manifold, and thus the obtained trajectory accurately represents a connecting orbit.

Even if the unstable manifold can be accurately approximated by the unstable tangent space, a systematic search for an initial condition that eventually reaches the destination equilibrium is a formidable task, especially for a chaotic system where nearby trajectories diverge exponentially with time.
When the unstable manifold at the origin equilibrium solution is two-dimensional, an exhaustive search strategy can be employed \cite{Gibson2008, Halcrow2009, Cvitanovic2010a, Suri2019}. In this case, the search space is a circle on the unstable tangent space with an angle being the only variable. However, when the unstable tangent space at the origin equilibrium has more than two dimensions, the search space is too large for an exhaustive search. To improve the dimensionality drawback, Farano \textit{et al.}\cite{Farano2019} propose an adjoint-based variational method for finding a state on an energy shell around the origin equilibrium whose trajectory reaches another energy shell around the destination equilibrium. They do not constrain the initial condition to be located on the unstable tangent space at the origin equilibrium, hence as a second step the trajectory is confirmed to shadow a connecting orbit by matching the endpoints of the trajectory against the linearized dynamics around the two equilibria. In all these methods determining the size of the hypersphere around the origin equilibrium solution is not a trivial task: the hypersphere should be small enough in order for the tangent space to accurately approximate the manifold, and large enough to let the required time integration intervals be feasibly short.

An alternative to the shooting-based methods which search for a single state on the connecting orbit is to search in the space of \textit{connecting curves}, i.e. all smooth curves in the state space which connect the two equilibria. Among all such curves, only connecting orbits are integral curves of the vector field induced by the governing equation. The idea is to start from a connecting curve pivoted on the two fixed points, then deform the curve until the tangent velocity coincides with the local field vector along the entire curve, and thus a connecting orbit is achieved. This approach has several advantages over the reviewed shooting-based methods for computing connecting orbits: First, there is no limitation on the dimensionality of the unstable manifold at the origin equilibrium because no exhaustive search is needed; Secondly, the approach does not suffer from the exponential separation of trajectories with time since the connecting curve is deformed locally and no time integration is required; And lastly, this approach yields the exact and the entire connecting orbit without requiring to truncate it.

Despite the conceptual advantages of searching in the space of connecting curves over the shooting-based alternatives, this approach is not extensively developed on the practical side. Liu \textit{et al.}\cite{liu1994} use rational Chebyshev basis functions for the spectral representation of variables along the infinite temporal direction probably for the first time in this context. They formulate the problem as a system of nonlinear equations by setting the temporal derivative equal to the right-hand side of the governing equation for every state variable at every temporal collocation point, and solve the system of equations using standard Newton iterations. Dong and Lan\cite{dong2014} extend the variational method of Lan and Cvitanovi\'c\cite{Lan2004}, originally developed for finding periodic orbits, to the problem of constructing connecting orbits. They view the problem of deforming connecting curves towards a connecting orbit as a minimization problem: a connecting orbit is found by minimizing a cost function which penalizes the deviation of a connecting curve from being an integral curve of the vector field. They employ an infinitesimal-step version of Newton iterations for continuously deforming the curve, and use finite differences for calculating the tangent velocity vector. In his PhD thesis, Pallantla\cite{Pallantla2018} employs the same spectral representation of variables in the temporal direction as in Ref.~\onlinecite{liu1994}, and deforms the curve in the direction of the steepest descent of the cost function. The common drawback of the aforementioned algorithms is that they all require explicit construction of the Jacobian matrix. In a system with $M$ temporal and $N$ spatial degrees of freedom the size of the Jacobian matrix scales as $\mathcal{O}(M^2N^2)$ which can be prohibitively large for high-dimensional dynamical systems such as three-dimensional fluid flows.

In order to transfer the advantages of searching in the space of connecting curves to high-dimensional dynamical systems, we propose a Jacobian-free variational method for constructing connecting orbits between two equilibrium solutions. The method employs an adjoint-based optimization technique to minimize a cost function which measures the deviation of a connecting curve between two equilibria from an integral curve of the vector field. We construct a globally contracting dynamical system in the space of connecting curves. Fixed points of this dynamical system are minima of the non-negative cost function, hence global minima of the cost function, taking zero value, correspond to connecting orbits of the original dynamical system. Connecting orbits are therefore found by integrating the dynamics in the space of connecting curves. Due to the explicit construction of the dynamical system in the space of connecting curves, the memory requirement scales as $\mathcal{O}(MN)$ which allows the proposed method to be applied to high-dimensional dynamical systems.

The remainder of the present article is organized as follows. In Sec.~\ref{sec:variational_method} the problem of constructing a connecting orbit is set up as a minimization problem, and in Sec.~\ref{sec:adjoint_optimization} the adjoint-based minimization technique is formulated for a general autonomous dynamical system. In Sec.~\ref{sec:spectral_representation} a spectral representation suitable for the discretization along the unbounded temporal domain is discussed. To demonstrate the robustness of the proposed variational method, in Sec.~\ref{sec:application_to_KSE} we consider the one-dimensional Kuramoto–Sivashinsky equation in a spatiotemporally chaotic regime, and show that several connecting orbits can be converged reliably. Finally, in Sec.~\ref{sec:summary}, the manuscript is summarized, and an outlook for future research is given.

\section{Variational method for finding connecting orbits}
\label{sec:variational_method}
We consider general autonomous dynamical systems of the form
\begin{equation}
    \label{eq:governing_eqn}
    \frac{\partial\mathbf{u}}{\partial t}=\mathbf{f}(\mathbf{u}),
\end{equation}
where the smooth nonlinear operator $\mathbf{f}$ governs the evolution of an $n$-dimensional real field $\mathbf{u}\in\mathcal{M}\subset\mathbb{R}^n$ defined over a $d$-dimensional spatial domain $\mathbf{x}\in\Omega\subset\mathbb{R}^d$ and time $t\in\mathbb{R}$ subject to time-independent boundary conditions (BCs) at $\partial\Omega$, the boundaries of the spatial domain $\Omega$.

A connecting orbit between two equilibrium solutions is a solution trajectory $\mathbf{u}(\mathbf{x},t)$ of the governing equation \eqref{eq:governing_eqn} such that the asymptotic conditions
\begin{equation}
    \label{eq:connecting_orbit_def}
    \lim_{t\to -\infty}\mathbf{u}=\mathbf{u}_-\;,\quad \lim_{t\to +\infty}\mathbf{u}=\mathbf{u}_+\;,\quad \mathbf{f}(\mathbf{u}_\pm)=\mathbf{0},
\end{equation}
are satisfied in the temporal direction. The connecting orbit is a heteroclinic connection if $\mathbf{u}_-\neq\mathbf{u}_+$, and a homoclinic connection if $\mathbf{u}_- = \mathbf{u}_+$ (while implicitly assuming that the entire orbit is not the equilibrium solution itself.)

\begin{figure}
    \centering
    \includegraphics[width = \linewidth]{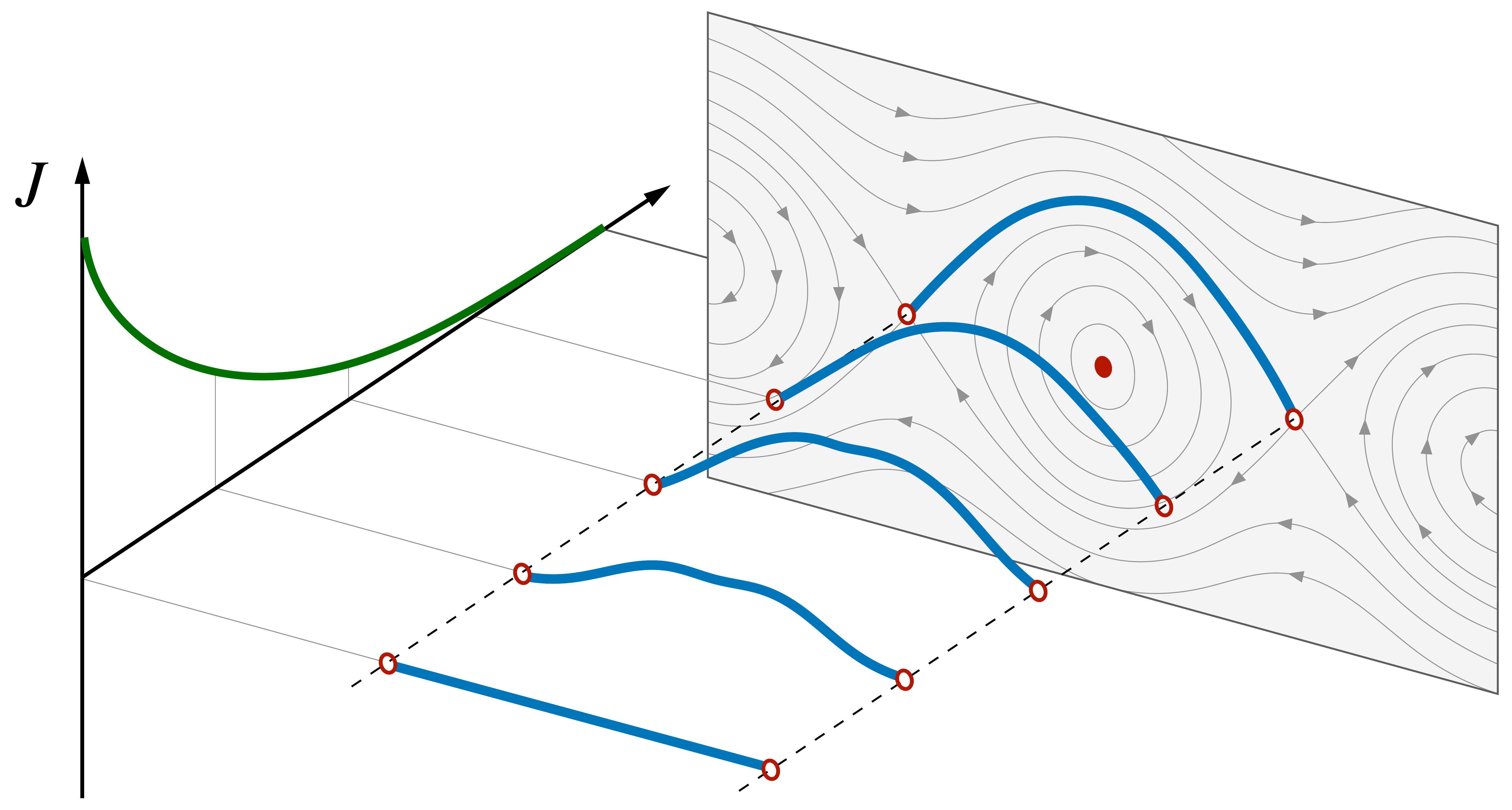}
    \caption{Schematic of the variational method for constructing a connecting orbit between two equilibrium solutions: A connecting curve pivoted on the two fixed points is deformed such that a cost function $J$ measuring the deviation of the connecting curve from being an integral curve of the vector field is minimized. For a connecting orbit the tangent velocity vector matches the field vector along the entire curve, and thus the global minimum of the cost function, $J=0$, is achieved.}
    \label{fig:idea}
\end{figure}

In the $(d+1)$-dimensional space-time domain of the dynamical system \eqref{eq:governing_eqn}, connecting orbits are solutions to a boundary value problem subject to the same BCs as Eq.~\eqref{eq:governing_eqn} in $d$ spatial directions, augmented by the asymptotic BCs \eqref{eq:connecting_orbit_def} in the temporal direction. The idea of the proposed variational method is to consider $C^\infty$ space-time fields that satisfy the boundary conditions in all $(d+1)$ directions, and vary the field until Eq.~\eqref{eq:governing_eqn} is satisfied at each and every space-time coordinate. Geometrically, $\mathbf{f}(\mathbf{u})$ is a vector field in the $n$-dimensional state space $\mathcal{M}$, $\mathbf{u}_-$ and $\mathbf{u}_+$ are two fixed points, and connecting orbits are integral curves of this vector field extending from $\mathbf{u}_-$ to $\mathbf{u}_+$. In this picture, the search space is the space of all smooth curves in the state space that connect the two fixed points. We define the space of connecting curves, denoted by $\mathscr{C}_g$, as
\begin{equation}
\label{eq:search_space}
    \mathscr{C}_g = \left\{\mathbf{u}(\mathbf{x},s) \;\;\Bigg|\;
    \begin{array}{l}
        \mathbf{u}:\Omega\times\mathbb{R}\to\mathcal{M}\\
        \lim_{s\to \pm\infty}\mathbf{u}=\mathbf{u}_\pm \\
        \mathbf{u}\text{ satisfies BCs at }\partial\Omega
    \end{array}
    \right\}.
\end{equation}
We parameterize connecting curves by $s\in\mathbb{R}$ in order to distinguish the evolution along a connecting curve from the evolution along a solution trajectory of the governing equation \eqref{eq:governing_eqn} which is parameterized by the physical time $t$. Connecting orbits form a subset $\mathscr{C}\subset\mathscr{C}_g$ in which the tangent velocity vector, $\partial\mathbf{u}/\partial s$, coincides with the local field vector, $\partial\mathbf{u}/\partial t=\mathbf{f}(\mathbf{u})$, along the entire connecting curve. As a measure of deviation of a connecting curve from being a connecting orbit we define the non-negative cost function $J^2$ as
\begin{equation}
    \label{eq:cost_function_general}
    J^2 = \int_{-\infty}^{+\infty}\int_\Omega{\mathbf{r}\cdot\mathbf{r}}\;\d\mathbf{x}\d s,
\end{equation}
where $\mathbf{r}$ is the local deviation of the tangent velocity vector from the field vector, or the residual of Eq.~\eqref{eq:governing_eqn}:
\begin{equation}
\label{eq:residual_definition}
    \mathbf{r}(\mathbf{u}) = \mathbf{f}(\mathbf{u})-\frac{\partial\mathbf{u}}{\partial s},
\end{equation}
and $\cdot$ indicates the standard Euclidean inner product.
The residual $\mathbf{r}$ is zero everywhere along a connecting orbit. Therefore, the cost function takes zero value for $\mathbf{u}\in\mathscr{C}$ while it takes a positive value for $\mathbf{u}\in\mathscr{C}_g\setminus\mathscr{C}$. The problem of finding connecting orbits can now be viewed as a minimization problem in $\mathscr{C}_g$: Absolute minima of $J^2$, for which $J=0$, correspond to connecting orbits $\mathbf{u}\in\mathscr{C}$. Fig.~\ref{fig:idea} schematically shows the idea of this approach: Minimizing the cost function $J$ deforms a curve connecting two fixed points of the vector field towards an integral curve of the vector field bounded between the two equilibria, thereby a connecting orbit.

\section{Adjoint-based minimization of the cost function}
\label{sec:adjoint_optimization}
We have recast the problem of computing connecting orbits into a minimization problem in the space of connecting curves extended between two equilibrium solutions. Absolute minima of the non-negative cost function $J^2$ with $J=0$ correspond to a connecting orbit. To solve the minimization problem, we employ an adjoint-based technique inspired by the recent works by Farazmand\cite{Farazmand2016} on constructing equilibria and travelling waves, and by and Azimi \textit{et al.}\cite{Azimi2022} on constructing periodic orbits of nonlinear dynamical systems. We construct a dynamical system in the space of connecting curves, $\mathscr{C}_g$, such that along its trajectories the cost function is guaranteed to decrease monotonically. Therefore, connecting orbits are found by integrating the constructed dynamics in $\mathscr{C}_g$ until a minimum of the cost function is reached. Parametrizing this dynamical system by a fictitious time $\tau$ we need to construct the operator $\mathbf{G}(\mathbf{u})$ such that evolution of $\mathbf{u}$ governed by
\begin{equation}
\label{eq:G_general}
    \frac{\partial\mathbf{u}}{\partial \tau}=\mathbf{G}(\mathbf{u}),
\end{equation}
guarantees
\begin{equation}
    \label{eq:monotonic_decrase_of_J_general}
    \frac{\partial J^2}{\partial \tau} \leq 0;\quad \forall \tau.
\end{equation}

We define the inner product space $\mathscr{C}_s\supset\mathscr{C}_g$
\begin{equation}
\label{eq:inner_product_space}
    \mathscr{C}_s = \left\{\mathbf{q}(\mathbf{x},s) \;\;\Bigg|\;
    \begin{array}{l}
        \mathbf{q}:\Omega\times\mathbb{R}\to\mathbb{R}^n\\
        \lim_{s\to \pm\infty}\mathbf{q}=\mathbf{v}_\pm\in\mathbb{R}^n
    \end{array}
    \right\},
\end{equation}
together with the real-valued inner product
\begin{equation}
    \begin{split}
        \left<\;,\;\right>&:\mathscr{C}_s\times\mathscr{C}_s\to\mathbb{R},\\
        \left<\mathbf{q}_1,\mathbf{q}_2\right>=&\int_{-\infty}^{+\infty}\int_\Omega{\mathbf{q}_1\cdot\mathbf{q}_2}\;\d\mathbf{x}\d s,
    \end{split}
\end{equation}
and $L_2$-norm
\begin{equation}
    \left\|\mathbf{q}\right\|=\sqrt{\left<\mathbf{q},\mathbf{q}\right>}.
\end{equation}
In contrast to the space of connecting curves $\mathcal{C}_g$, the elements of $\mathcal{C}_s$ have arbitrary asymptotic states $\mathbf{v}_\pm\in\mathbb{R}^n$. The rate of change of the cost function $J^2=\left\|\mathbf{r}\right\|^2=\left<\mathbf{r},\mathbf{r}\right>$ is obtained by the inner product of $\mathbf{r}(\mathbf{u})$ with its directional derivative along the to-be-determined operator $\partial\mathbf{u}/\partial\tau=\mathbf{G}(\mathbf{u})$:
\begin{equation}
    \label{eq:rate_of_J_substituted}
    \frac{\partial J^2}{\partial \tau}=2\left<\left(\nabla_\mathbf{u}\mathbf{r}\right)\frac{\partial\mathbf{u}}{\partial\tau}\;,\;\mathbf{r}\right>.
\end{equation}
The directional derivative of $\mathbf{r}(\mathbf{u})$ along $\mathbf{G}$ is defined as
\begin{equation}
\label{eq:directional_derivative}
    \pmb{\mathscr{L}}(\mathbf{u};\mathbf{G}) = \lim_{\epsilon\to 0}\frac{\mathbf{r}(\mathbf{u}+\epsilon\mathbf{G})-\mathbf{r}(\mathbf{u})}{\epsilon}.
\end{equation}
Using the adjoint of the directional derivative we can write Eq.~\eqref{eq:rate_of_J_substituted} as
\begin{equation}
    \label{eq:rate_of_J_with_adjoint}
    \frac{\partial J^2}{\partial \tau}=2\left<\pmb{\mathscr{L}}^\dagger(\mathbf{u};\mathbf{r}),\mathbf{G}\right>,
\end{equation}
where $\pmb{\mathscr{L}}^\dagger$ is the adjoint operator of $\pmb{\mathscr{L}}$, with
\begin{equation}
    \label{eq:adjoint_definition}
    \left<\pmb{\mathscr{L}}(\mathbf{u};\mathbf{G}),\mathbf{r}\right>=\left<\mathbf{G},\pmb{\mathscr{L}}^\dagger(\mathbf{u};\mathbf{r})\right>,
\end{equation}
for all connecting curves $\mathbf{u}\in\mathscr{C}_g$. The residual $\mathbf{r}$ (defined in Eq.~\eqref{eq:residual_definition}) and the operator $\mathbf{G}$ (defined in Eq.~\eqref{eq:G_general}) are functions of $\mathbf{u}$, and belong to the inner product space $\mathcal{C}_s$ with certain properties that are detailed shortly. By choosing $\mathbf{G}(\mathbf{u})=-\pmb{\mathscr{L}}^\dagger(\mathbf{u};\mathbf{r})$ the monotonic decrease of the cost function is guaranteed:
\begin{eqnarray}
    \frac{\partial J^2}{\partial \tau}&&=2\left<\pmb{\mathscr{L}}^\dagger(\mathbf{u};\mathbf{r}),-\pmb{\mathscr{L}}^\dagger(\mathbf{u};\mathbf{r})\right>\nonumber\\
    &&=-2\left\|\pmb{\mathscr{L}}^\dagger(\mathbf{u};\mathbf{r})\right\|^2\leq 0.
\end{eqnarray}
The dynamical system $\partial\mathbf{u}/\partial\tau=\mathbf{G}(\mathbf{u})=-\pmb{\mathscr{L}}^\dagger(\mathbf{u};\mathbf{r})$ is globally contracting: All trajectories are eventually attracted to stable fixed points at which $\partial \mathbf{u}/\partial\tau=\mathbf{0}$ and $J^2$ takes a minimum value. Although the monotonic decrease of the cost function is guaranteed along trajectories of the dynamics in $\mathcal{C}_g$, reaching the global minimum is not. To find a connecting orbit, therefore, the dynamics in the space of connecting curves is integrated until a fixed point is reached. Those fixed points of $\partial\mathbf{u}/\partial\tau=\mathbf{G}(\mathbf{u})$ which correspond to the global minimum of the cost function, $J=0$, are connecting orbits of the original dynamical system $\partial\mathbf{u}/\partial t=\mathbf{f}(\mathbf{u})$, and those corresponding to $J>0$ are rejected.

The dynamical system $\partial\mathbf{u}/\partial\tau=\mathbf{G}(\mathbf{u})$ is constructed in the space of connecting curves $\mathcal{C}_g$ defined in Eq.~\eqref{eq:search_space}. This imposes certain BCs on the residual $\mathbf{r}(\mathbf{u})$ and the operator $\mathbf{G}(\mathbf{u})$. In the temporal direction, $\lim_{s\to\pm\infty}\mathbf{r}=\mathbf{0}$ since $\mathbf{u}$ satisfies the correct asymptotic BCs for all $\tau$, and $\lim_{s\to\pm\infty}\mathbf{G}=\mathbf{0}$ since the correct asymptotic values of $\mathbf{u}$ must be preserved. In space, $\mathbf{u}$ satisfies the correct BCs at $\partial\Omega$ for all $\tau$; consequently, the spatial BCs of $\mathbf{r}$ and $\mathbf{G}$ are determined following similar arguments. For example, $\mathbf{r}$ and $\mathbf{G}$ will be periodic in directions where $\mathbf{u}$ is periodic, will take zero value where $\mathbf{u}$ satisfies Dirichlet boundary conditions, and so forth. These properties must be taken into account while deriving the adjoint operator from the definition \eqref{eq:adjoint_definition}. Derivation of the adjoint operator for the Kuramoto-Sivashinsky system, introduced in Sec. \ref{sec:application_to_KSE}, is presented in Appendix \ref{sec:KSE_adj_derivation} where the zero asymptotic values of $\mathbf{r}$ and $\mathbf{G}$ in the temporal direction and their periodicity in space enable us to derive the adjoint operator as an explicit function of the space-time field $\mathbf{u}$.

Both heteroclinic and homoclinic connections can be constructed using the introduced variational method. In the case of a homoclinic connection to an equilibrium solutions, zero variation in time, i.e. the equilibrium solution itself, is a trivial solution satisfying the definition \eqref{eq:connecting_orbit_def}. Therefore, depending on the initial connecting curve from which the integration starts, a trivial or a nontrivial solution with $J=0$ can be obtained. The definition of a heteroclinic connection does not have any trivial solution.

On an abstract level, we construct the operator $\mathbf{G}$ following the same logic as that in Refs.~\onlinecite{Azimi2022} and \onlinecite{Farazmand2016}. However, in the different contexts the form of the operator differs as it acts on different objects and the dynamical system guaranteeing the monotonic decrease of the cost function evolves objects representing the specific sought-after invariant solution: Farazmand\cite{Farazmand2016} converges equilibrium solutions, and thus constructs $\mathbf{G}$ for evolving spatial fields, i.e. points in the state space; Azimi \textit{et al.}\cite{Azimi2022} converge periodic orbits, hence they construct $\mathbf{G}$ for evolving space-time fields that are periodic in the temporal direction, i.e. closed loops in the state space; and here we converge connecting orbits, thus we construct $\mathbf{G}$ for evolving space-time fields satisfying the asymptotic conditions \eqref{eq:connecting_orbit_def} in the temporal direction, i.e. connecting curves between two fixed points in the state space.

\section{Spectral representation in time}
\label{sec:spectral_representation}
An efficient implementation of the proposed adjoint-based variational method is aided by an accurate spectral representation of a space-time field $\mathbf{q}(\mathbf{x},s)\in\mathscr{C}_s$ in the $s$ direction, such that the asymptotic conditions at $s\to\pm\infty$ are directly enforced by the chosen expansion.
The spectral accuracy significantly reduces the number of time sections, and thereby memory, required for an accurate representation of connecting orbits. We use rational Chebyshev basis functions for the spectral representation in the temporal direction (see Chapter 17 of Ref.~\onlinecite{Boyd2000} for details).

Rational Chebyshev functions, $R_n(s)$, are given by
\begin{equation}
    \label{eq:basis_function_def}
        R_n(s) = \cos(n\theta);\quad \;n\in\mathbb{W},
\end{equation}
where $\theta\in(0,\pi)$ and $s\in\mathbb{R}$ are related via
\begin{equation}
    \label{eq:Rational_domain}
    s = s_0 + S\cot(\theta) \iff \theta=\cot^{-1}\left(\frac{s-s_0}{S}\right),
\end{equation}
with $s_0\in\mathbb{R}$ and $S\in\mathbb{R}^+$ being mapping parameters.

Rational Chebyshev collocation points are obtained by a uniform discretization of $\theta$. Therefore, $M$ interior collocation points are
\begin{equation}
    \label{eq:t_grid}
    s_j=s_0+S\cot{\left(\frac{j\pi}{M+1}\right)};\quad j=1,2,\dots,M,
\end{equation}
with $j=0$ and $j=M+1$ being reserved for the asymptotic values $s\to +\infty$ and $s\to -\infty$, respectively. The uniform discretization of $\theta$ results in a non-uniform distribution of grid points in $s$. Collocation points are denser around $s_0$, the center of the distribution, and become sparser further away from the center. The spacing between successive grid points is linearly scaled by $S$.

A real function $q(s)$ with $s\in\mathbb{R}$ and constant asymptotic values is approximated by the truncated expansion in a rational Chebyshev basis, $q(s)\approx\sum_{k=0}^{M+1}{c_kR_k(s)}$, where the expansion coefficients are

\begin{equation}
\label{eq:rational_coeffs}
    c_k=\dfrac{2}{(M+1)\Bar{c}_k}\sum_{m=0}^{M+1}{\dfrac{1}{\Bar{c}_m}q(s_m)\cos{\left(\dfrac{mk\pi}{M+1}\right)}},
\end{equation}
with grid points $s_m$ defined in equation (\ref{eq:t_grid}) and
\begin{equation}
    \label{eq:c_bar}
    \Bar{c}_j =
    \begin{cases}
        2, & \text{if } j=0 \text{ or } M+1,\\
        1, & \text{otherwise}.
    \end{cases}
\end{equation}

Having a grid function $q(s_j)$ with $j=0,1,\dots,M+1$ over rational Chebyshev grid points (\ref{eq:t_grid}), the differentiation matrix $D_t$ is constructed as:
\begin{widetext}
\begin{equation}
\label{eq:rational_diff_matrix}
    {D_t}_{\;j,m} = \frac{2}{S(M+1)}\sin^2{\left(\frac{j\pi}{M+1}\right)}\sum_{k=0}^{M+1}{\frac{k}{\Bar{c}_m\Bar{c}_k
    }\cos{\left(\frac{mk\pi}{M+1}\right)}\sin{\left(\frac{kj\pi}{M+1}\right)}};\quad j,m=0,1,\dots,M+1.
\end{equation}
\end{widetext}

The expansion in a rational Chebyshev basis allows us to represent the space-time objects in the unbounded temporal direction, and we can expect spectral accuracy with fast convergence as a function of the expansion's truncation order. Rational Chebyshev functions form a generic basis for the spectral representation of functions over the entire real axis with constant asymptotic values and are thus a suitable expansion for connecting orbits for any studied physical system.

\section{Application to Kuramoto-Sivashinsky equation}
\label{sec:application_to_KSE}
As a proof of concept, we apply the introduced method for constructing connecting orbits to the one-dimensional Kuramoto-Sivashinsky equation (KSE)\cite{Kuramoto1976, Sivashinsky1977}. The KSE is a nonlinear partial differential equation which emerges in various physical contexts such as flame propagation\cite{Sivashinsky1977}, plasma physics\cite{Laquey1975}, or interfacial fluids instability\cite{Hooper1985}. The KSE is also commonly used as a model system for examining new methods developed for chaotic fluid flows and transitional turbulence since it exhibits spatiotemporally chaotic behavior and displays some similar features to the Navier-Stokes equations.

The one-dimensional KSE for a real field $u(x,t)$ on the periodic spatial domain $0 \leq x < L$ is
\begin{equation}
\label{eq:KSE}
    \frac{\partial u}{\partial t} = -u\frac{\partial u}{\partial x} - \frac{\partial^2 u}{\partial x^2} - \nu\frac{\partial^4 u}{\partial x^4},
\end{equation}
with constant positive damping parameter $\nu$. The dynamics of the KSE is controlled by the single dimensionless group $\mathbb{L}=L/\sqrt{\nu}$. Here, we fix $\nu=1$ and consider the domain size $L$ as the control parameter. For $L<2\pi$, the trivial equilibrium solution $u(x,t)=\mathrm{const.}$ is linearly stable, and is the global attractor of the dynamics. By increasing $L$, solutions of the KSE undergo a series of bifurcations, and for a sufficiently large domain size the dynamics can exhibit spatiotemporally chaotic behavior \cite{Smyrlis1996}. We demonstrate the application of the proposed method by constructing connecting orbits between equilibrium solutions of the KSE for $L=22$. This domain size is large enough for the KSE to exhibit spatiotemporally chaotic dynamics, yet small enough to have low-dimensional unstable manifolds at the equilibria found, over which an exhaustive search for possible connecting orbits is practical. The state space geometry of the KSE for this parameter value has previously been explored in detail by Cvitanovi\'c and collaborators\cite{Cvitanovic2010a}. They identified several connecting orbits using the shooting method described in Section \ref{sec:introduction}. We construct a complete set of connecting orbits between all known equilibrium solutions of this system; complete in the sense that at least one connecting orbit between any pair of equilibrium solutions is computed, or it is confirmed by the exhaustive search in Ref.~\onlinecite{Cvitanovic2010a} that no connecting orbit exists between the two equilibria.

\subsection{Formulation of the adjoint-based variational method for the  KSE}
The KSE \eqref{eq:KSE} has the form of the general dynamical system \eqref{eq:governing_eqn} with $n=d=1$ and $\Omega=[0,L)$. The residual field, defined in Eq.~\eqref{eq:residual_definition}, for the KSE is
\begin{equation}
\label{eq:KSE_residual}
    r = -\frac{\partial u}{\partial s} -u\frac{\partial u}{\partial x} - \frac{\partial^2 u}{\partial x^2} - \frac{\partial^4 u}{\partial x^4}.
\end{equation}
The dynamical system along whose trajectories the cost function decreases monotonically is derived based on the adjoint operator of the directional derivative of $r$. The adjoint operator for the KSE system is constructed by a series of integrations by part (see Appendix \ref{sec:KSE_adj_derivation} for details):
\begin{equation}
    \mathscr{L}^\dagger(u;r) = \dfrac{\partial r}{\partial s}+u\dfrac{\partial r}{\partial x}-\dfrac{\partial^2 r}{\partial x^2}-\dfrac{\partial^4 r}{\partial x^4}.
\end{equation}
Therefore, the dynamical system in the space of connecting curves, $u(x,s;\tau)\in\mathcal{C}_g$, that minimizes the cost function $J^2$ is
\begin{equation}
\label{eq:KSE_G}
    \dfrac{\partial u}{\partial \tau} = -\mathscr{L}^\dagger(u;r) = -\dfrac{\partial r}{\partial s}-u\dfrac{\partial r}{\partial x}+\dfrac{\partial^2 r}{\partial x^2}+\dfrac{\partial^4 r}{\partial x^4}.
\end{equation}

\subsection{Symmetry preservation}
The KSE \eqref{eq:KSE} is equivariant under continuous translations in the $x$-direction
\begin{equation}
\label{eq:translation}
    \gamma(\alpha)u(x,t)=u(x+\alpha L,t);\quad \alpha\in[0,1),
\end{equation}
and under inversions about the origin
\begin{equation}
\label{eq:reflection}
    \sigma u(x,t)=-u(-x,t).
\end{equation}
The translation operator $\gamma(\alpha)$ and inversion operator $\sigma$ commute with the residual \eqref{eq:KSE_residual} of the KSE. Consequently, the dynamics in the space of connecting curves, Eq.~\eqref{eq:KSE_G}, is equivariant under the action of $\gamma(\alpha)$ and $\sigma$. This means that if the integration of Eq.~\eqref{eq:KSE_G} starts from an initial space-time field that is invariant under the action of $\sigma\circ\gamma(\alpha)$, the dynamics preserves the resulting point-inversion symmetry, and therefore the constructed connecting orbit belongs to the same symmetric subspace of the state space $\mathcal{M}$.

The KSE \eqref{eq:KSE} preserves the spatial mean value of the evolving field. Consequently, the spatial mean along a connecting orbit is constant and the same as the end point equilibrium solutions. We consider the dynamics of the KSE in the subspace of fields with zero spatial mean. The zero mean value is not enforced during the evolution of a connecting curve towards a connecting orbit. However, since the two end point equilibria do have zero spatial mean, a converged connecting orbit with $J=0$ takes zero mean value as well.

\subsection{Numerical implementation}
\label{sec:numerical_implementation}
\subsubsection{Spectral discretization}
A connecting curve $u(x,s)$ is discretized in the temporal direction using $M+2$ time sections (including the end point equilibria) over the rational Chebyshev grid while each time section is represented by $N$ Fourier modes in space:
\begin{equation}
    u(x_n,s_m) = \sum_{j=-\frac{N}{2}}^{\frac{N}{2}-1}\,\hat{u}_j(s_m)\exp{\left(j\frac{2\pi x_n}{L}i\right)},
\end{equation}
where $x_n=nL/N$ with indices $0\leq n<N$ are the uniform grid points in space; $s_m$ with indices $0\leq m\leq M+1$ are the non-uniform rational Chebyshev collocation points in time; $\hat{u}_j(s_m)$ is the $j$th Fourier coefficient of the time section at $s_m$; and $i$ is the imaginary unit.

In spectral space, the connecting curve $u$ is represented by an $(M+2)\times N$ matrix of complex numbers $\hat{u}_{m,j}=\hat{u}_j(s_m)$. The derivative of order $q\in\mathbb{W}$ of this space-time field with respect to $x$ is obtained by the Hadamard product $D_x^{(q)}\odot \hat{u}$ where ${D_x^{(q)}}_{m,j}=(2\pi j i/L)^q$, and its derivative of order $q\in\mathbb{W}$ with respect to $s$ is obtained by multiplying $\hat{u}$ from the left by $D_t^q$, where the temporal differentiation matrix $D_t$ is defined in Eq.~\eqref{eq:rational_diff_matrix}.
The residual $r$ and the descent direction $G$ are discretized in the same way with the only difference that their time sections at $s_0$ and $s_{M+1}$ (corresponding to $s\to +\infty$ and $s\to -\infty$, respectively) are identically zero (see Sec.~\ref{sec:adjoint_optimization}).
The nonlinear terms are calculated in physical space where products are of elementwise Hadamard type. Transforming back and forward between physical and spectral representations of the space-time fields requires one-dimensional forward or backward discrete Fourier transformation of each time section.

\subsubsection{Initialization}
The initial connecting curve is chosen as a convex combination of the equilibrium solutions $u_-$ and $u_+$, plus a symmetry breaking term:
\begin{equation}
\label{eq:KSE_initial_curve}
\begin{split}
    u_0(x,s;a) = &\frac{1}{2}\left[\left(1+\tanh{(s)}\right)u_+(x)+\left(1-\tanh{(s)}\right)u_-(x)\right]\\
    &+a\exp{(-s^2)}v(x);\qquad a\in\{0,1\},
\end{split}
\end{equation}
with $x\in[0,L)$ and $s\in\mathbb{R}$. If $u_-(x)$ and $u_+(x)$ both are inversion symmetric about the same point $x=x_0$, then $a=0$ results in an initial space-time field for which all time sections are invariant under the same inversion symmetry. Since the proposed variational dynamics preserves the inversion symmetry, we can set $a=0$ in order to search in the inversion-symmetric subspace of connecting trajectories. In order to break such a symmetry, we add the second line, i.e. set $a\neq0$, where $v(x)$ is a field which does not have the inversion symmetry shared between $u_-(x)$ and $u_+(x)$.

\subsubsection{Time stepping}
The defined dynamical system $\partial u/\partial \tau=G$ is globally contracting and we are only concerned about the asymptotic state $u=u_0+\int_0^\infty G \d\tau$. Consequently, we select the numerical integration scheme based on simplicity and stability rather than accuracy. We use semi-implicit forward Euler time-stepping scheme which has first-order accuracy in $\tau$, and treats the linear terms of $G$ in $u$ implicitly and the nonlinear terms explicitly. The code was developed in C++ with OpenMP parallelization of local calculations.

\subsection{Results and discussion}
\label{sec:results_and_discussion}
\begin{figure}
    \centering
    \begin{subfigure}{\linewidth}
        \includegraphics[width = 0.55\linewidth]{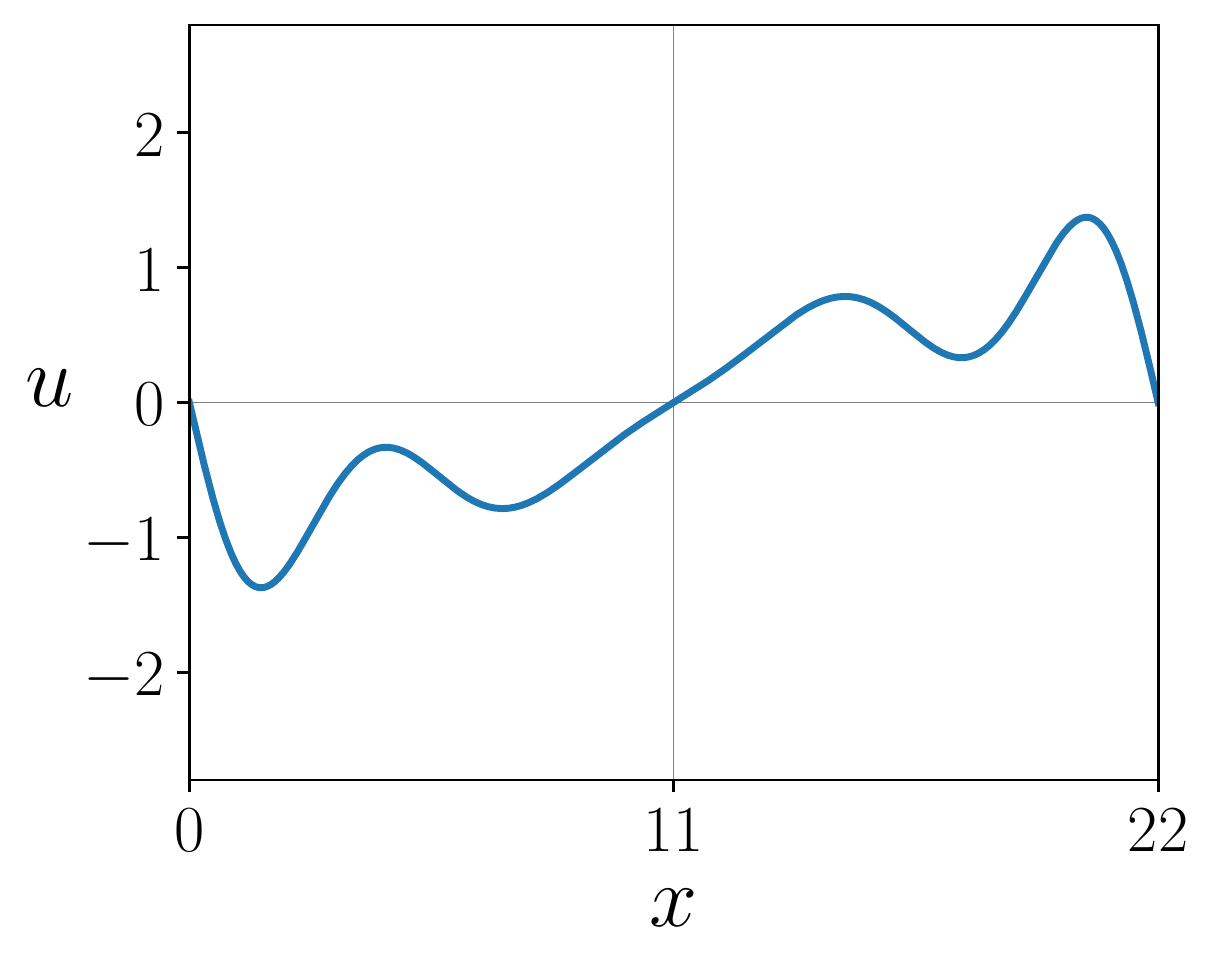}
        \caption{$E_1$}
    \end{subfigure}\\
    \begin{subfigure}{\linewidth}
        \includegraphics[width = 0.55\linewidth]{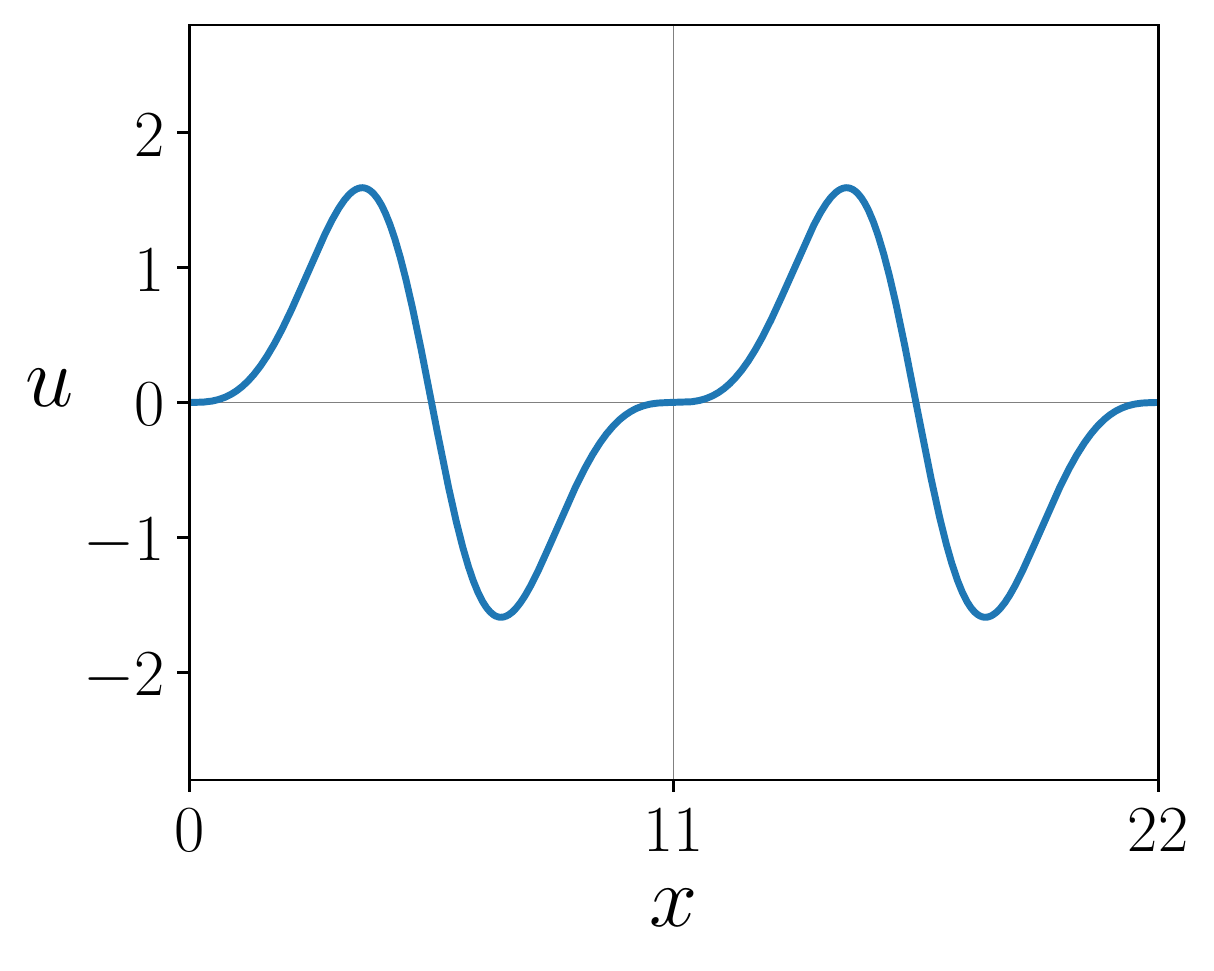}
        \caption{$E_2$}
    \end{subfigure}\\
    \begin{subfigure}{\linewidth}
        \includegraphics[width = 0.55\linewidth]{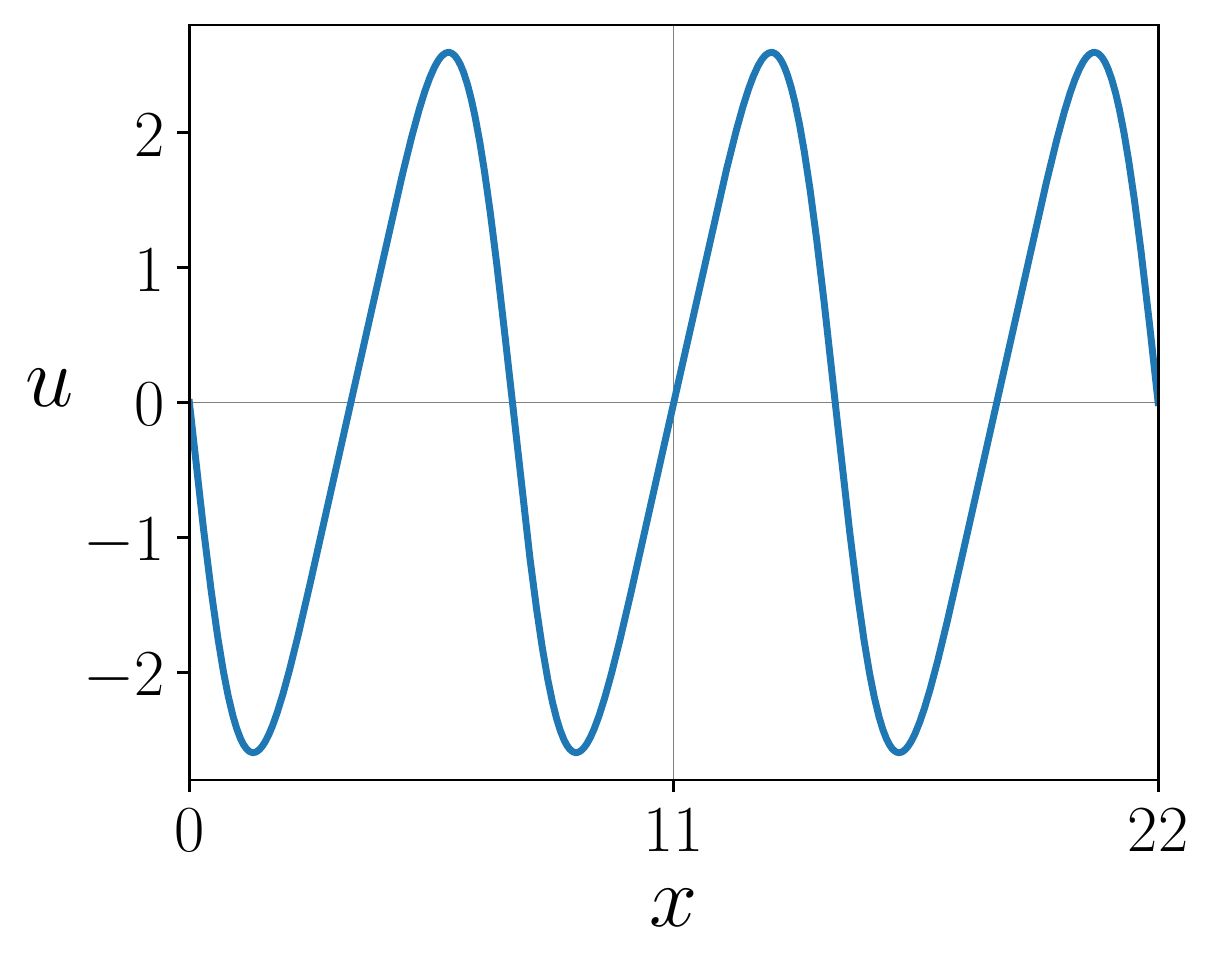}
        \caption{$E_3$}
    \end{subfigure}
    
    \caption{Nontrivial equilibrium solutions of the KSE for $L=22$. $E_1$, $E_2$ and $E_3$ are symmetric under inversion about the origin. $E_2$ and $E_3$ are also symmetric under discrete shift by $L/2$ and $L/3$, respectively.}
    \label{fig:KSE_FPs}
\end{figure}

In the subspace of fields with zero spatial mean, the KSE with $L=22$ has four known equilibrium solutions including the trivial solution $u=0$. Hereafter we denote the trivial equilibrium solution by $E_0$, and the nontrivial ones by $E_1$, $E_2$ and $E_3$ as shown in Fig.~\ref{fig:KSE_FPs}. We construct these equilibrium solutions following the adjoint-based variational method of Farazmand\cite{Farazmand2016}. $E_1$, $E_2$ and $E_3$ are invariant under inversion about the origin, $\sigma$. $E_2$ and $E_3$ are also symmetric under discrete shifts $\gamma(1/2)$ and $\gamma(1/3)$, respectively. Therefore, in addition to the inversion about $x=0$ and $L/2$, $E_2$ is symmetric under inversion about $x=L/4$ and $3L/4$, and $E_3$ is symmetric under inversion about $x=L/6$, $L/3$, $2L/3$, and $5L/6$ as well.
The repelling eigenvalues of all four equilibrium solutions are listed in Table \ref{tab:KSE_eigenvals}, and their associated eigenvectors are shown in Figs.~\ref{fig:eigenvectors_E0} to \ref{fig:eigenvectors_E3} in Appendix \ref{sec:KSE_eigenvecotrs}.

\begin{table}
    \centering
    \caption{Repelling eigenvalues of the equilibria of the KSE for $L=22$. The rest of the eigenvalues, except one zero eigenvalue for $E_1$, $E_2$ and $E_3$, have negative real part.}
        \begin{tabular}{p{0.25\linewidth}p{0.35\linewidth}}
        \toprule
             Solution & Unstable eigenvalues\\
        \midrule
             \multirow{3}{*}{$E_0$} & $\lambda_{1,2}\!=\!0.2198$\\
             & $\lambda_{3,4}\!=\!0.1952$\\
             & $\lambda_{5,6}\!=\!0.0749$ \vspace{2.5mm}\\
             \multirow{2}{*}{$E_1$} & $\lambda_{1,2}\!=\!0.1308\!\pm\!0.3341i$\\
             & $\lambda_{3,4}\!=\!0.0824\!\pm\!0.3402i$ \vspace{2.5mm}\\
             $E_2$ & $\lambda_{1,2}\!=\!0.1390\!\pm\!0.2384i$ \vspace{2.5mm}\\
             $E_3$ & $\lambda_{1,2}\!=\!0.0933$\\
        \bottomrule
        \end{tabular}
    \label{tab:KSE_eigenvals}
\end{table}

Connecting orbits are converged by integrating Eq.~\eqref{eq:KSE_G} until a fixed point in the vector field of $G$, corresponding to a minimum $J$, is achieved. Connecting orbits correspond to the global minima of $J$, for which $J=0$. In order to monitor the convergence, we define the arc length weighted cost function
\begin{equation}
\label{eq:J_arc}
    J_\mathrm{arc}=\dfrac{\displaystyle\int_{-\infty}^{+\infty}{\left|\mathbf{r}\right|\left|\dfrac{\partial\mathbf{u}}{\partial s}\right|}\d s}{\displaystyle\int_{-\infty}^{+\infty}{\left|\dfrac{\partial\mathbf{u}}{\partial s}\right|}\d s},
\end{equation}
with $\left|\;\cdot\;\right|$ being
\begin{equation}
    \left|\mathbf{q}\right|=\sqrt{\int_\Omega\mathbf{q}\cdot\mathbf{q}\;\d\mathbf{x}}\;;\quad\mathbf{q}\in\mathscr{C}_s.
\end{equation}
Obviously, $J_\mathrm{arc}=0$ if and only if $J=0$. However, the numerical evaluation of $J_\mathrm{arc}$ is not subject to the error accumulation associated with the numerical evaluation of the improper integral \eqref{eq:cost_function_general} that defines $J$. Moreover, since the trivial solution to the definition of a homoclinic connection has zero arc length, $J_\mathrm{arc}$ is undefined when the trivial solution is achieved, while $J=0$ for either trivial or nontrivial solutions. We consider the algorithm converged when $J_\mathrm{arc}<10^{-12}$.

Due to the continuous translational symmetry of the KSE, $E_i$ with $i=1,2,3$ represent its so-called group orbit of all symmetry related states, i.e. $\gamma(\alpha)E_i$ where $\alpha\in[0,1)$. Every connecting orbit, therefore, has infinite dynamically equivalent copies corresponding to similar translations of the origin and the destination equilibrium solutions. We construct connecting orbits of certain relative phase between the two end points by fixing the origin equilibrium and shifting the destination equilibrium solution when constructing the initial connecting curve using Eq.~\eqref{eq:KSE_initial_curve}. In the following, we first demonstrate the application of the introduced method by constructing a connecting orbit from $E_1$ to $E_2$. We then present converged connecting orbits between other equilibrium solutions, and compare to the same orbits obtained from other methods reported in the literature if applicable.

The search for a heteroclinic connection from $E_1$ to $E_2$ is initialized by a connecting curve constructed using Eq.~\eqref{eq:KSE_initial_curve} in the inversion-symmetric subspace of $\mathcal{M}$ ($a=0$). We discretize the space-time domain by $N=64$ Fourier modes in space and $M=550$ rational Chebyshev grid points in time. The scaling of the temporal discretization is set to $S=55$, and the center of the distribution to $s_0=0$. For this system, the integration scheme described in section \ref{sec:numerical_implementation} is stable for $\Delta\tau=0.01$.

\begin{figure}
    \centering
    \includegraphics[width=\linewidth]{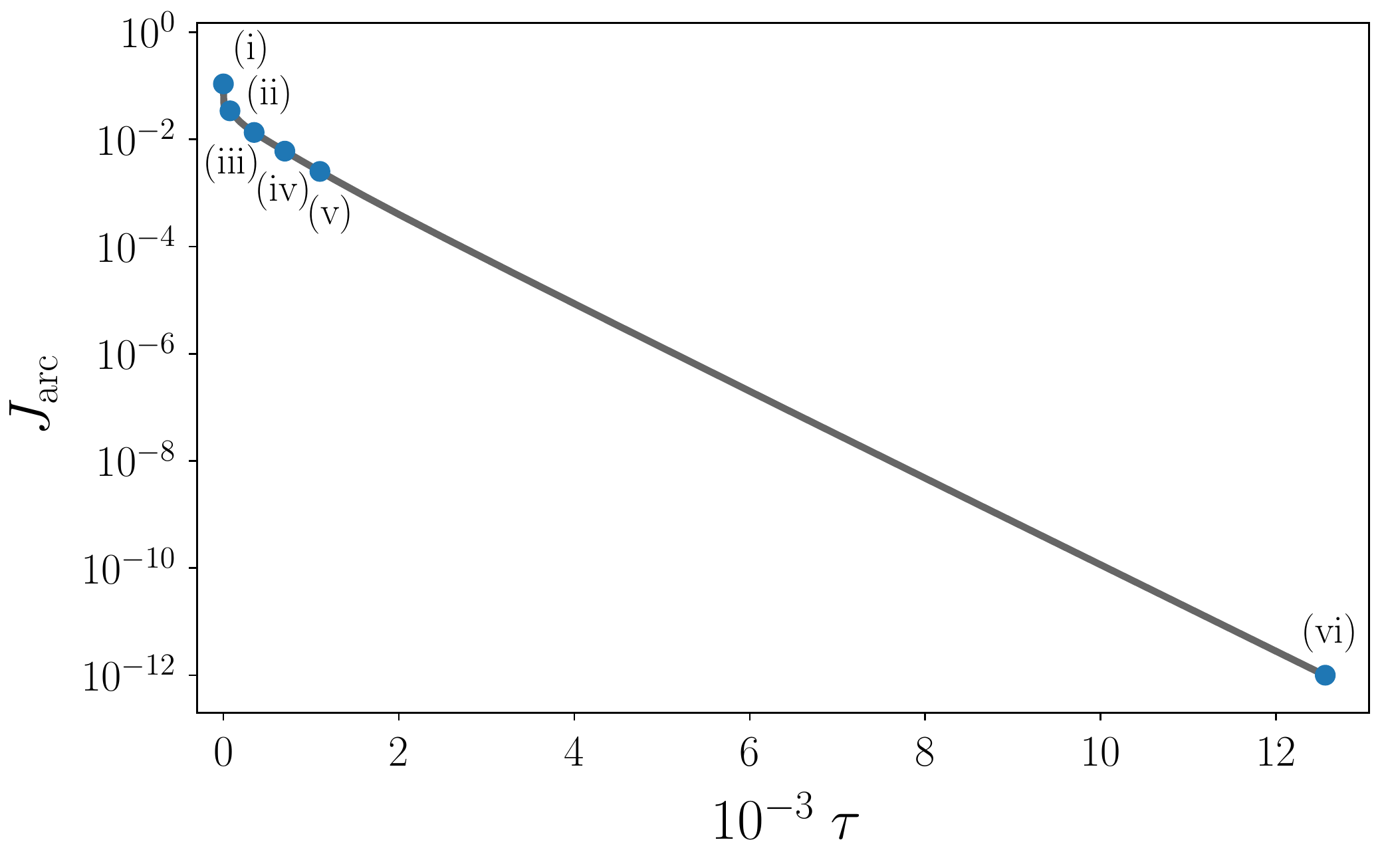}
    \caption{Monotonic decrease of the arc length cost function $J_\mathrm{arc}$ against the fictitious time $\tau$ as the dynamics in the space of connecting curves evolves an initial connecting curve towards a connecting orbit for which $J=0$. A three dimensional projection of the state space corresponding to the marked times (i) to (vi) is shown in Fig.~\ref{fig:deformation}.}
    \label{fig:residual}
\end{figure}

After a sharp initial decrease, the arc length cost function decays exponentially with the fictitious time, as shown in Fig.~\ref{fig:residual}, and reaches the convergence criterion, $J_\mathrm{arc}=10^{-12}$, at $\tau\approx1.25\times 10^4$. In the vector field induced by $G$, heteroclinic connections are attracting fixed points. The exponential decay of the cost function suggests that when the evolving connecting curve gets close enough to the connecting orbit, the dynamics is dominated by the leading, i.e. the slowest, eigendirection of the linearized dynamics in the vicinity of the fixed point of $\partial\mathbf{u}/\partial\tau=G$.

\begin{figure}
    \centering
    \includegraphics[width = \linewidth]{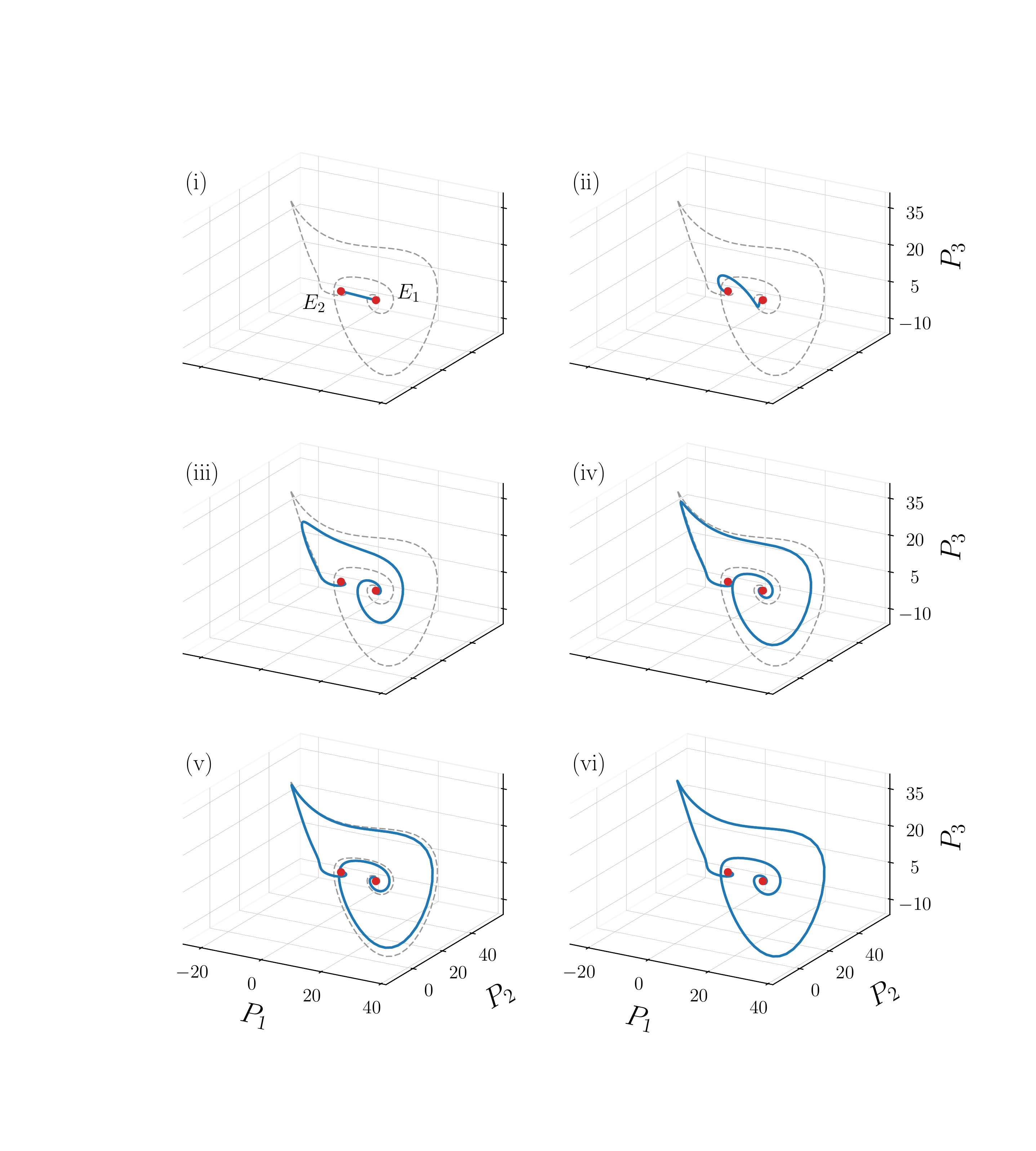}
    \caption{Continuous deformation of a connecting curve by the dynamics constructed in the space of connecting curves towards a heteroclinic connection from the fixed point $E_1$ to $E_2$. The solid blue line is the evolving connecting curve at the times marked on Fig.~\ref{fig:residual}, and the dashed line is the converged heteroclinic connection. The state space is projected on $P_k(s)=\Im\{\hat{u}_k(s)\};\;k=1,2,3$.}
    \label{fig:deformation}
\end{figure}

\begin{figure}
\centering
    \begin{subfigure}{\linewidth}\centering
        \includegraphics[width=0.8\linewidth]{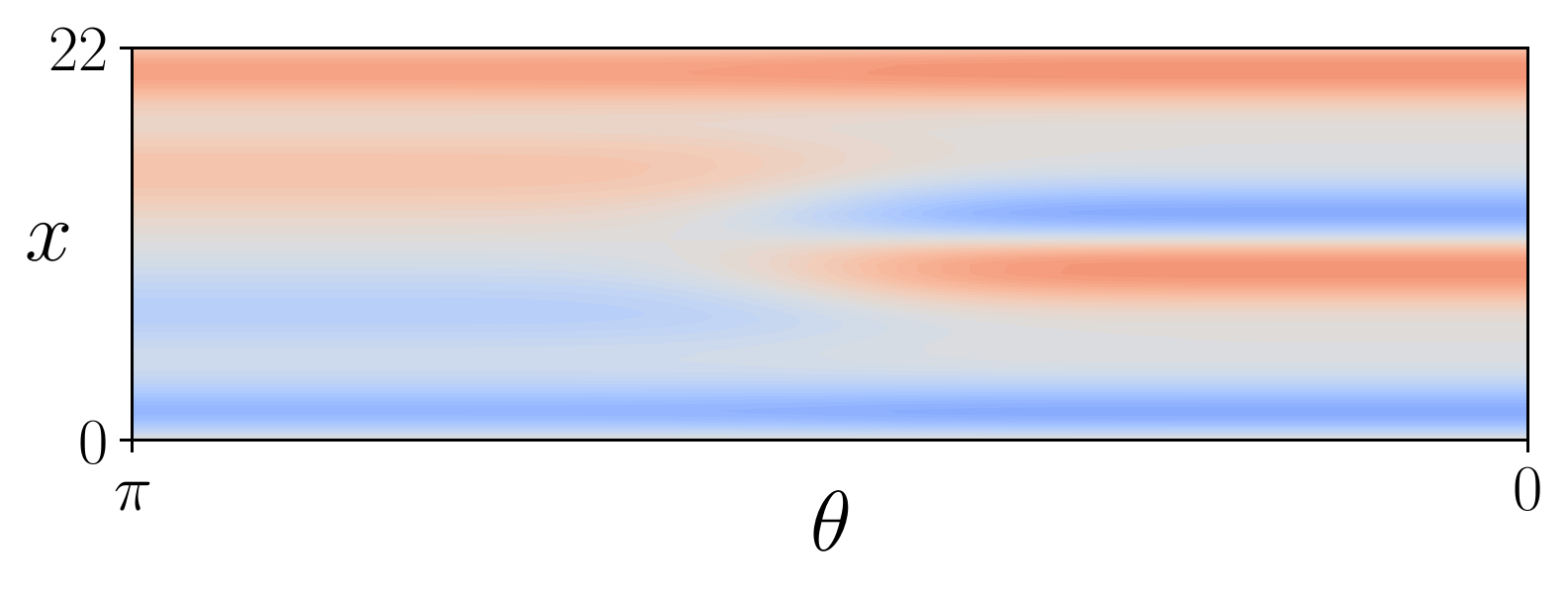}
        \caption{The initial connecting curve at $\tau=0$. See marker (i) on Fig.~\ref{fig:residual} and panel (i) of Fig.~\ref{fig:deformation}.}
    \end{subfigure}\\
    \begin{subfigure}{\linewidth}\centering
            \includegraphics[width=0.8\linewidth]{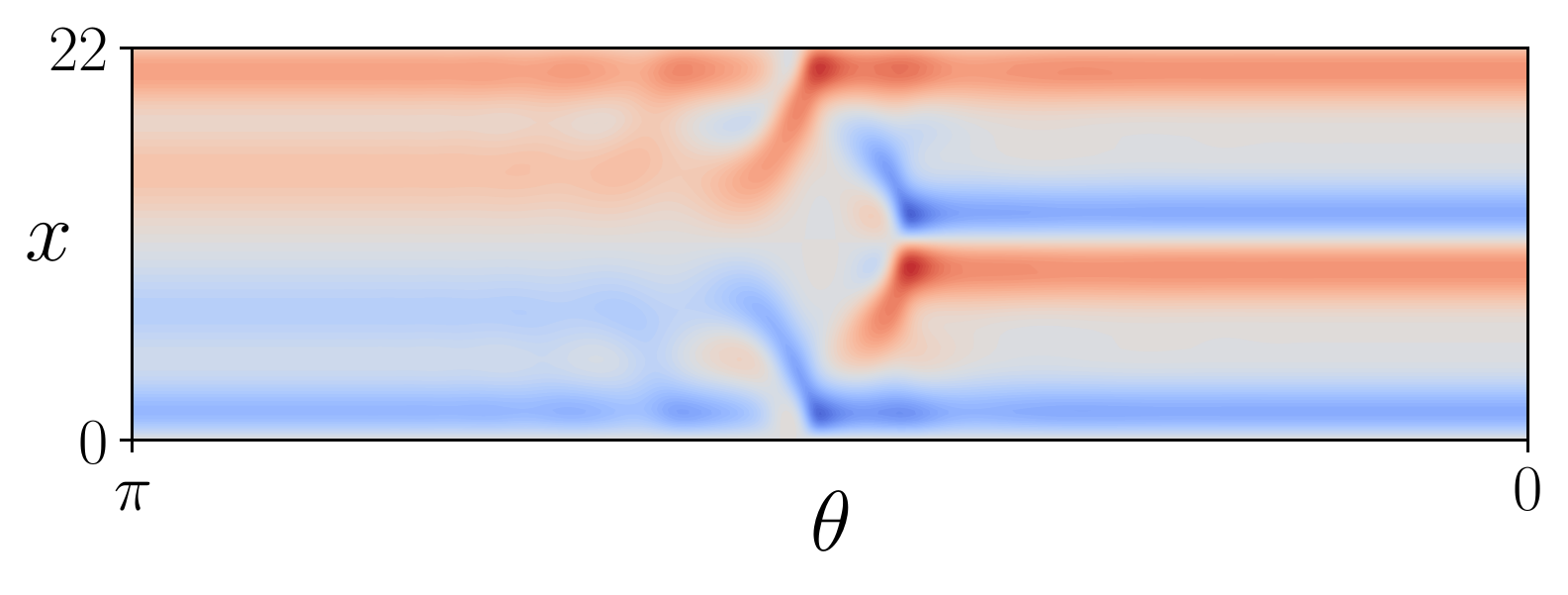}
        \caption{The converged connecting orbit at $\tau=1.25\times 10^4$. See marker (vi) on Fig.~\ref{fig:residual} and panel (vi) of Fig.~\ref{fig:deformation}.}
    \end{subfigure}
    \caption{The space-time contour of the initial connecting curve and the converged connecting orbit from the equilibrium solution $E_1$ to $E_2$. The initial connecting curve is symmetric under inversion about the origin. Since the dynamics in the space of connecting curves preserves the center symmetry, the converged connecting orbit belongs to center-symmetric subspace as well. The temporal dimension is mapped on the uniformly discretized finite interval $[\pi,0]$ where $\theta=\pi$ and $\theta=0$ correspond to $s\to-\infty$ and $s\to+\infty$, respectively (see Eq.~\eqref{eq:t_grid}).}
    \label{fig:initial_final_contours}
\end{figure}

Fig.~\ref{fig:deformation} shows six snapshots of the continuous deformation of the connecting curve from $E_1$ to $E_2$ governed by the dynamics in the space of connecting curves \eqref{eq:KSE_G} towards a heteroclinic connection. A substantial deformation towards the final shape of the connecting orbit takes place in the beginning of the evolution. The major remaining part of the integration time is spent on the slight remaining deviation from the final orbit. The space-time field corresponding to the initial connecting curve (snapshot (i) in Fig.~\ref{fig:deformation}) and the converged connecting orbit (snapshot (vi) in Fig.~\ref{fig:deformation}) are displayed in panels (a) and (b) of Fig.~\ref{fig:initial_final_contours}, respectively.

\begin{figure}
    \centering
    \includegraphics[width = \linewidth]{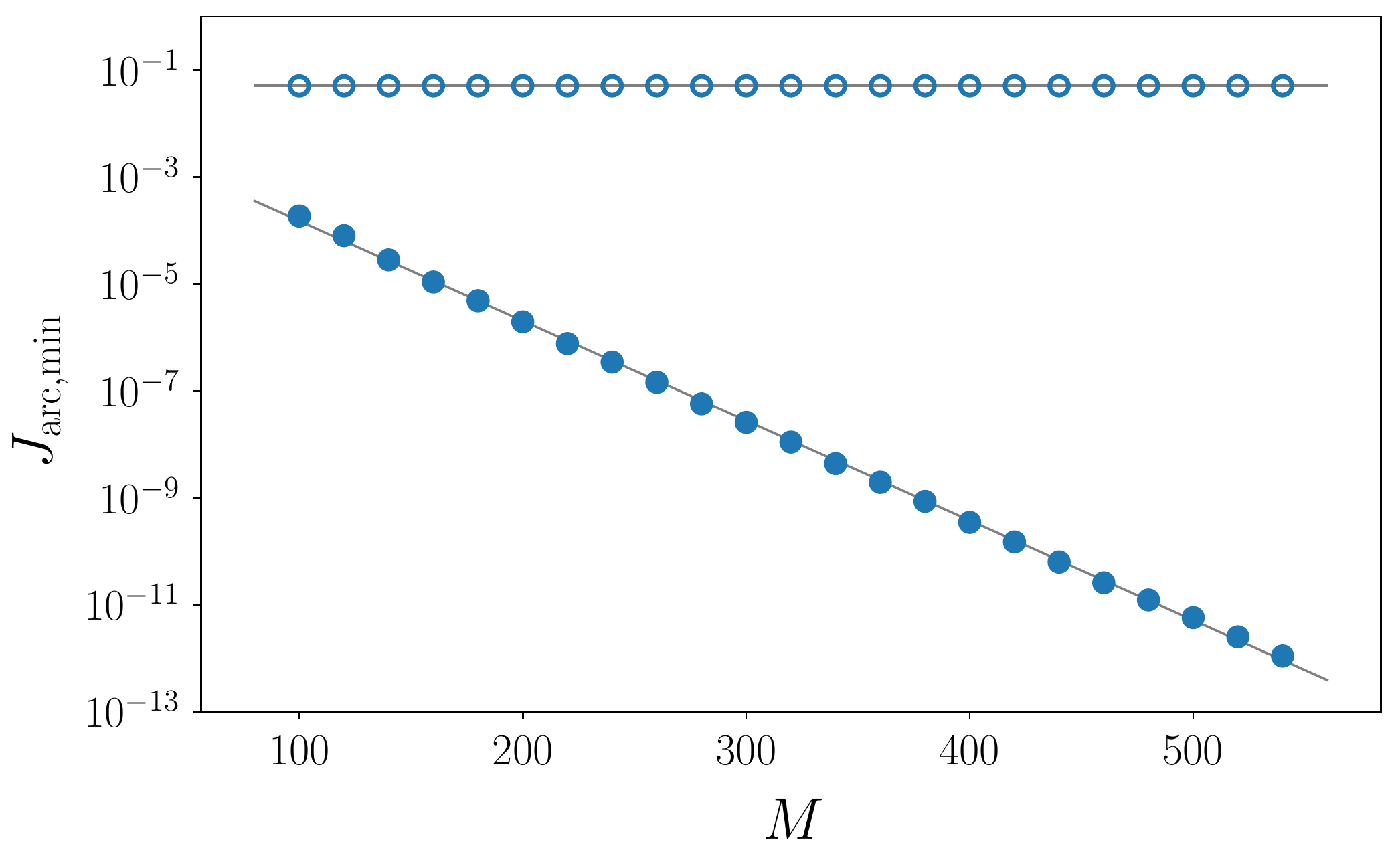}
    \caption{Variation of the asymptotic value of the arc length cost function $J_\mathrm{arc}$ by refining the temporal resolution $M$. Filled circles: Exponential decrease of $J_\mathrm{arc,min}$ to zero in successfully converging to a connecting orbit from $E_1$ to $E_2$. Open circles: The cost function getting stuck in a local minimum in the failed search for a connecting orbit from $E_2$ to $\gamma(1/4)E_2$ in an over-constrained subspace.}
    \label{fig:spectral_accuracy}
\end{figure}

The spatial resolution is chosen by monitoring the energy spectrum of spatial Fourier modes in a direct numerical simulation of the KSE for $L=22$. The spatial resolution $N=64$ ensures at least six orders of magnitude drop in the modulus of spatial Fourier coefficients at all times. The converged connecting orbit from $E_1$ to $E_2$, as an equilibrium solution to Eq.~\eqref{eq:KSE_G}, is structurally stable for a wide range of temporal resolutions $M$. However, the accuracy of the spectral representation in time, and therefore the minimum achieved value of the cost function, $J_\mathrm{arc,min}:=\lim_{\tau\to\infty}J_\mathrm{arc}(\tau)$, varies with $M$. Fig.~\ref{fig:spectral_accuracy} show the spectral convergence of $J_\mathrm{arc,min}$ with $M$. Notice that $J_\mathrm{arc,min}$ can be considerably higher than the convergence criterion when $M$ is not large enough. If a local minimum of the cost function is reached, in contrast, $J_\mathrm{arc,min}$ does not improve as the temporal resolution is increased. As an example of a failing search, we try to converge a connecting orbit between $E_2$ and $\gamma(1/4)E_2$ from an initial connecting curve constructed using Eq.~\eqref{eq:KSE_initial_curve} with $a=0$ (see Sec.~\ref{sec:orbits_from_E2} why such connection cannot exist). The integration from this initial connecting curve does not reach a global minimum but approaches a local minimum with $J_\mathrm{arc,min}=5.1\times10^{-2}$. As shown on Fig.~\ref{fig:spectral_accuracy}, the minimum value does not decrease as the temporal discretization is refined, confirming that a converged local minimum has been identified and no connecting orbit was found. 

\subsubsection{Connecting orbits originating from $E_0$: Six-dimensional unstable manifold}
\begin{figure}
\centering
    \begin{subfigure}{\linewidth}
        \includegraphics[width=0.65\linewidth]{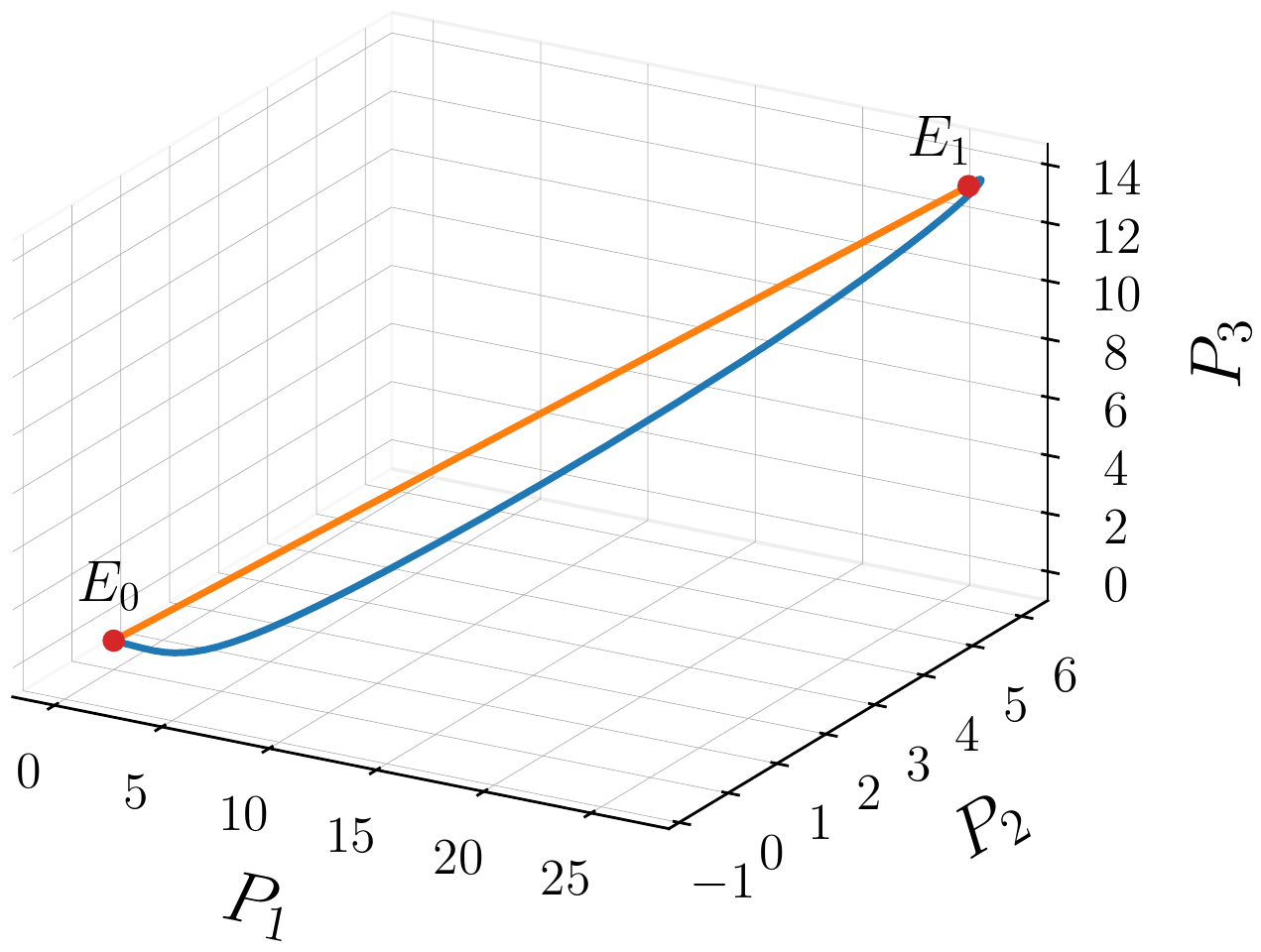}
        \caption{From $E_0$ to $E_1$}
    \end{subfigure}\\
    \begin{subfigure}{\linewidth}
        \includegraphics[width=0.65\linewidth]{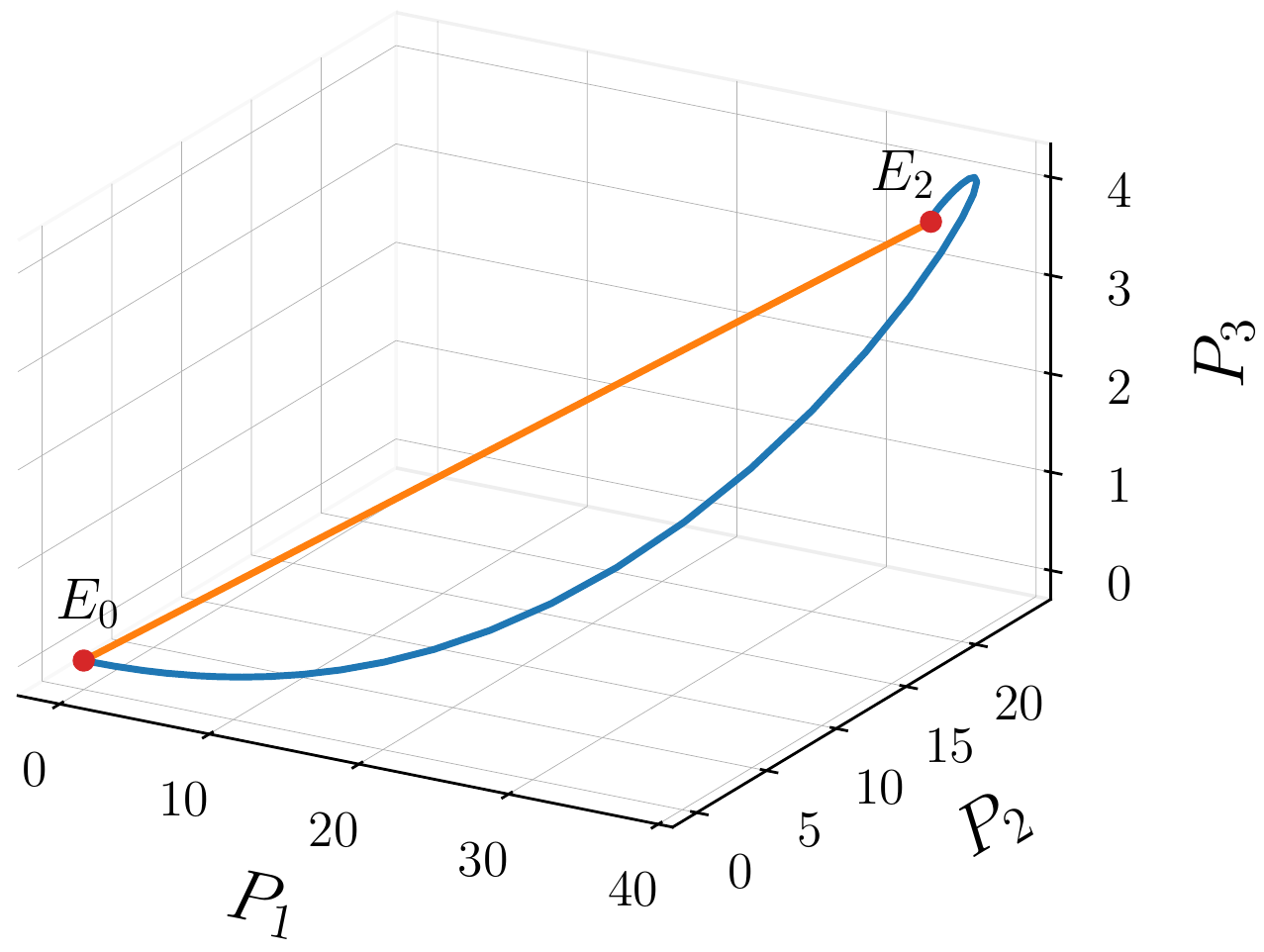}
        \caption{From $E_0$ to $E_2$}
    \end{subfigure}\\
    \begin{subfigure}{\linewidth}
        \includegraphics[width=0.65\linewidth]{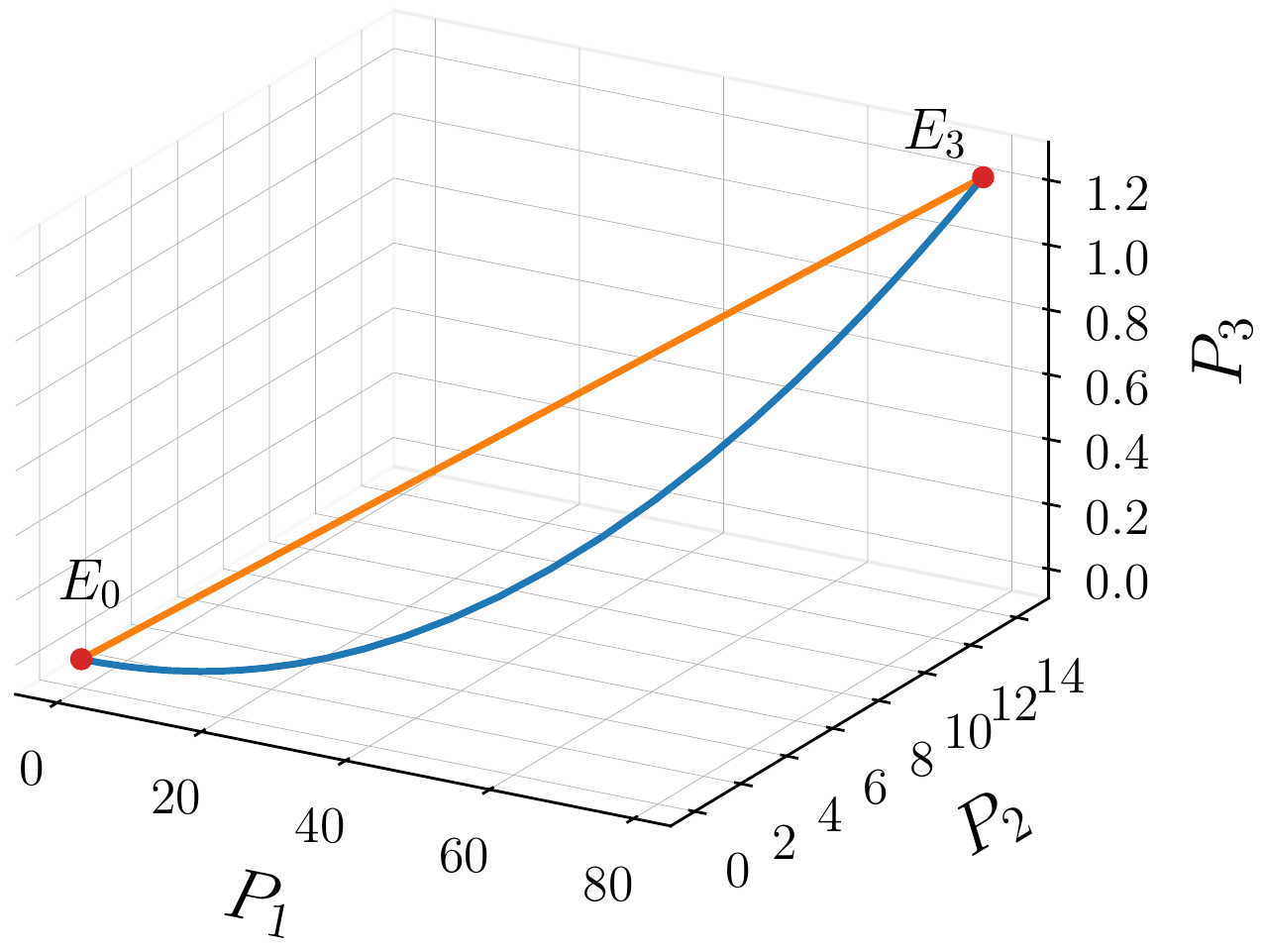}
        \caption{From $E_0$ to $E_3$}
    \end{subfigure}
    
    \caption{Connecting orbits from $E_0$ to the other three equilibrium solutions in the center-symmetric subspace. The orange line shows the initial connecting curve, and the blue line shows the converged connecting orbit. The state space is projected on $P_k(s)=\Im\{\hat{u}_{bk}(s)\};\;k=1,2,3$ with $b=1$ in (a), $b=2$ in (b), and $b=3$ in (c).}
    \label{fig:state_space_from_E0}
\end{figure}

We converge a heteroclinic connection from $E_0$ to $E_1$, $E_2$ and $E_3$ from an initial connecting curve constructed using Eq.~\eqref{eq:KSE_initial_curve} with $a=0$. A three-dimensional state space projection and the space-time contour of heteroclinic connections from $E_0$ to the other three equilibrium solutions are exhibited in Figs.~\ref{fig:state_space_from_E0} and \ref{fig:contour_from_E0}, respectively. The algorithm settings are presented in Appendix \ref{sec:parameters}.

The unstable manifold of $E_0$ is six-dimensional. Each of the repeated unstable eigenvalues of $E_0$, Table \ref{tab:KSE_eigenvals}, is associated to one eigenvector symmetric under reflection across $x=0$ and another one symmetric under inversion about the origin (see Fig.~\ref{fig:eigenvectors_E0}). An exhaustive search in the unstable tangent space at $E_0$ is not practical even in the inversion-symmetric subspace of the KSE where the reflection-symmetric eigenvectors do not exist, and the unstable manifold is three-dimensional. Dong and Lan\cite{dong2014} have computed a heteroclinic connection from $E_0$ to $E_1$ using their variational method which employs finite differences for calculating tangent velocity vectors. They have used \numprint{6000} sections to discretize this connecting orbit in time, and obtain residuals of order $\mathcal{O}(10^{-6})$. To achieve this value of $J_\mathrm{arc}$ (and similarly the suprimum norm of the residual $r$), $M=25$ interior time sections suffice for the proposed variational method.

\begin{figure}
\centering
    \begin{subfigure}{\linewidth}
        \includegraphics[width=0.8\linewidth]{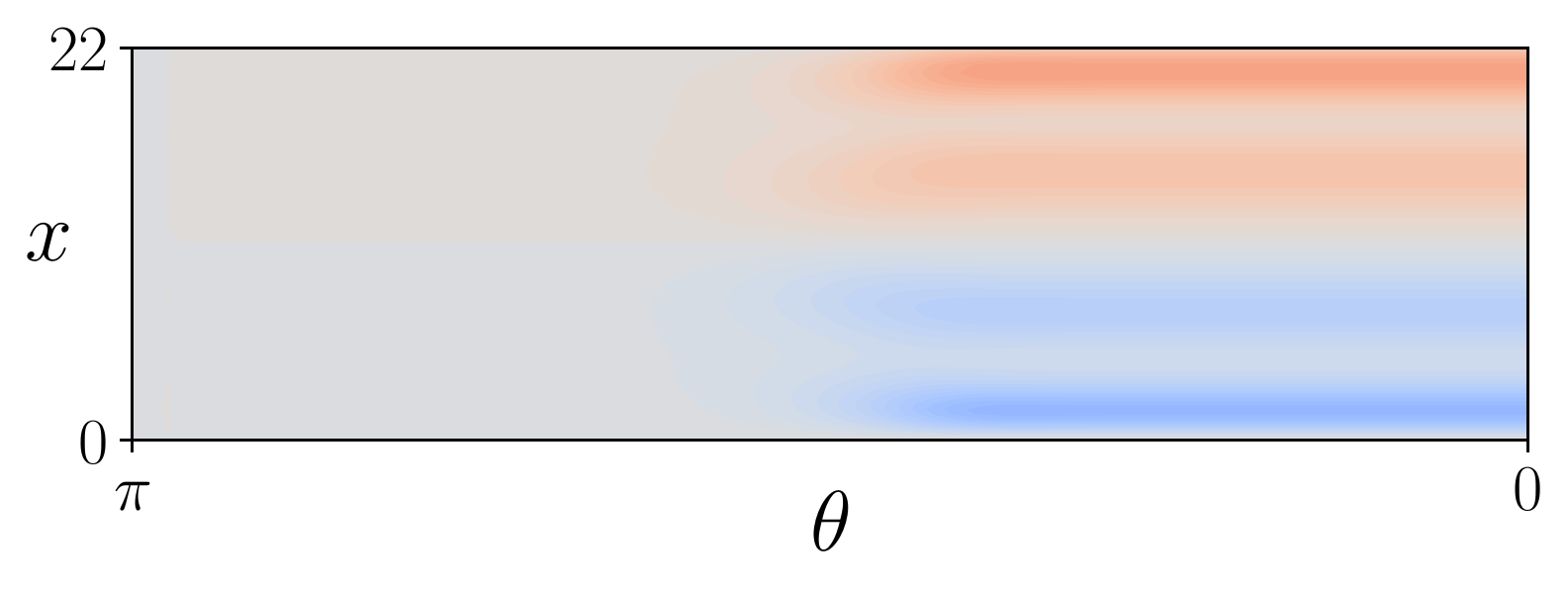}
        \caption{From $E_0$ to $E_1$}
    \end{subfigure}\\
    \begin{subfigure}{\linewidth}
        \includegraphics[width=0.8\linewidth]{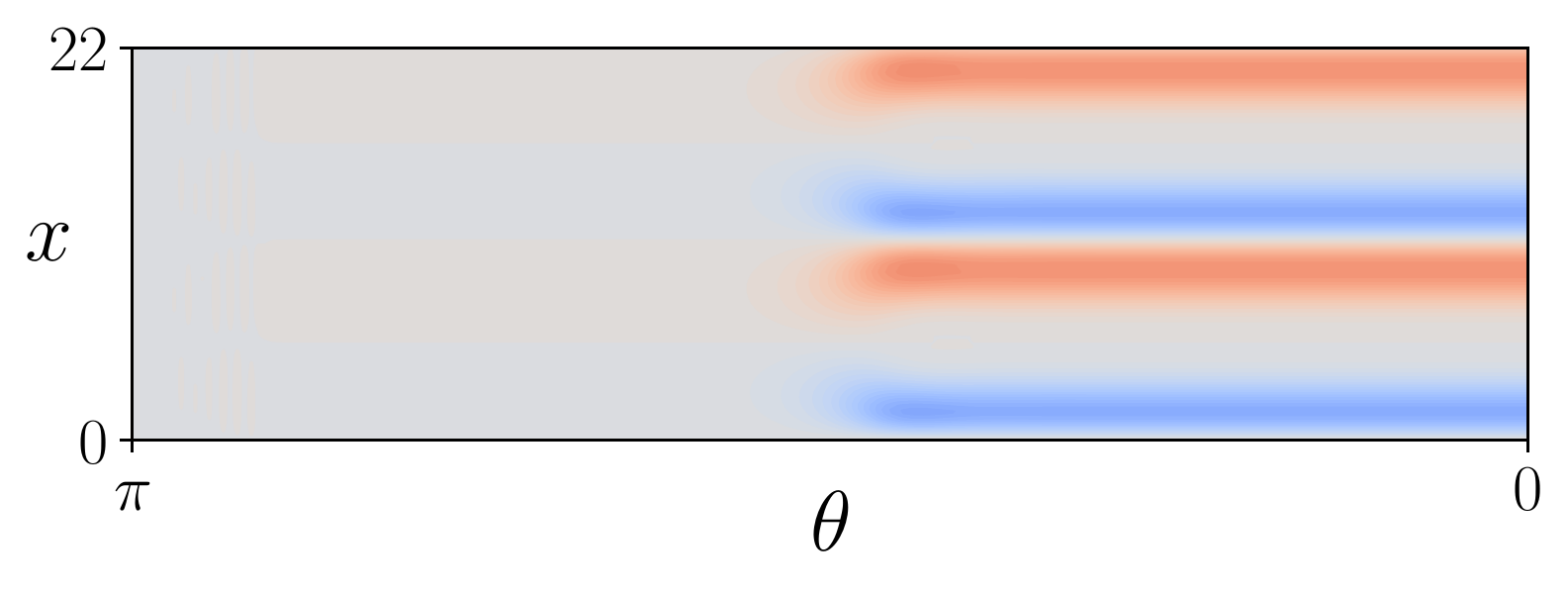}
        \caption{From $E_0$ to $E_2$}
    \end{subfigure}\\
    \begin{subfigure}{\linewidth}
        \includegraphics[width=0.8\linewidth]{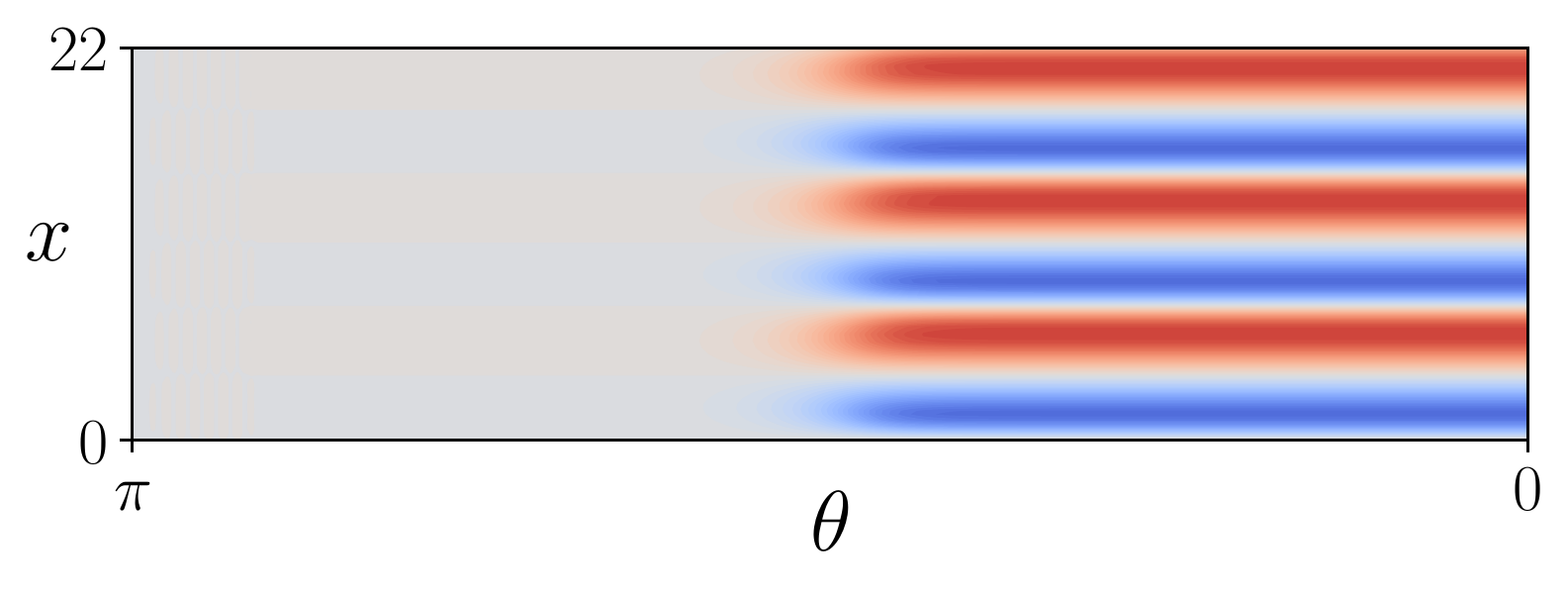}
        \caption{From $E_0$ to $E_3$}
    \end{subfigure}
    
    \caption{The space-time contour of the converged connecting orbits from $E_0$ to the other three equilibrium solutions in the center-symmetric subspace.}
    \label{fig:contour_from_E0}
\end{figure}

\subsubsection{Connecting orbits originating from $E_1$: Four-dimensional unstable manifold}
\begin{figure}
\centering
    \includegraphics[width=0.65\linewidth]{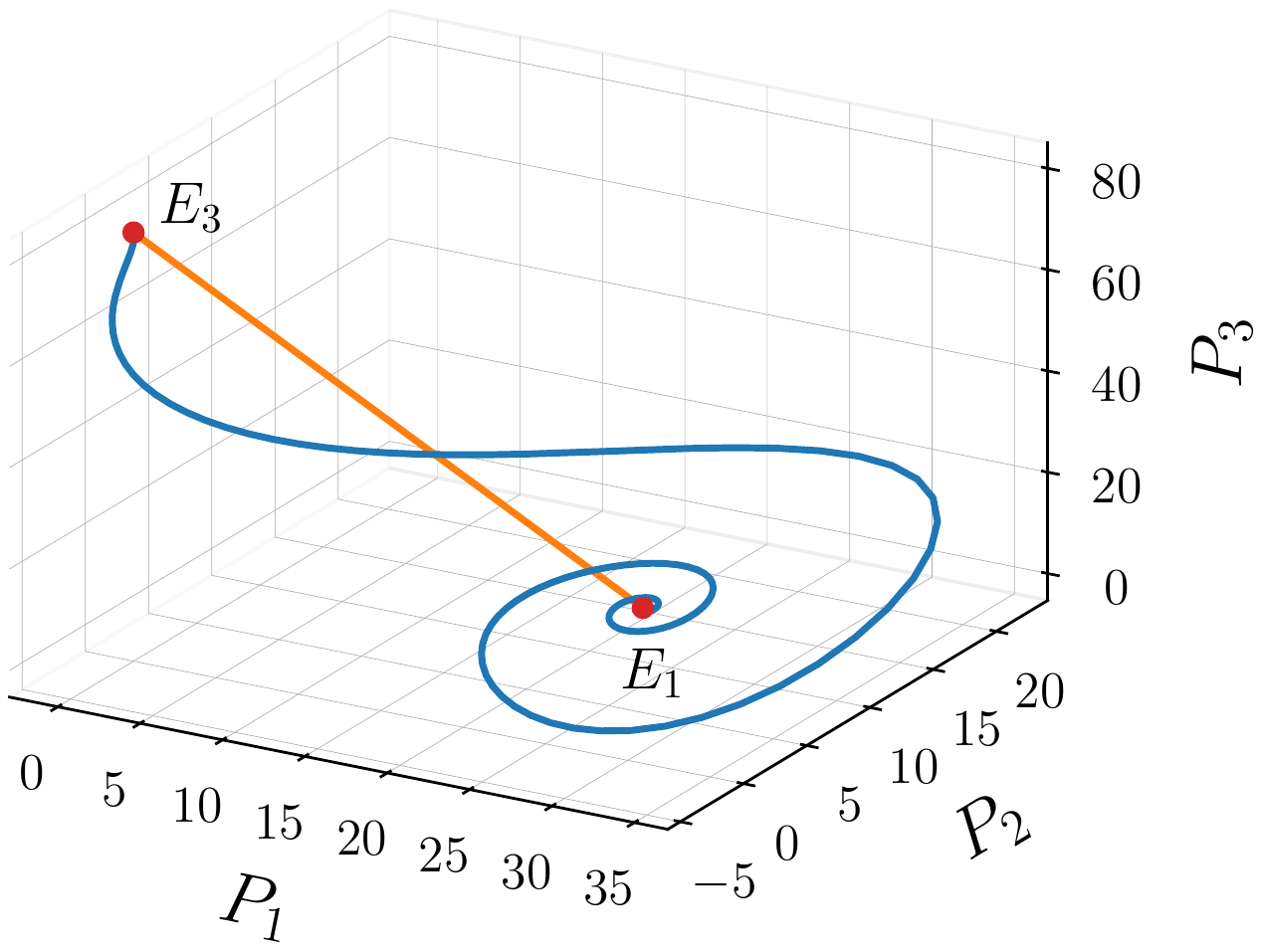}
    \caption{Connecting orbit from $E_1$ to $E_3$ in the center-symmetric subspace. The orange line shows the initial connecting curve, and the blue line shows the converged connecting orbit. The state space is projected on $P_k(s)=\Im\{\hat{u}_{k}(s)\};\;k=1,2,3$.}
    \label{fig:state_space_from_E1}
\end{figure}

\begin{figure}
\centering
    \includegraphics[width=0.8\linewidth]{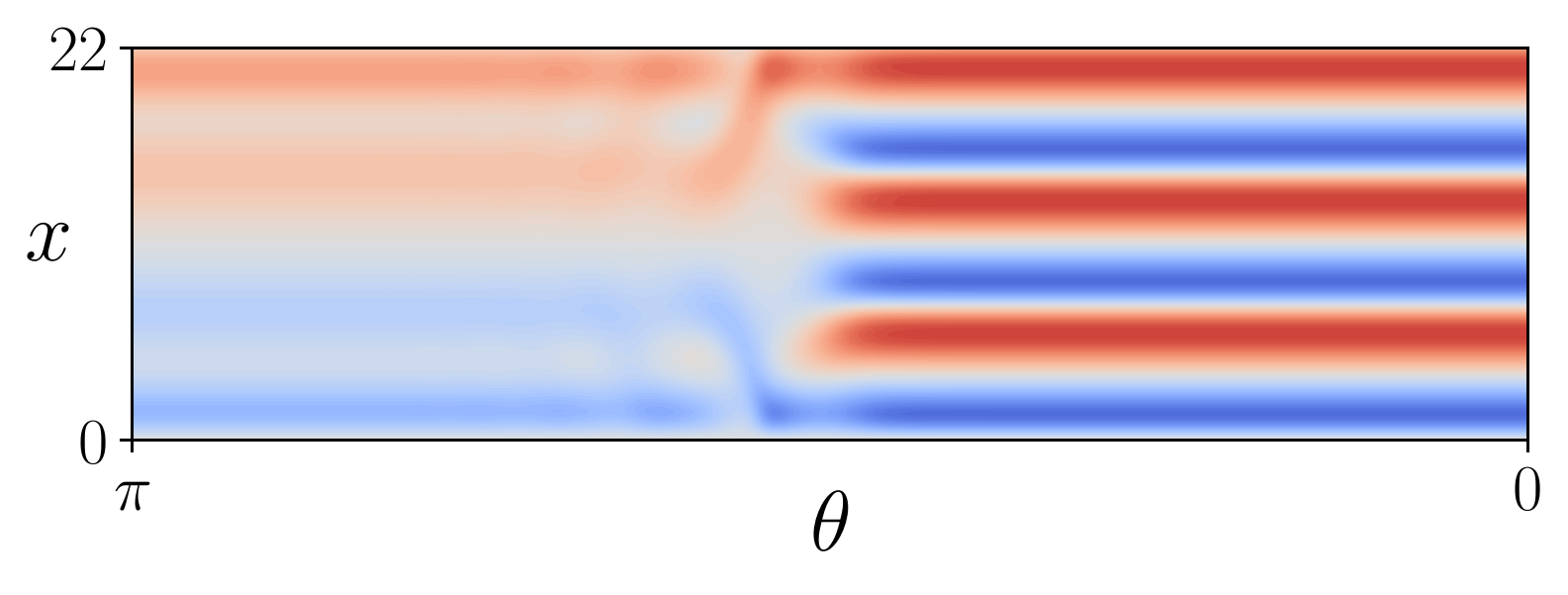}
    \caption{The space-time contour of the converged connecting orbit from $E_1$ to $E_3$ in the center-symmetric subspace.}
    \label{fig:contour_from_E1}
\end{figure}

We demonstrated the details of converging a heteroclinic connection from $E_1$ to $E_2$ in the beginning of this section (see Figs.~\ref{fig:residual} to \ref{fig:spectral_accuracy}). We also converge a heteroclinic connection from $E_1$ to $E_3$ from an initial connecting curve constructed using Eq.~\eqref{eq:KSE_initial_curve} with $a=0$. Figs.~\ref{fig:state_space_from_E1} and \ref{fig:contour_from_E1} show a three-dimensional state space projection and the space-time contour plot of the converged heteroclinic connection from $E_1$ to $E_3$, respectively. The algorithm settings are presented in Appendix \ref{sec:parameters}.

The unstable manifold of $E_1$ is four-dimensional. One pair of complex conjugate unstable eigenvalues of $E_1$, Table \ref{tab:KSE_eigenvals}, is associated to eigenvectors invariant under reflection across $x=0$, while the other pair is associated to eigenvectors invariant under inversion about the origin (see Fig.~\ref{fig:eigenvectors_E1}). An exhaustive search in the four-dimensional unstable tangent space at $E_1$ is not practical. Cvitanovi\'c \textit{et al.}\cite{Cvitanovic2010a} perform an exhaustive search in the two-dimensional plane spanned by the reflection-symmetric eigenvectors at $E_1$, and show that all trajectories starting from that plane are chaotic, and do not reach any of the equilibrium solutions. They perform another exhaustive search in the two-dimensional plane spanned by the inversion-symmetric eigenvectors, and show that trajectories starting from that plane form a one-parameter family of heteroclinic connections from $E_1$ to $E_2$, except one bordering orbit that converges to $E_3$.  

\subsubsection{Connecting orbits originating from $E_2$: Two-dimensional unstable manifold}
\label{sec:orbits_from_E2}
\begin{figure}
\centering
    \begin{subfigure}{\linewidth}
        \includegraphics[width=0.65\linewidth]{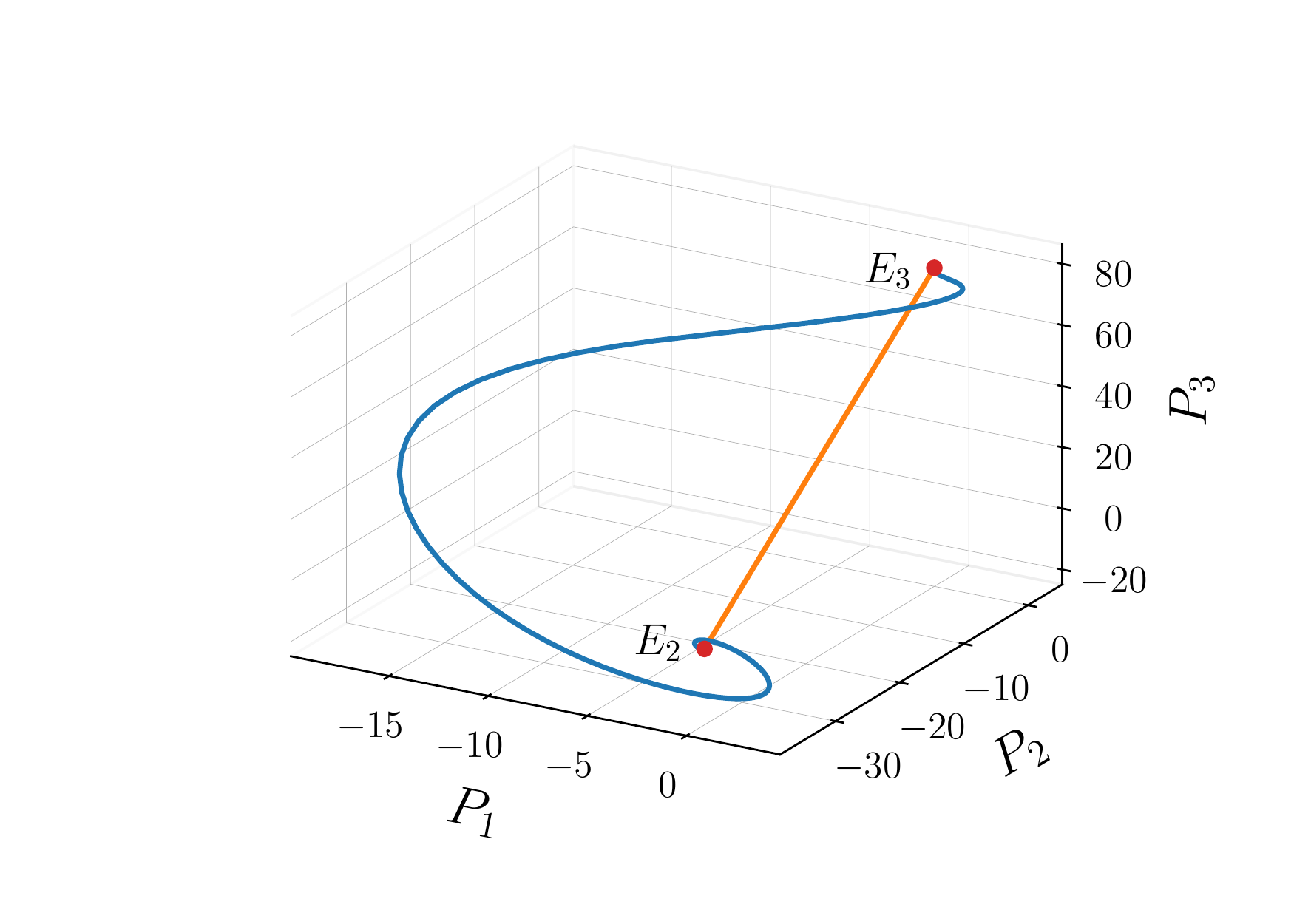}
        \caption{From $E_2$ to $E_3$}
    \end{subfigure}\\
    \begin{subfigure}{\linewidth}
        \includegraphics[width=0.65\linewidth]{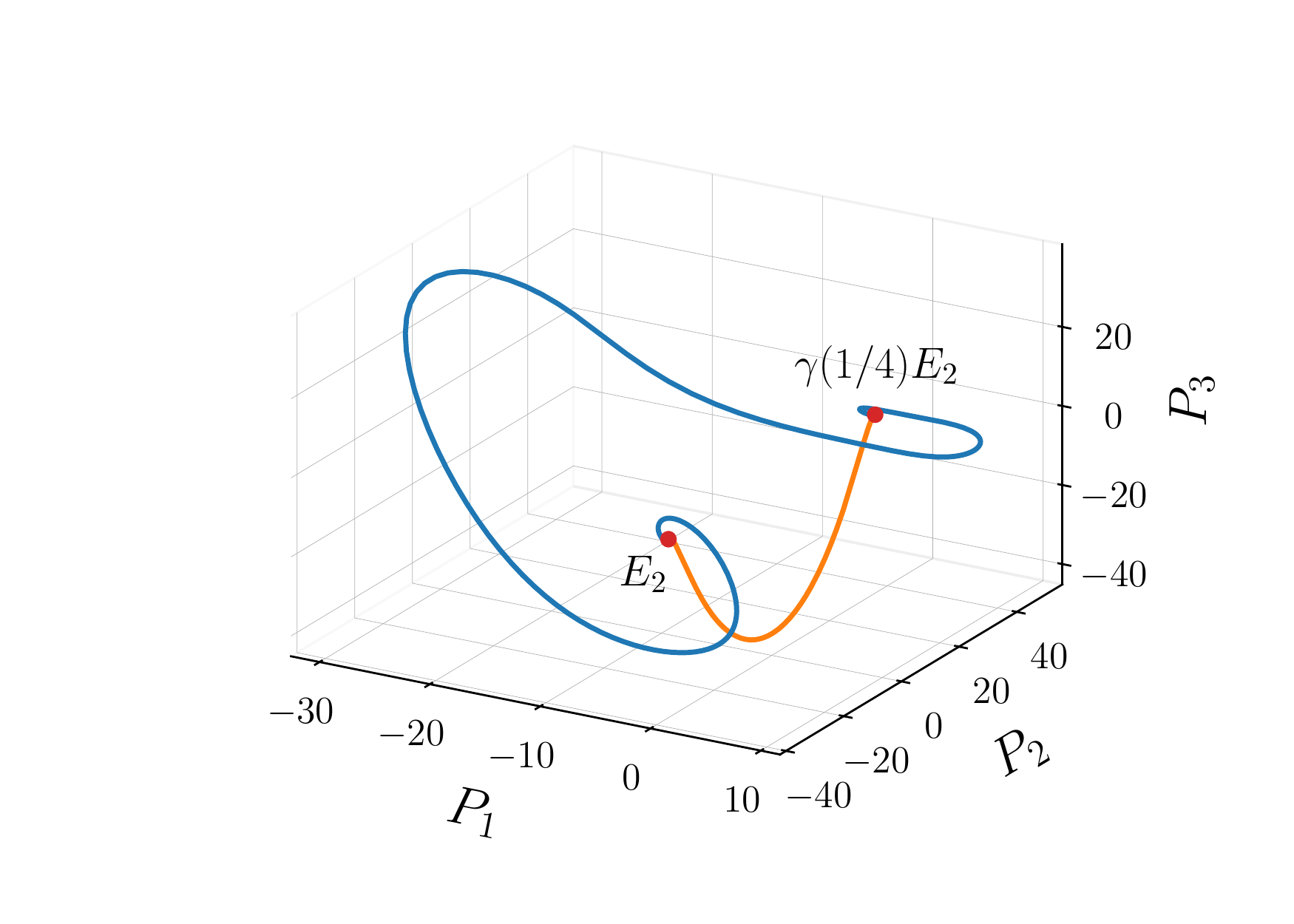}
        \caption{From $E_2$ to $\gamma(1/4)E_2$}
    \end{subfigure}
    
    \caption{Connecting orbits from $E_2$ to $E_3$ and $\gamma(1/4)E_2$ in the center-symmetric subspace. The orange line shows the initial connecting curve, and the blue line shows the converged connecting orbit. The state space is projected on $P_k(s)=\Im\{\hat{u}_{k}(s)\};\;k=1,2,3$.}
    \label{fig:state_space_from_E2}
\end{figure}

We converge two heteroclinic connections from $E_2$ to $E_3$ and $\gamma(1/4)E_2$. The initial conditions are constructed using Eq.~\eqref{eq:KSE_initial_curve} by setting $a=0$ for the connecting orbit between $E_2$ and $E_3$, and setting $a=-1$ and $v=\Re\{v_{1,2}\}$ for the connecting orbit between $E_2$ and $\gamma(1/4)E_2$ where $\Re\{v_{1,2}\}$ is the real part of the complex conjugate unstable eigenvectors at $E_2$ (see Fig.~\ref{fig:eigenvectors_E2}). In the latter, adding the symmetry breaking term ($a\neq0$) is necessary because $E_2$ and $\gamma(1/4)E_2$ are both symmetric under inversion about $x=kL/4$ with $k=0,1,2,3$, thus an initial connecting curve constructed by setting $a=0$ is symmetric under inversion about all these points. The dynamics \eqref{eq:KSE_G} preserves all the four inversion symmetries while no connecting orbit can exist in such subspace of $\mathcal{M}$, because the unstable eigenvectors of $E_2$ are symmetric only about $x=0$ and $L/2$, meaning that as soon as a trajectory of the KSE leaves $E_2$, the inversion symmetries about $x=L/4$ and $3L/4$ are broken. Consequently, as shown on Fig.~\ref{fig:spectral_accuracy}, $a=0$ results in getting stuck in a local minimum of the cost function as the dynamics \eqref{eq:KSE_G} is integrated. A three-dimensional state space projection and the space-time contour plot of the connecting orbits from $E_2$ to $E_3$ and $\gamma(1/4)E_2$ are shown in Figs.~\ref{fig:state_space_from_E2} and \ref{fig:contour_from_E2}, respectively. The algorithm settings are presented in Appendix \ref{sec:parameters}.

By an exhaustive search in the two-dimensional unstable tangent space at $E_2$, Cvitanovi\'c \textit{et al.}\cite{Cvitanovic2010a} show that the unstable manifold of $E_2$ is a one-parameter family of connecting orbits that converge to $\gamma(1/4)E_2$, except one orbit that connects $E_2$ to $E_3$.

\begin{figure}
\centering
    \begin{subfigure}{\linewidth}
        \includegraphics[width=0.8\linewidth]{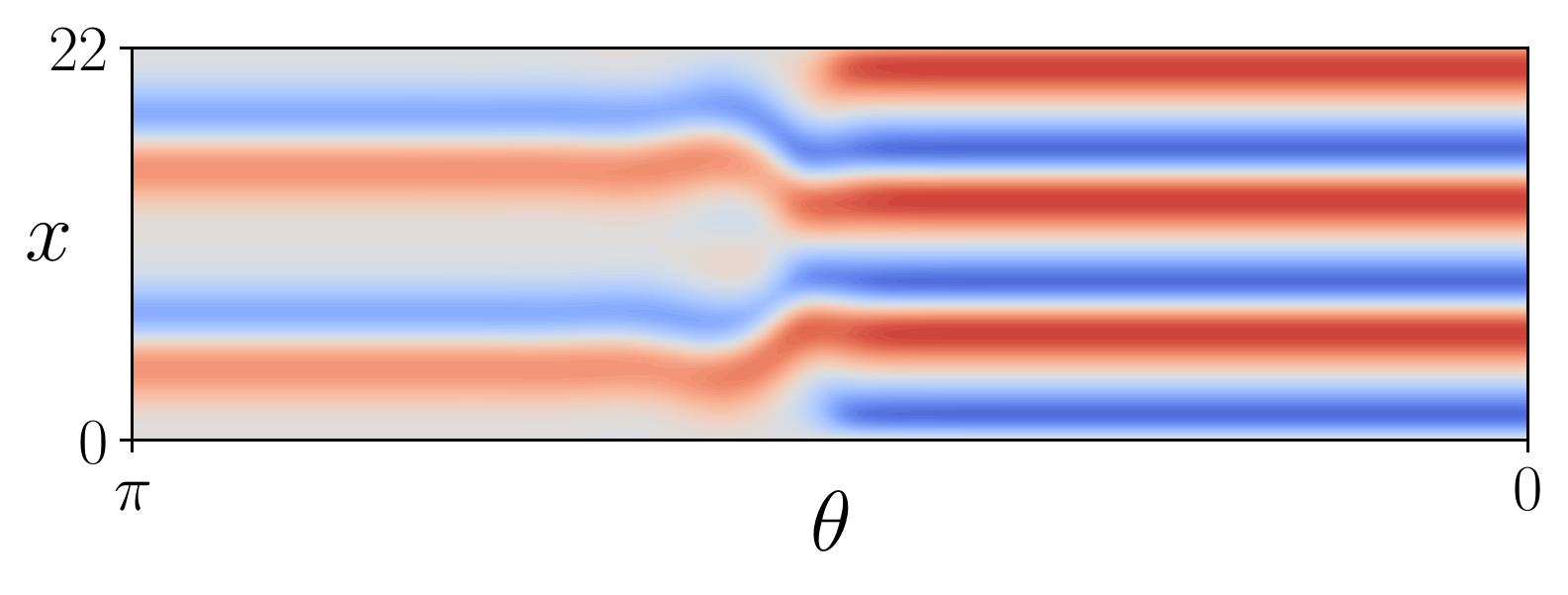}
        \caption{From $E_2$ to $E_3$}
    \end{subfigure}\\
    \begin{subfigure}{\linewidth}
        \includegraphics[width=0.8\linewidth]{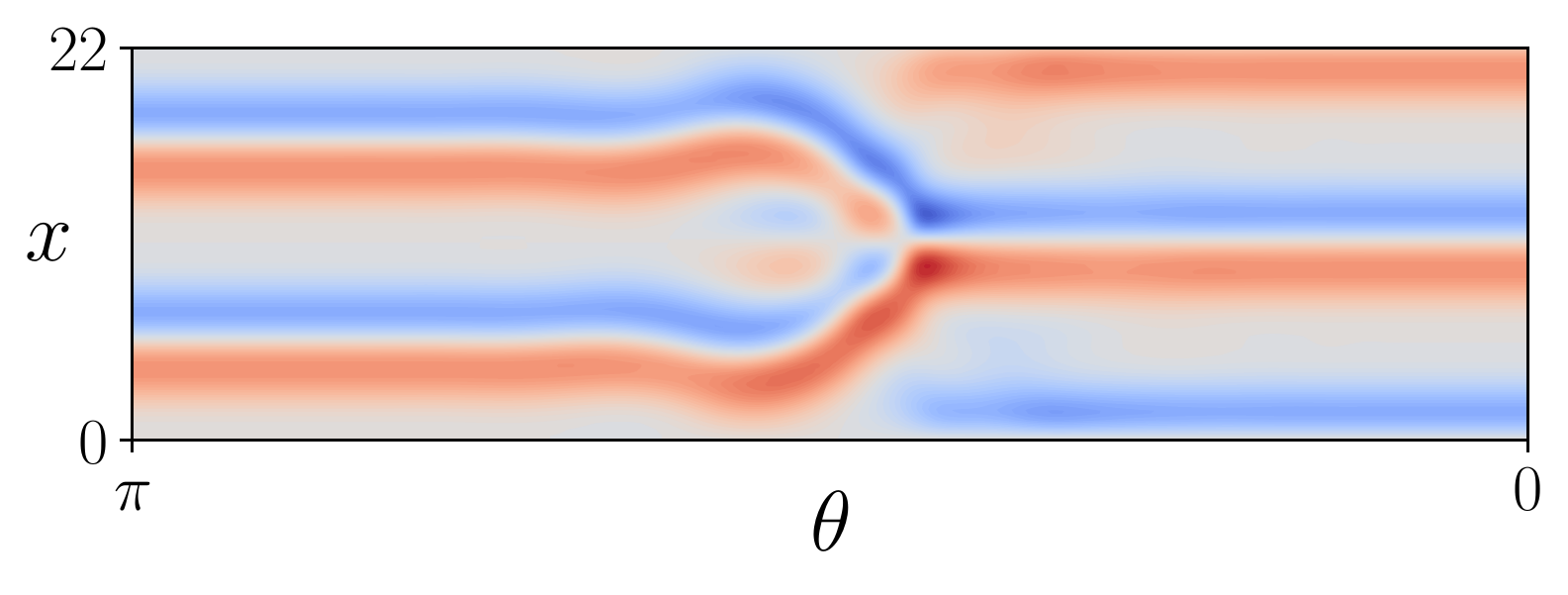}
        \caption{From $E_2$ to $\gamma(1/4)E_2$}
    \end{subfigure}
    
    \caption{The space-time contour of the converged connecting orbits from $E_1$ to $E_3$ and $\gamma(1/4)E_2$ in the center-symmetric subspace.}
    \label{fig:contour_from_E2}
\end{figure}

\subsubsection{Connecting orbits originating from $E_3$: Two-dimensional unstable manifold}
\begin{figure}
\centering
    \begin{subfigure}{\linewidth}
        \includegraphics[width=0.65\linewidth]{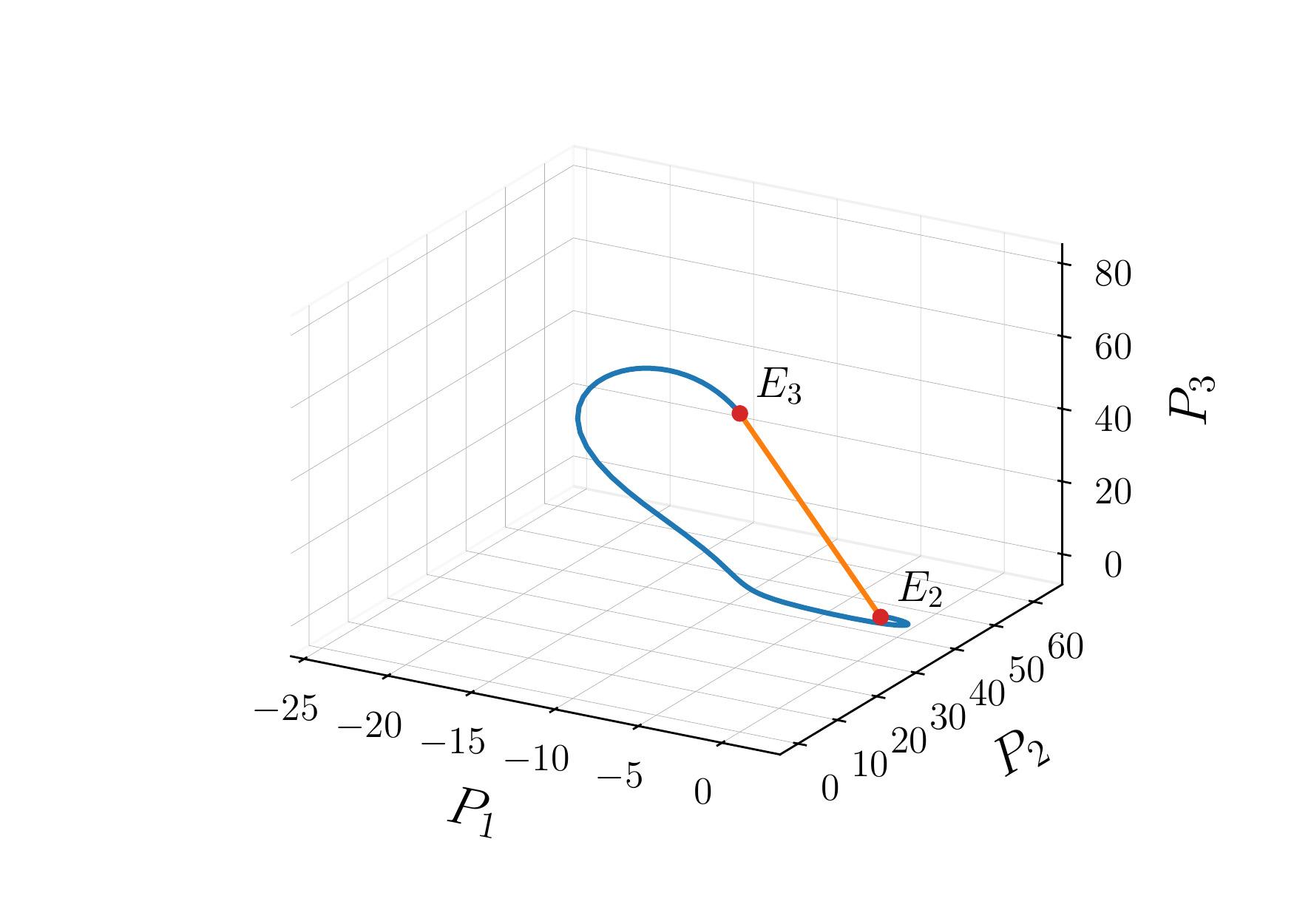}
        \caption{Orbit 1: The initial connecting curve is constructed via Eq.~\eqref{eq:KSE_initial_curve} by setting $a=0$.}
    \end{subfigure}\\
    \begin{subfigure}{\linewidth}
        \includegraphics[width=0.65\linewidth]{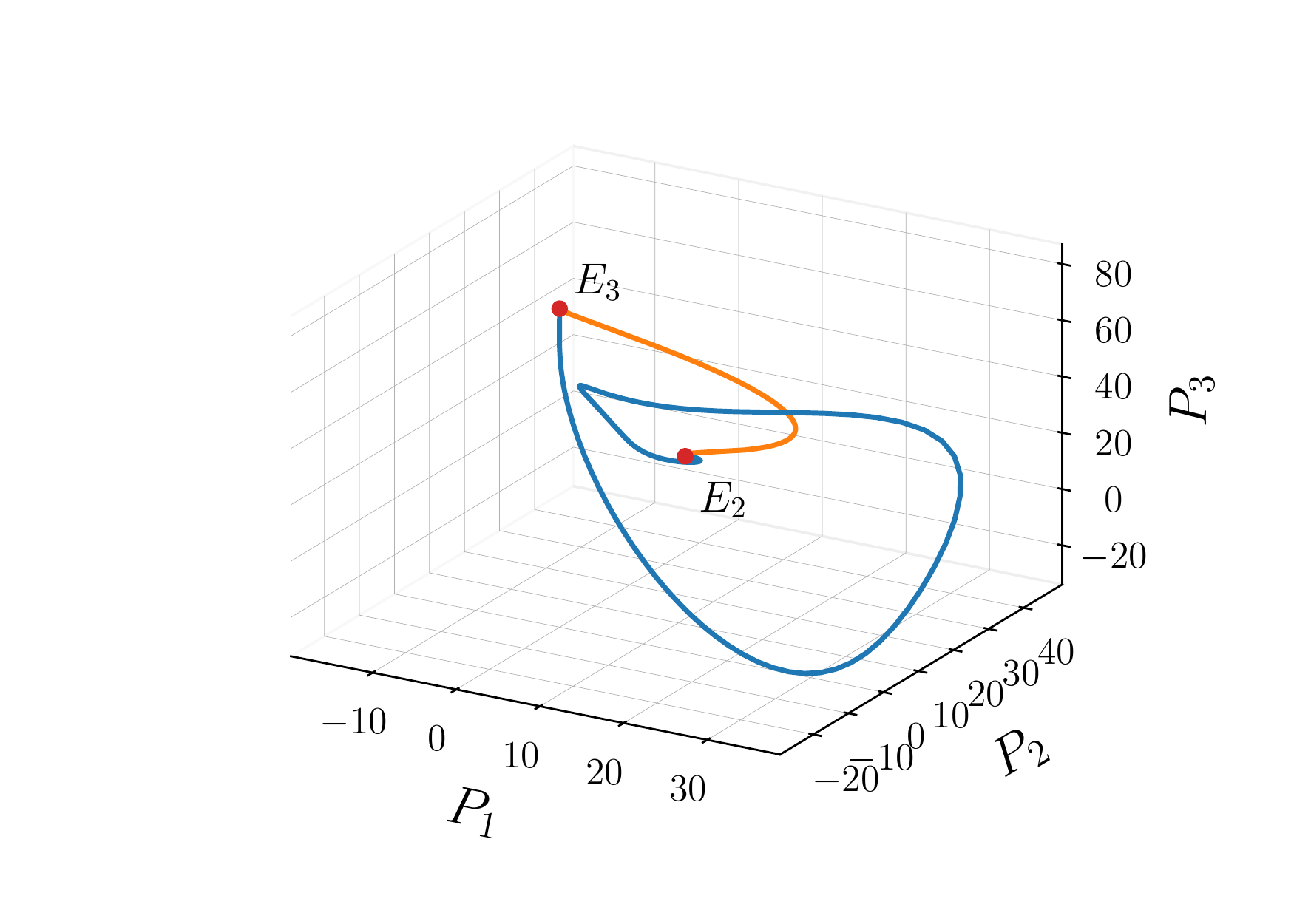}
        \caption{Orbit 2: The initial connecting curve is constructed via Eq.~\eqref{eq:KSE_initial_curve} by setting $a=-1$ and $v=\sin{(x)}$.}
    \end{subfigure} \hfill
    
    \caption{Two connecting orbits from $E_3$ to $E_2$ in the center-symmetric subspace. The orange line shows the initial connecting curve, and the blue line shows the converged connecting orbit. The state space is projected on $P_k(s)=\Im\{\hat{u}_{k}(s)\};\;k=1,2,3$.}
    \label{fig:state_space_from_E3}
\end{figure}

We converge two heteroclinic connections from $E_3$ to $E_2$. The initial conditions are constructed using Eq.~\eqref{eq:KSE_initial_curve} by setting $a=0$ in one, and $a=-1$ and $v=\sin{(x)}$ in the other. A three-dimensional state space projection and the space-time contour of these connecting orbits are shown in Figs.~\ref{fig:state_space_from_E3} and \ref{fig:contour_from_E3}, respectively. The algorithm settings are presented in Appendix \ref{sec:parameters}.

The unstable manifold of $E_3$ is two-dimensional. The repeated positive eigenvalue of $E_3$, Table \ref{tab:KSE_eigenvals}, is associated to one eigenvector symmetric under reflection across $x=0$, and another eigenvector symmetric under inversion about the origin (see Fig.~\ref{fig:eigenvectors_E3}). Cvitanovi\'c \textit{et al.}\cite{Cvitanovic2010a} conduct an exhaustive search in the two-dimensional unstable tangent space at $E_3$, and identify two heteroclinic connections from $E_3$ to $E_2$ corresponding to the perturbation of $E_3$ along the inversion-symmetric eigenvector and its opposite direction. Fixing $E_3$ and shifting $E_2$ in space by $L/3$ and $2L/3$ puts the translated copy of $E_2$ in the same relative phase to $E_3$ as the original configuration. Therefore, the exhaustive search identifies two other pairs of heteroclinic connections from $E_3$ to the group orbit of $E_2$, which are copies of the first pair of connecting orbits shifted by $L/3$ and $2L/3$ in the $x$-direction.

\begin{figure}
\centering
    \begin{subfigure}{\linewidth}
        \includegraphics[width=0.8\linewidth]{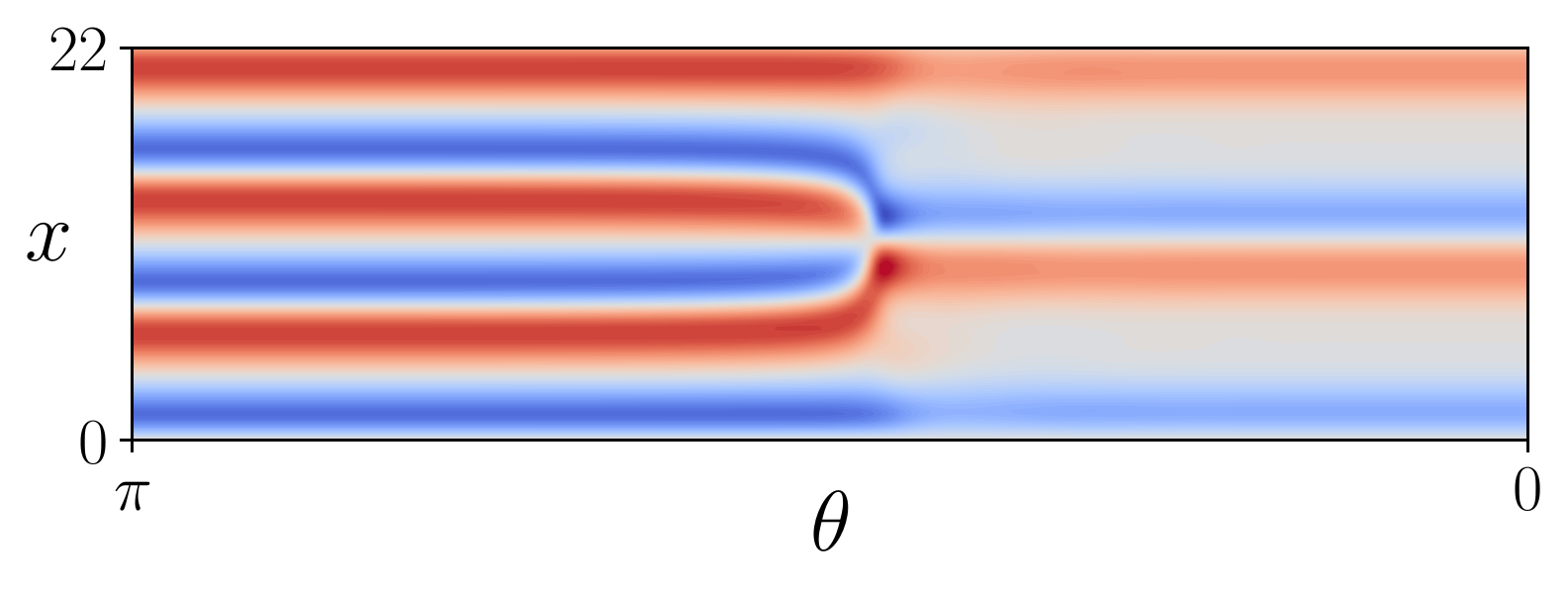}
        \caption{Orbit 1.}
    \end{subfigure}\\
    \begin{subfigure}{\linewidth}
        \includegraphics[width=0.8\linewidth]{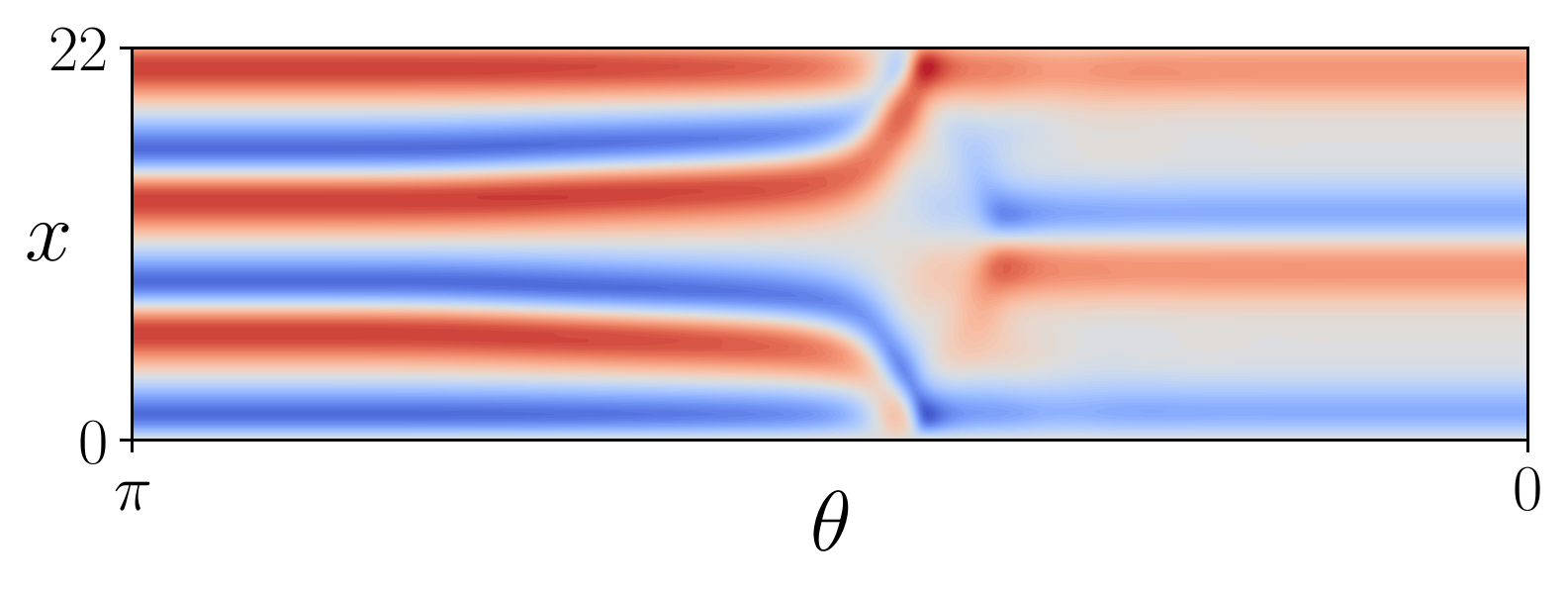}
        \caption{Orbit 2.}
    \end{subfigure}
    
    \caption{The space-time contour of the two converged connecting orbits from $E_3$ to $E_2$ in the center-symmetric subspace.}
    \label{fig:contour_from_E3}
\end{figure}

\section{Summary and concluding remarks}
\label{sec:summary}
Connecting orbits are of significant importance for studying spatiotemporally chaotic dynamical systems in terms of their invariant state space structures. We introduce a variational method for computing connecting orbits between two equilibrium solutions by searching in the space of all smooth curves in the state space that connect the two equilibria. In this method, the deviation of a connecting curve from an integral curve of the vector field is penalized by a non-negative cost function. A dynamical system in the space of connecting curves is set up such that along its trajectories the cost function is guaranteed to decrease monotonically. All trajectories of this dynamical system eventually converge to an equilibrium, which corresponds to a minimum of the cost function. Global minima of the cost function, taking zero value, correspond to the connecting orbits of the original dynamics.
This method is not limited by the dimensionality of the unstable manifold at the origin equilibrium solution, does not suffer from exponential separation of trajectories, and does not require any domain truncation. The introduced method is Jacobian-free, and its memory requirement scales linearly with the number of degrees of freedom, which allows this method to be applied to high-dimensional dynamical systems including three-dimensional fluid dynamics problems.

As a proof of concept, we apply the introduced variational method to the one-dimensional KSE, and compute several connecting orbits between known equilibrium solutions of the system with domain size $L=22$. The set of converged solutions contains at least one connecting orbit between any two equilibrium solutions unless it is known from an exhaustive search in the unstable manifold of the origin equilibrium solution that they are not connected.

After demonstrating the feasibility of the introduced method for computing connecting orbits between equilibrium solutions of the one-dimensional KSE, we are extending the present work in two directions: One is applying this method to the three-dimensional wall-bounded fluid flows governed by the Navier-Stokes equations (NSE). The challenge in applying this method to the wall-bounded NSE lies not only in dealing with a dynamical system of considerably larger size, but also in handling the incompressibility constraint and the pressure field: Pressure is not governed by an explicit evolution equation, but by the so-called pressure Poisson equation to adapt itself to the velocity such that the velocity field remains divergence-free. Construction of the pressure field associated to an instantaneous divergence-free velocity field in a wall-bounded domain is not a trivial task\cite{Rempfer2006}, let alone the derivation of the adjoint operator in the presence of this nonlocal, nonlinear operator. The second direction is developing methods following a similar idea for computing connecting orbits between invariant solutions of other types, including between two periodic orbits and eventually between invariant tori.
Together with improved methods for constructing invariant solutions\cite{Azimi2022,Parker2022a,Parker2023}, the proposed methodology for computing connecting orbits represents a step towards a more complete characterization of the state-space structures supporting spatiotemporally chaotic dynamics. Eventually, the characterization of connecting orbits mediating transitions between invariant solutions may allow for efficient forecasting of chaos even in high-dimensional systems including fluid turbulence.  

\appendix
\section{Derivation of the adjoint operator for the KSE}
\label{sec:KSE_adj_derivation}
The directional derivative of the residual of the KSE, defined in Eq.~\eqref{eq:KSE_residual}, along $G$ is obtained by the definition \eqref{eq:directional_derivative} as
\begin{equation}
    \label{eq:KSE_directional_derivative}
    \mathscr{L}(u;G)=-\dfrac{\partial G}{\partial s}-\dfrac{\partial (uG)}{\partial x}-\dfrac{\partial^2 G}{\partial x^2}-\dfrac{\partial^4 G}{\partial x^4}.
\end{equation}
In order to find the adjoint operator, we expand the inner product of  $\mathscr{L}(u;G)$ and $r$
\begin{eqnarray}
\label{eq:KSE_inner_product_expansion}
    \big<\mathscr{L}(u&&;G),r\big>\nonumber\\
    =&&\int_{-\infty}^{+\infty}\int_{0}^{L}{\left(-\dfrac{\partial G}{\partial s}-\dfrac{\partial (uG)}{\partial x}-\dfrac{\partial^2 G}{\partial x^2}-\dfrac{\partial^4 G}{\partial x^4}\right)r}\d x \d s\nonumber\\
    =&&-\int_{0}^{L}\left\{\int_{-\infty}^{+\infty}\dfrac{\partial G}{\partial s}r\d s\right\}\d x\nonumber\\
    &&-\int_{-\infty}^{+\infty}\left\{\int_{0}^{L}\left(\dfrac{\partial (uG)}{\partial x}+\dfrac{\partial^2 G}{\partial x^2}+\dfrac{\partial^4 G}{\partial x^4}\right)r\d x\right\}\d s
\end{eqnarray}
Integrating by parts we can write the first and the second integral as follows
\begin{eqnarray*}
    &&\int_{0}^{L}\left\{\int_{-\infty}^{+\infty}\dfrac{\partial G}{\partial s}r\d s\right\}\d x = \nonumber\\ &&\qquad\int_{0}^{L}{\left\{\lim_{T\to\infty} \Big[Gr\Big]_{s=-T}^{s=T}-\int_{-\infty}^{+\infty}G\dfrac{\partial r}{\partial s}\d s\right\}}\d x,
\end{eqnarray*}

\begin{widetext}
\begin{eqnarray*}
    \int_{-\infty}^{+\infty}\Bigg\{\int_{0}^{L}\Bigg(\dfrac{\partial (uG)}{\partial x}+\dfrac{\partial^2 G}{\partial x^2}+\dfrac{\partial^4 G}{\partial x^4}\Bigg)r\d x\Bigg\}\d s = \int_{-\infty}^{+\infty}\Bigg\{&&\Big[uGr\Big]_{x=0}^{x=L}-\int_{0}^{L}uG\dfrac{\partial r}{\partial x}\d x+\Bigg[\dfrac{\partial G}{\partial x}r-G\dfrac{\partial r}{\partial x}\Bigg]_{x=0}^{x=L}+\int_{0}^{L}G\dfrac{\partial^2 r}{\partial x^2}\d x\nonumber\\
    &&+\Bigg[\dfrac{\partial^3G}{\partial x^3}r-\dfrac{\partial^2 G}{\partial x^2}\dfrac{\partial r}{\partial x}+\dfrac{\partial G}{\partial x}\dfrac{\partial^2 r}{\partial x^2}-G\dfrac{\partial^3 r}{\partial x^3} \Bigg]_{x=0}^{x=L}+\int_{0}^{L}G\dfrac{\partial^4 r}{\partial x^4}\d x\Bigg\}\d s.
\end{eqnarray*}
\end{widetext}
In the limit $T\to\infty$ the boundary term $[Gr]_{s=-T}^{s=T}$ vanishes since both $G$ and $r$ are asymptotically zero. All boundary terms $[\,\cdot\,]_{x=0}^{x=L}$ vanish too due to periodicity of $u$, $r$ and $G$ in $x$. Therefore, Eq.~\eqref{eq:KSE_inner_product_expansion} becomes
\begin{equation}
\label{eq:KSE_inner_product_left}
    \left<\mathscr{L}(u;G),r\right>=\int_{-\infty}^{+\infty}\int_{0}^{L}{\left(\dfrac{\partial r}{\partial s}+u\dfrac{\partial r}{\partial x}-\dfrac{\partial^2 r}{\partial x^2}-\dfrac{\partial^4 r}{\partial x^4}\right)G}\d x\d s.
\end{equation}
From the definition of the adjoint operator \eqref{eq:adjoint_definition}, this inner product equals
\begin{equation}
\label{eq:KSE_inner_product_right}
    \left<\mathscr{L}^\dagger(u;r),G\right> = \int_{-\infty}^{+\infty}\int_0^L\mathscr{L}^\dagger G \d x\d s.
\end{equation}
Comparing Eqs.~\eqref{eq:KSE_inner_product_left} and \eqref{eq:KSE_inner_product_right}, $\mathscr{L}^\dagger(u;r)$ is given by
\begin{equation}
    \mathscr{L}^\dagger(u;r) = \dfrac{\partial r}{\partial s}+u\dfrac{\partial r}{\partial x}-\dfrac{\partial^2 r}{\partial x^2}-\dfrac{\partial^4 r}{\partial x^4}.
\end{equation}

\section{Unstable eigenvectors of the equilibria of the KSE}
\label{sec:KSE_eigenvecotrs}
\begin{figure}
\centering
    \begin{subfigure}{0.49\linewidth}
        \includegraphics[width=\linewidth]{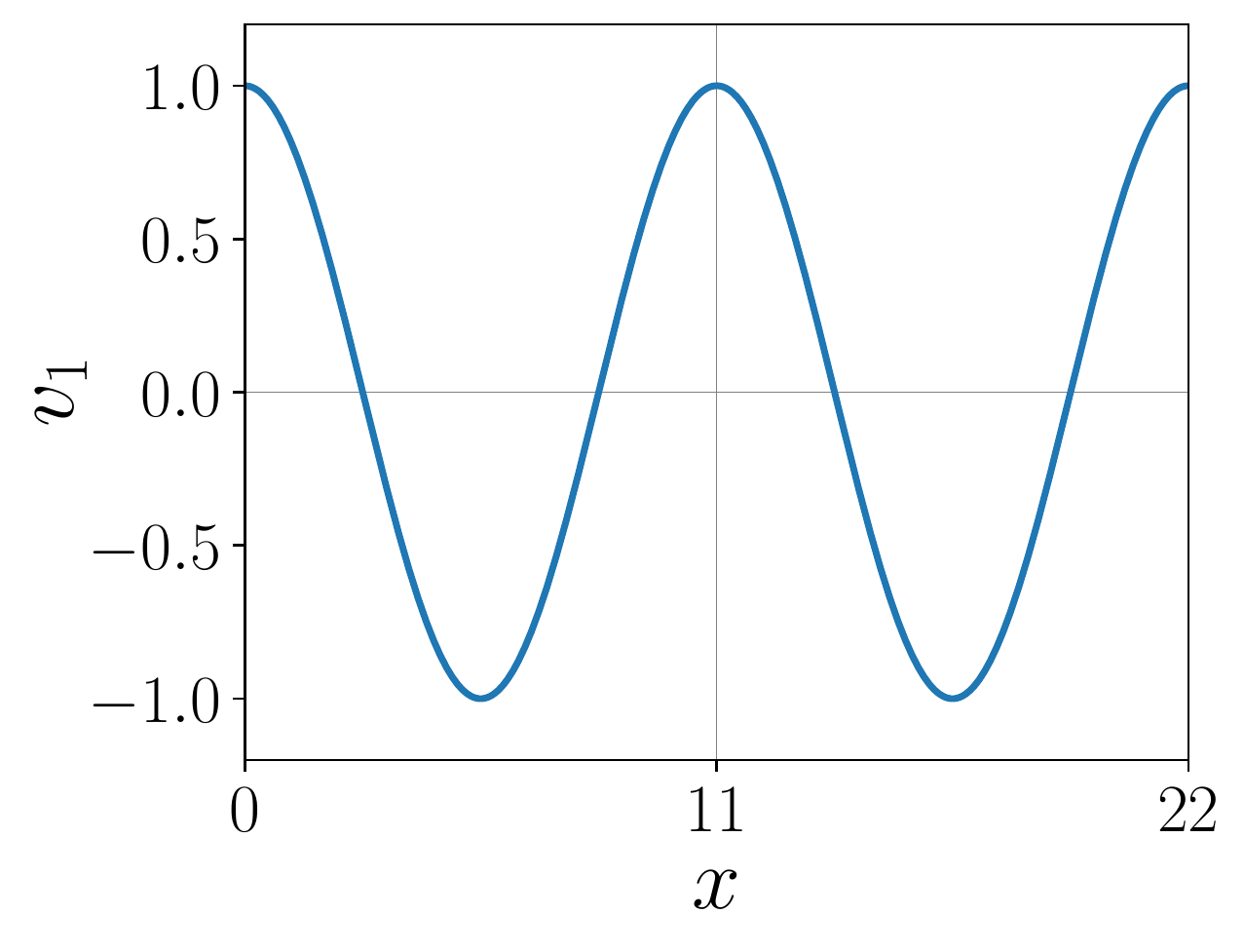}
        \caption{$\lambda_1=0.2198$}
    \end{subfigure}\hfill
    \begin{subfigure}{0.49\linewidth}
        \includegraphics[width=\linewidth]{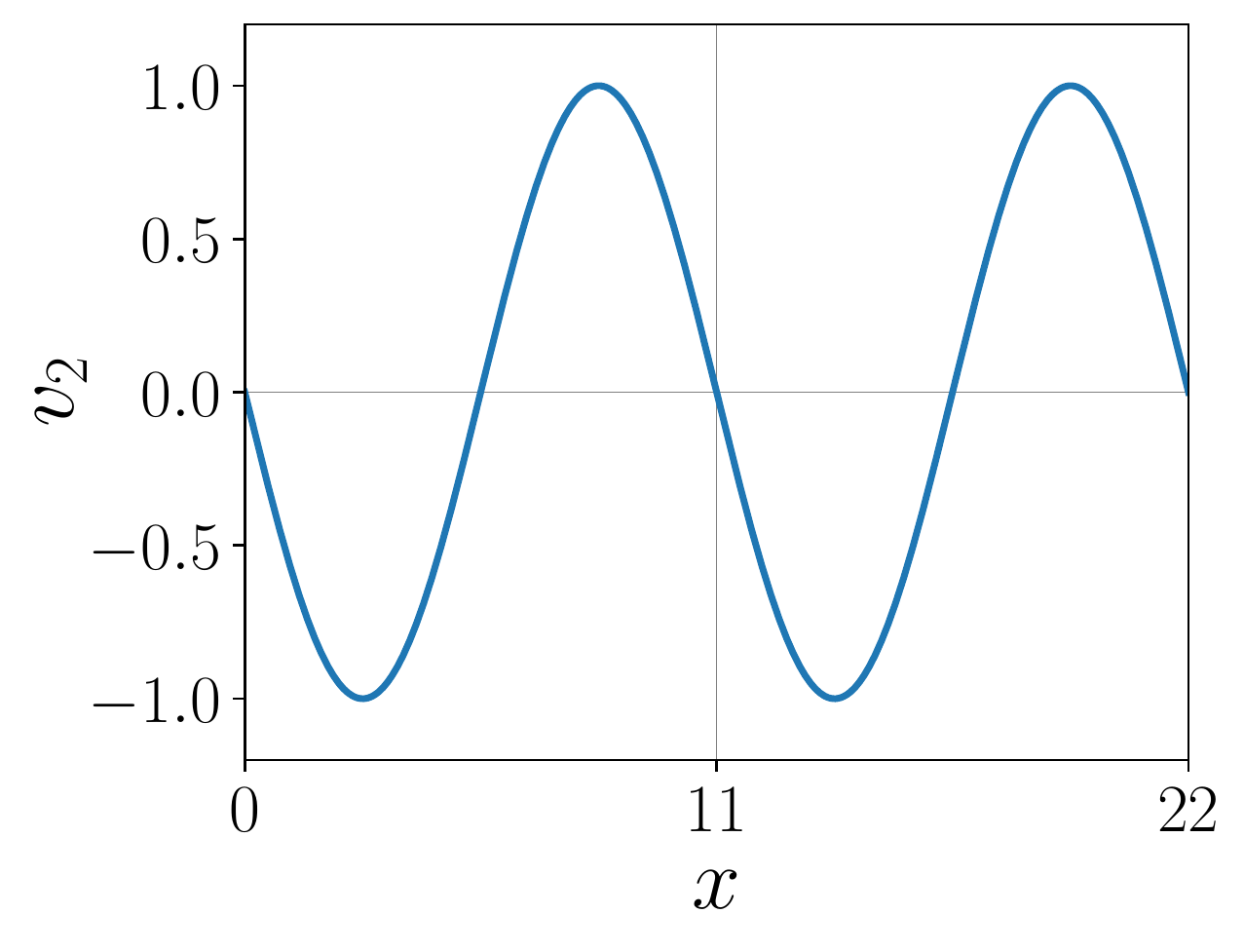}
        \caption{$\lambda_2=0.2198$}
    \end{subfigure}
    
    \begin{subfigure}{0.49\linewidth}
        \includegraphics[width=\linewidth]{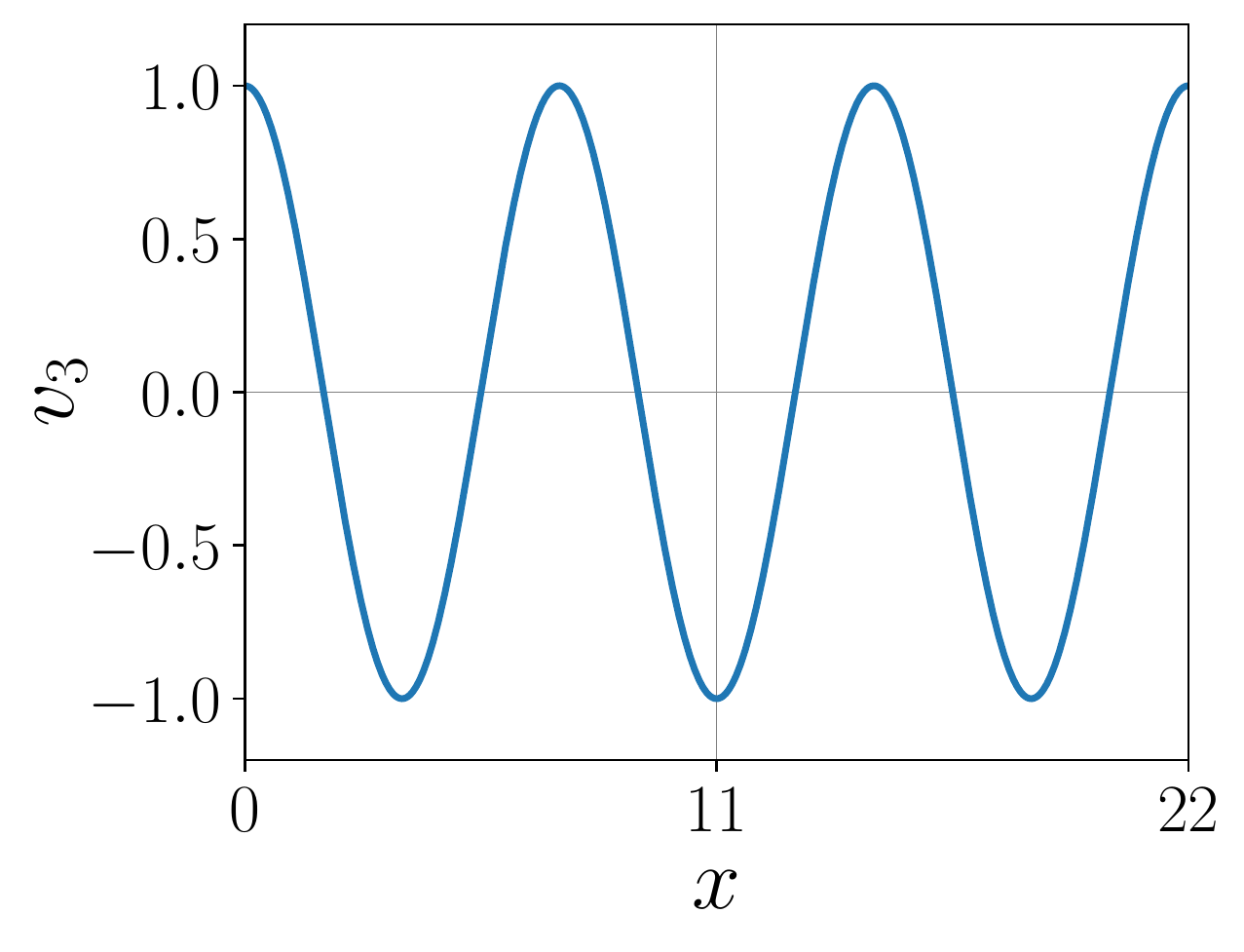}
        \caption{$\lambda_3=0.1952$}
    \end{subfigure}\hfill
    \begin{subfigure}{0.49\linewidth}
        \includegraphics[width=\linewidth]{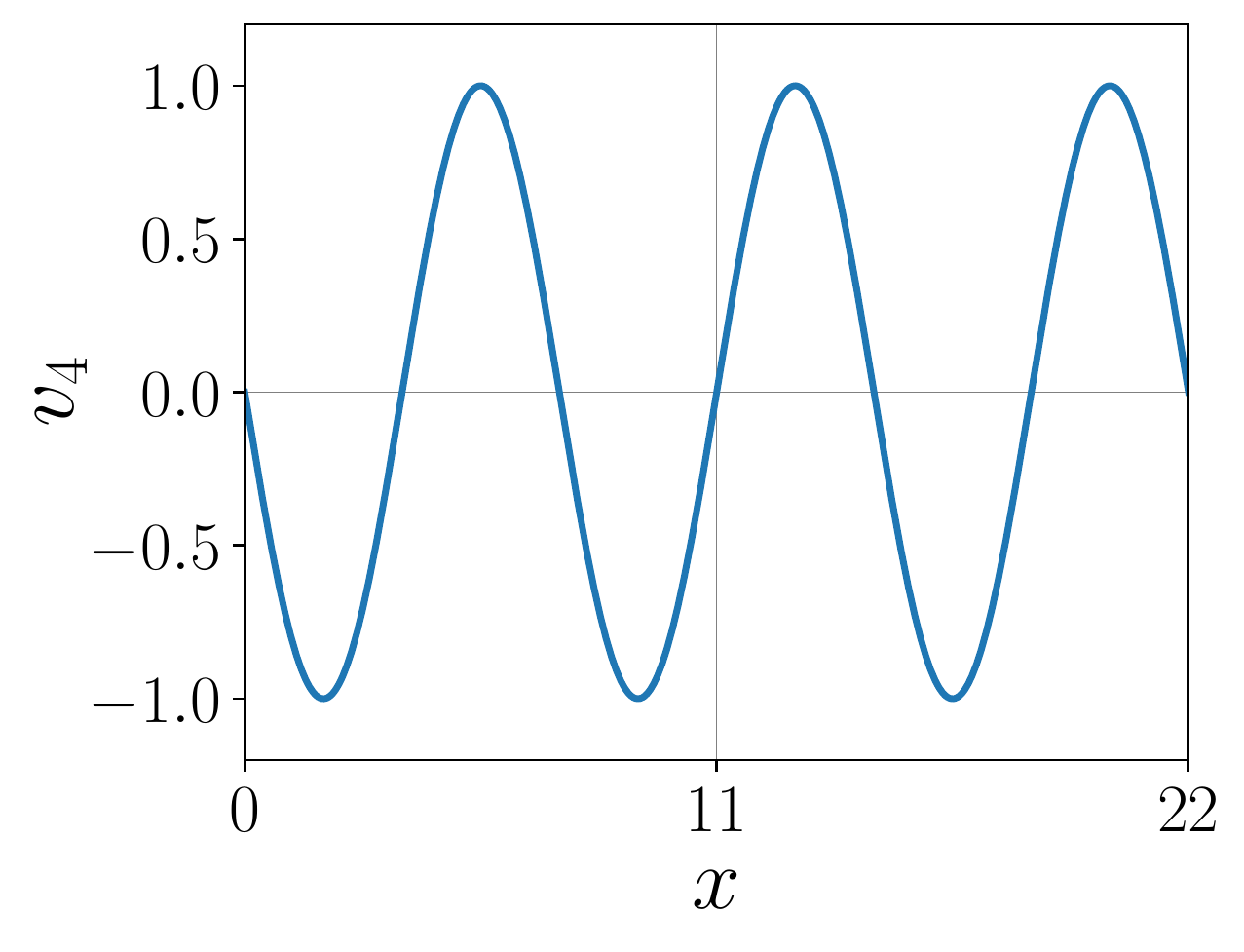}
        \caption{$\lambda_4=0.1952$}
    \end{subfigure}
    
    \begin{subfigure}{0.49\linewidth}
        \includegraphics[width=\linewidth]{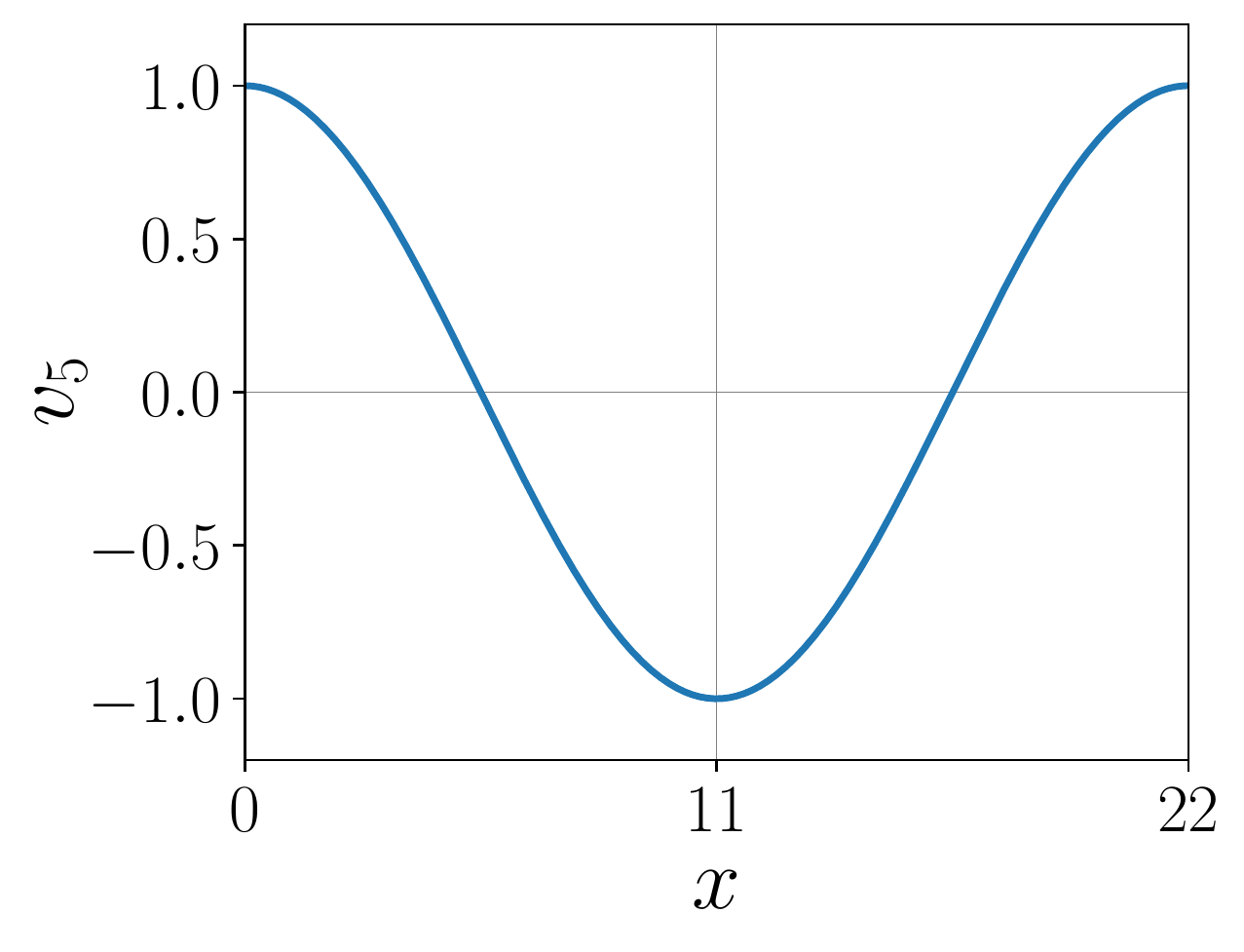}
        \caption{$\lambda_5=0.0749$}
    \end{subfigure}\hfill
    \begin{subfigure}{0.49\linewidth}
        \includegraphics[width=\linewidth]{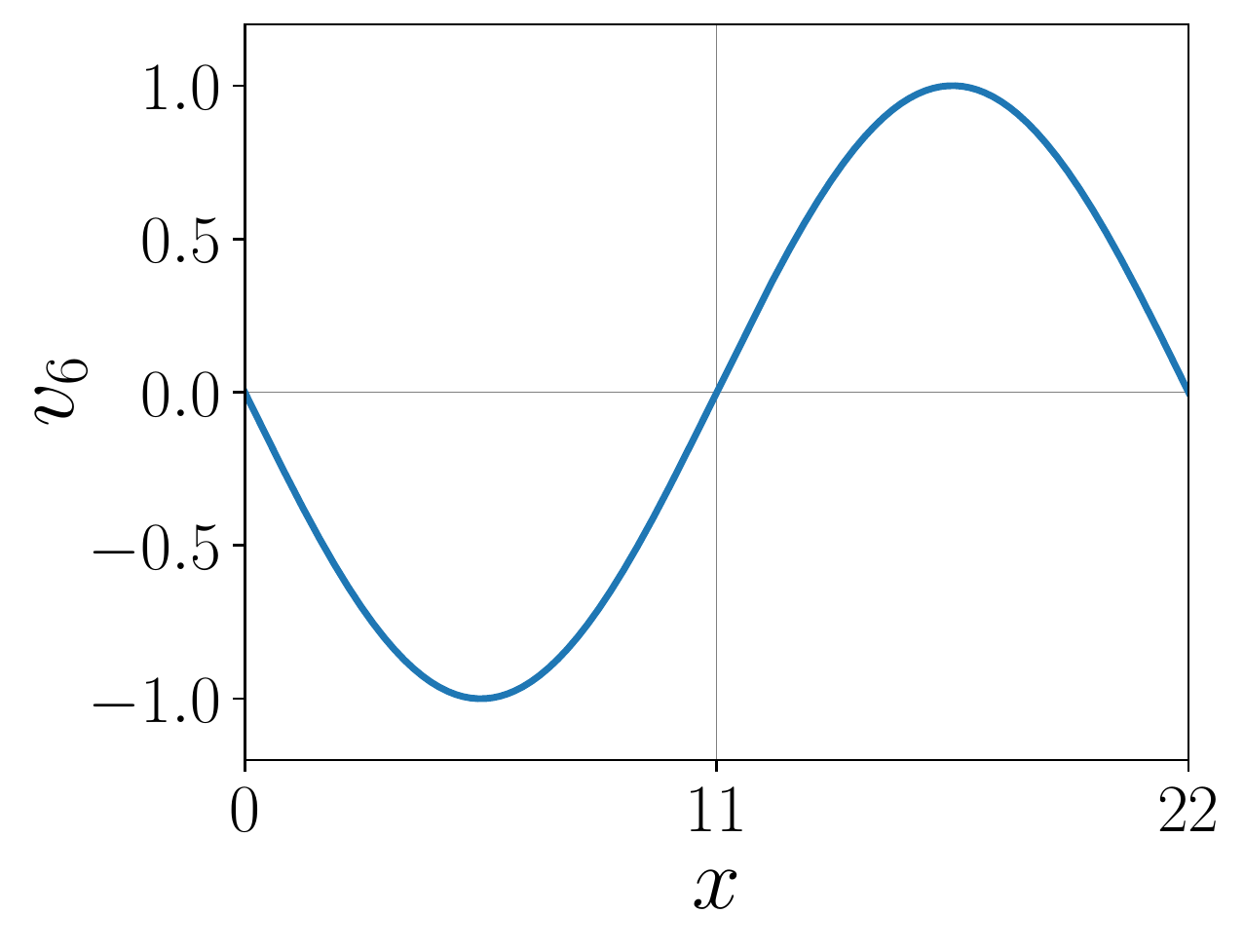}
        \caption{$\lambda_6=0.0749$}
    \end{subfigure}
    \caption{Unstable eigenvectors $v_i$ of the trivial equilibrium solution $E_0$. $\lambda_i$ is the eigenvalue associated with $v_i$.}
    \label{fig:eigenvectors_E0}
\end{figure}

\begin{figure}
\centering
    \begin{subfigure}{0.49\linewidth}
        \includegraphics[width=\linewidth]{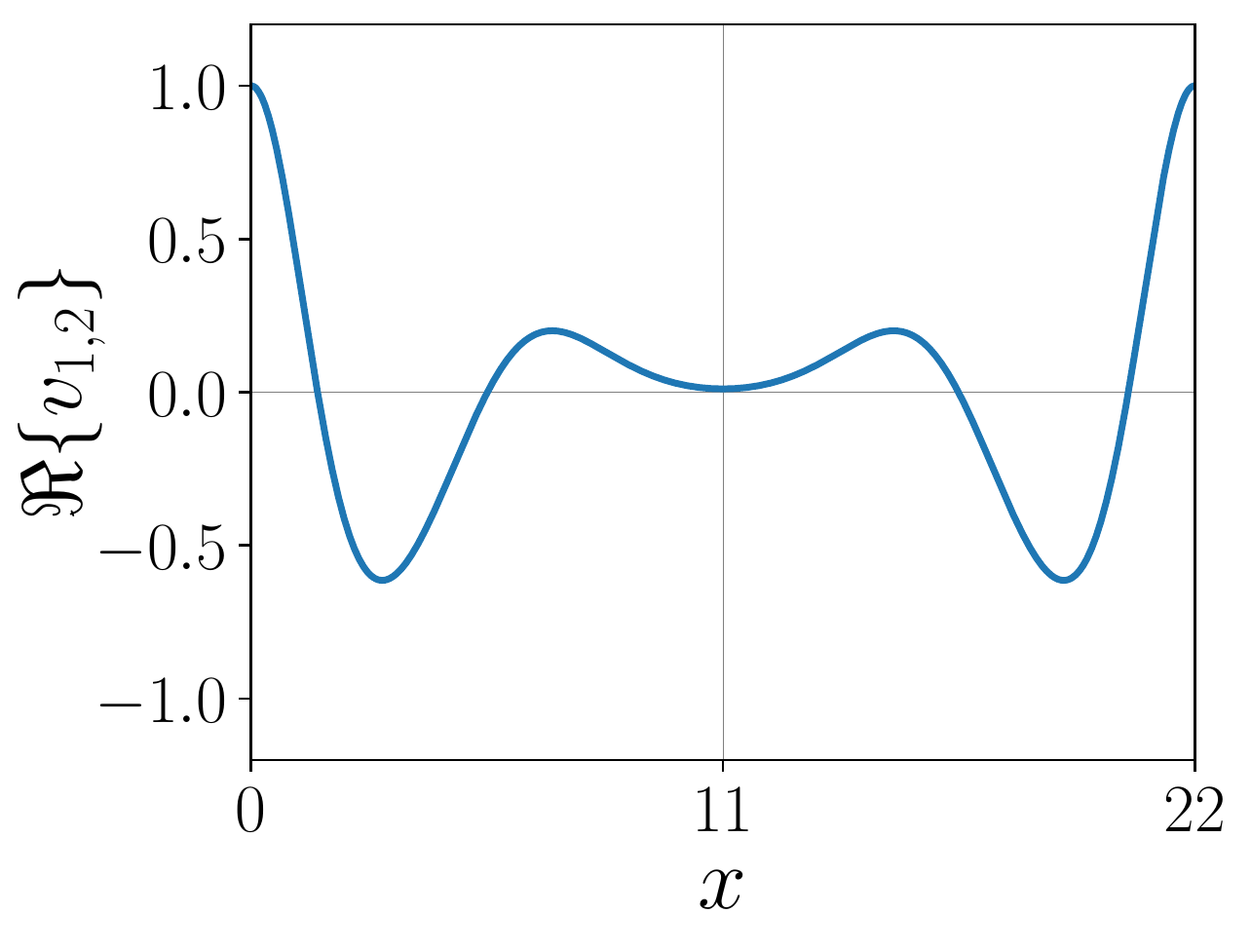}
        \caption{$\lambda_{1,2}=0.1308\pm0.3341i$}
    \end{subfigure} \hfill
    \begin{subfigure}{0.49\linewidth}
        \includegraphics[width=\linewidth]{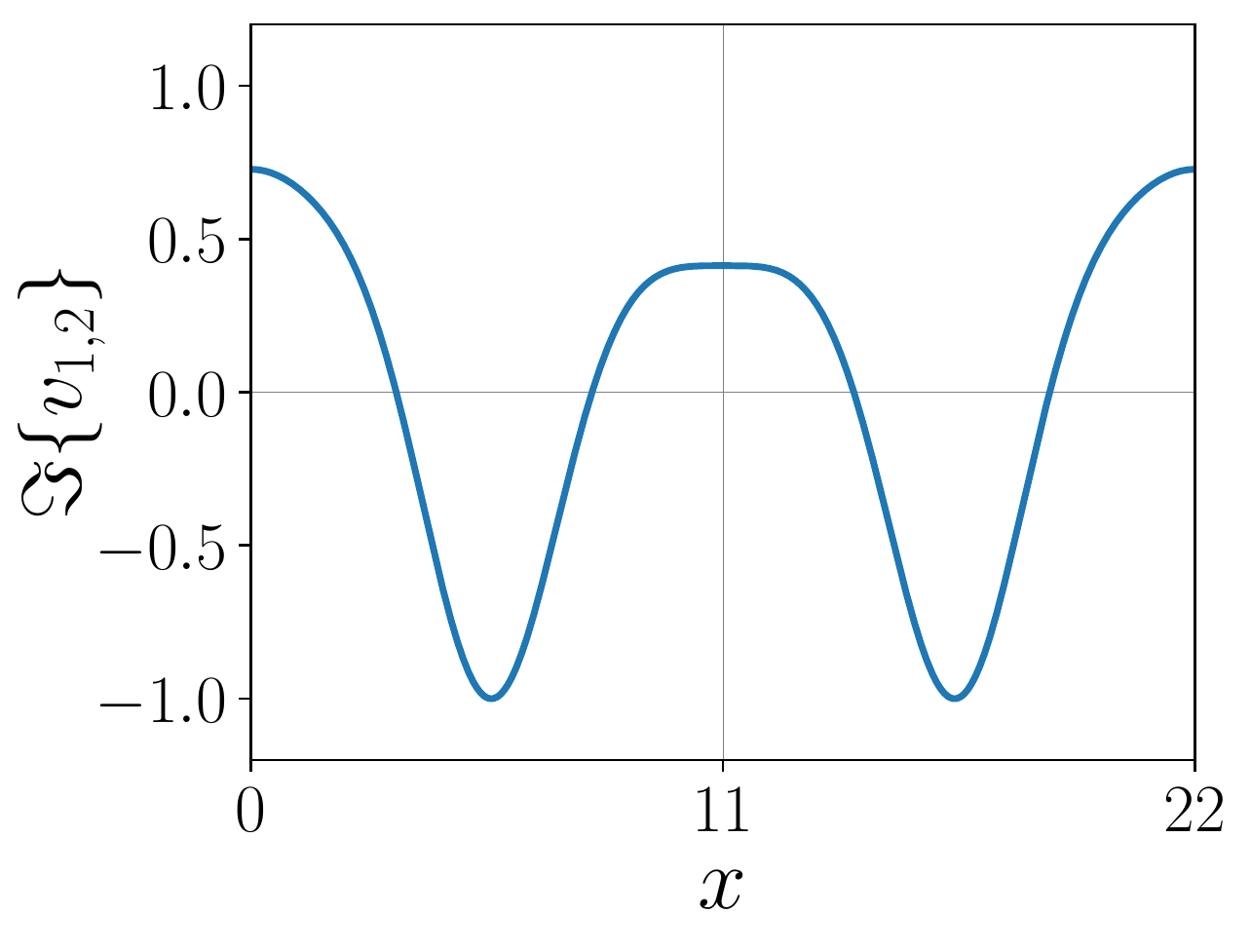}
        \caption{$\lambda_{1,2}=0.1308\pm0.3341i$}
    \end{subfigure}
    
    \begin{subfigure}{0.49\linewidth}
        \includegraphics[width=\linewidth]{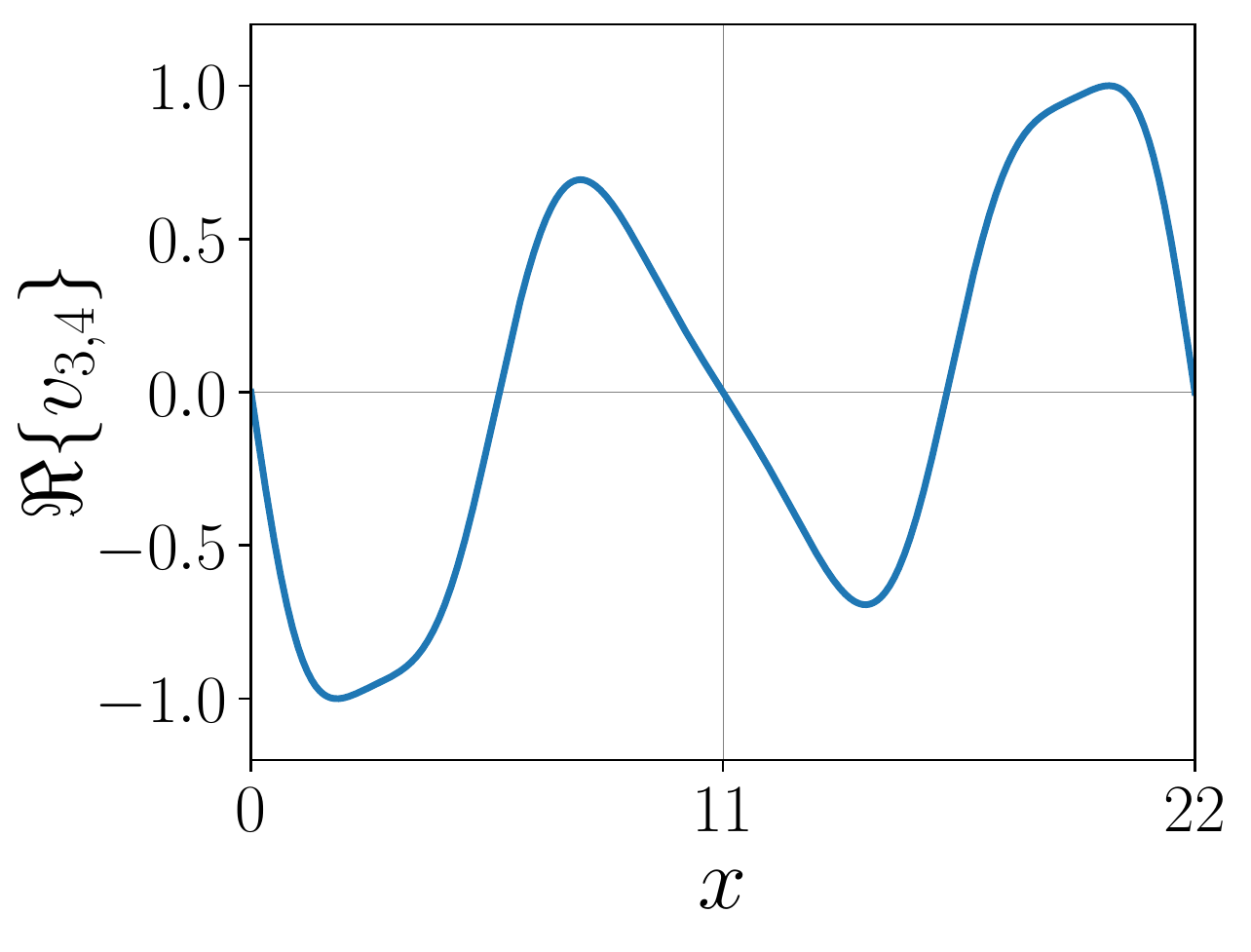}
        \caption{$\lambda_{3,4}=0.0824\pm0.3402i$}
    \end{subfigure} \hfill
    \begin{subfigure}{0.49\linewidth}
        \includegraphics[width=\linewidth]{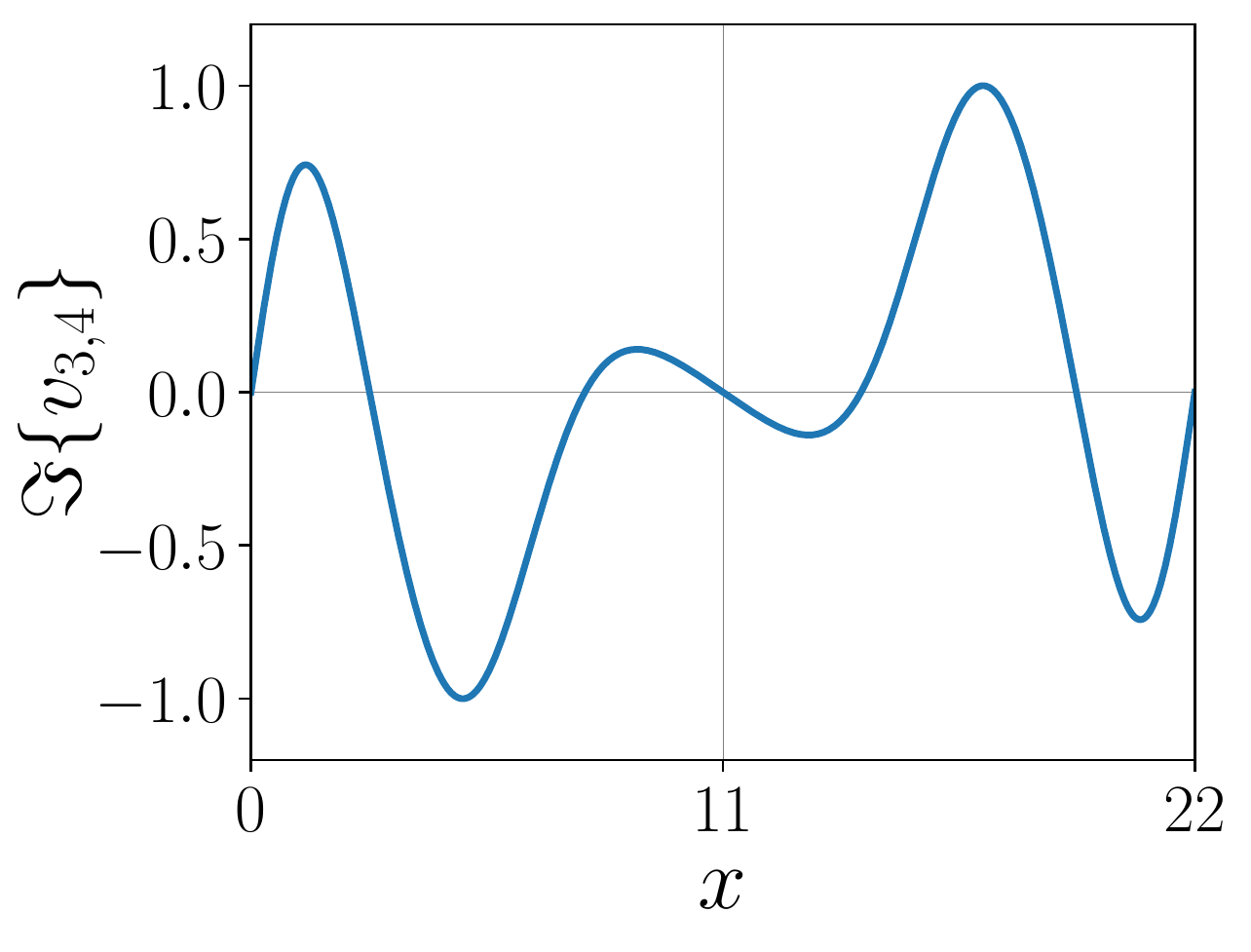}
        \caption{$\lambda_{3,4}=0.0824\pm0.3402i$}
    \end{subfigure}
    \caption{Unstable eigenvectors $v_i$ of the equilibrium solution $E_1$. $\lambda_i$ is the eigenvalue associated with $v_i$.}
    \label{fig:eigenvectors_E1}
\end{figure}

\begin{figure}
\centering
    \begin{subfigure}{0.49\linewidth}
        \includegraphics[width=\linewidth]{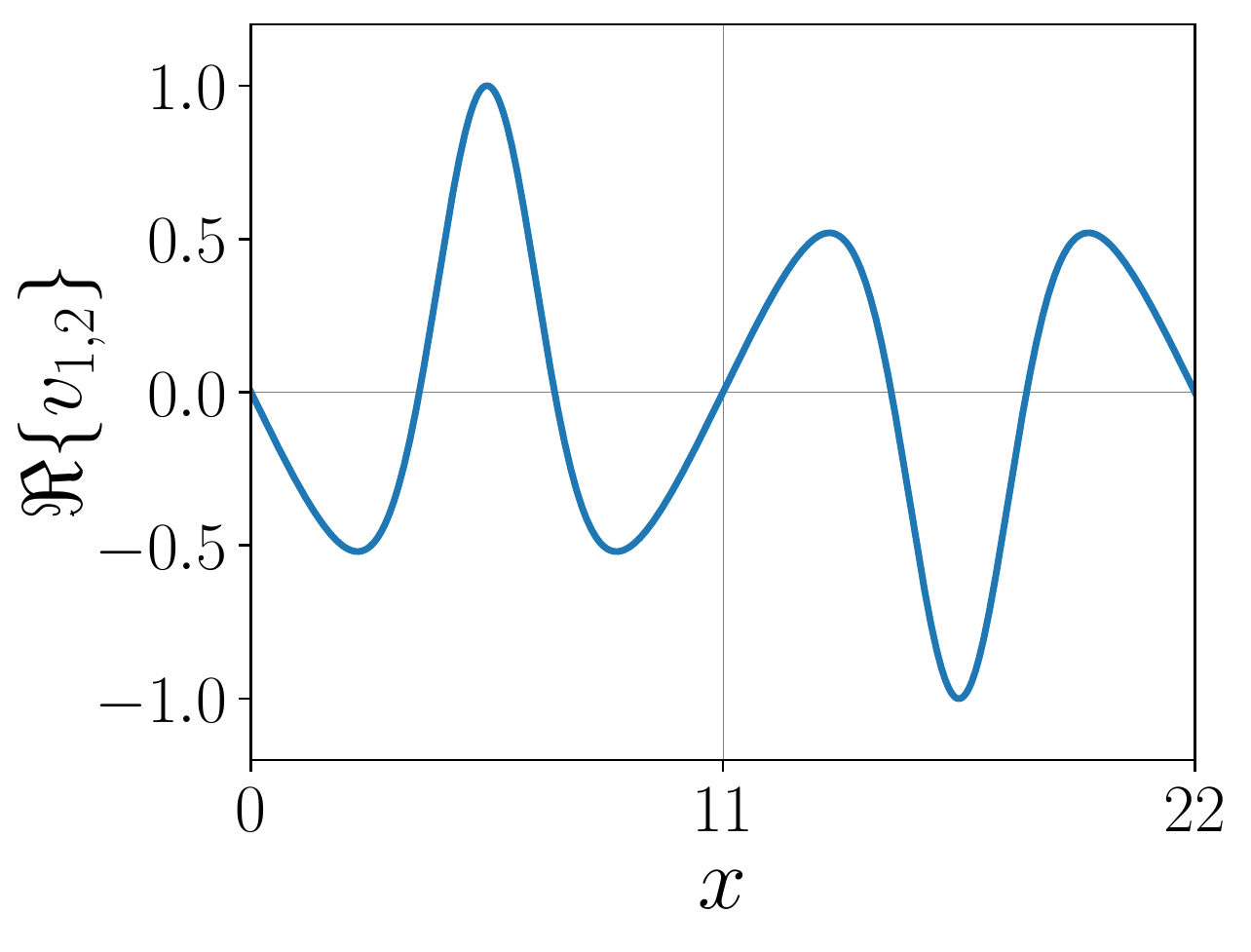}
        \caption{$\lambda_{1,2}=0.1390\pm0.2384i$}
    \end{subfigure} \hfill
    \begin{subfigure}{0.49\linewidth}
        \includegraphics[width=\linewidth]{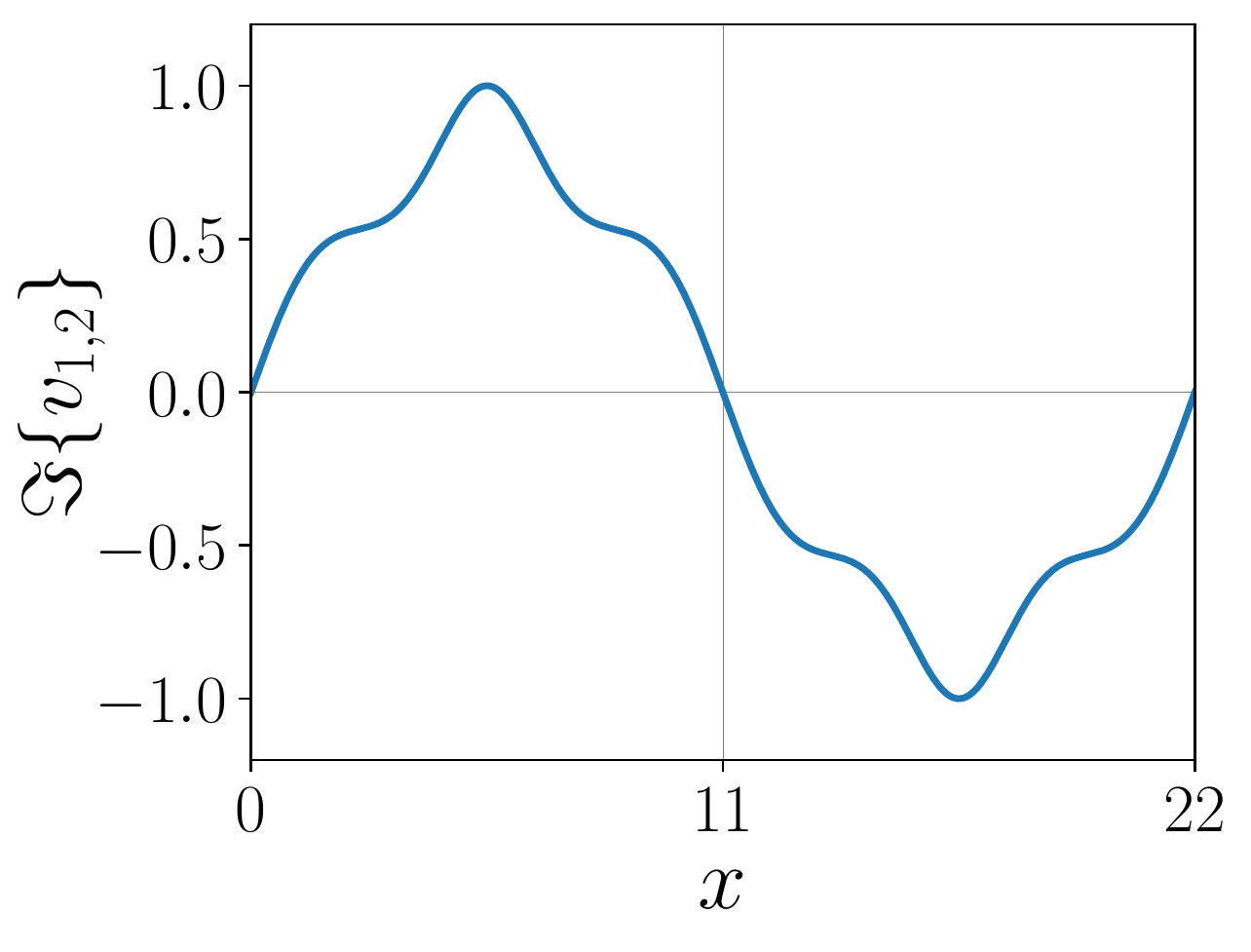}
        \caption{$\lambda_{1,2}=0.1390\pm0.2384i$}
    \end{subfigure}
    \caption{Unstable eigenvectors $v_i$ of the equilibrium solution $E_2$. $\lambda_i$ is the eigenvalue associated with $v_i$.}
    \label{fig:eigenvectors_E2}
\end{figure}

\begin{figure}
\centering
    \begin{subfigure}{0.49\linewidth}
        \includegraphics[width=\linewidth]{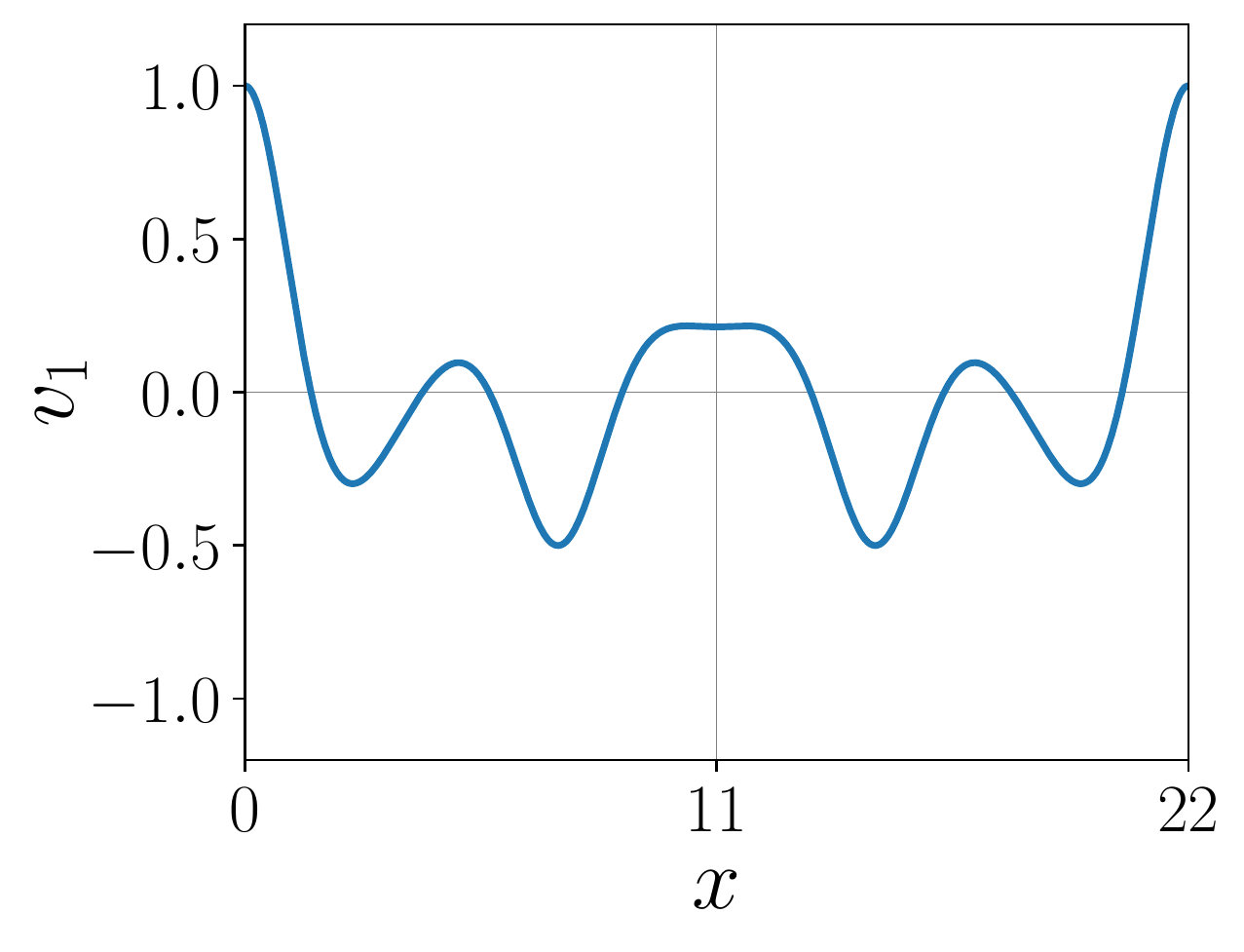}
        \caption{$\lambda_1=0.0933$}
    \end{subfigure} \hfill
    \begin{subfigure}{0.49\linewidth}
        \includegraphics[width=\linewidth]{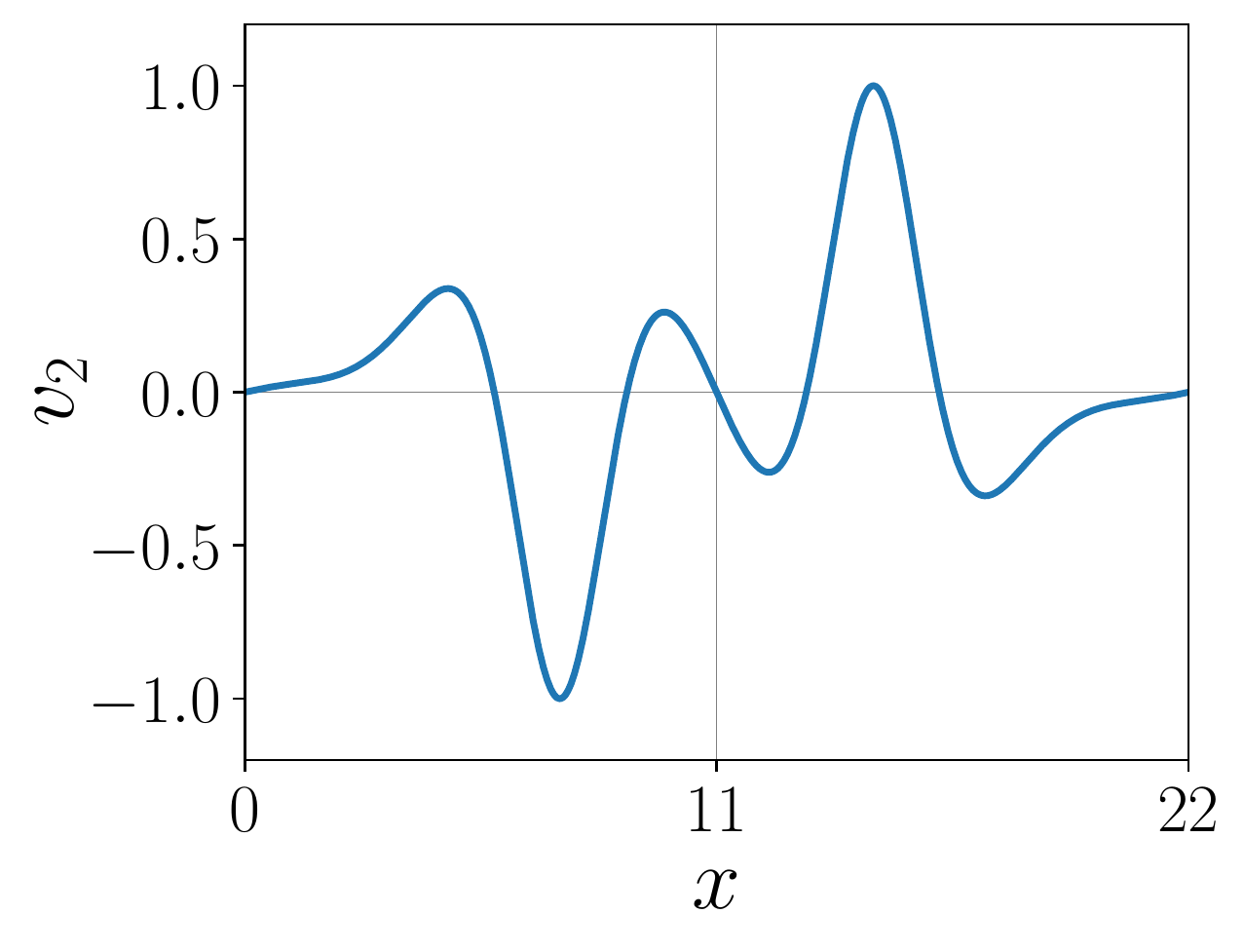}
        \caption{$\lambda_2=0.0933$}
    \end{subfigure}
    \caption{Unstable eigenvectors $v_i$ of the equilibrium solution $E_3$. $\lambda_i$ is the eigenvalue associated with $v_i$.}
    \label{fig:eigenvectors_E3}
\end{figure}

The KSE with $L=22$ has four known equilibrium solutions including the trivial solution $E_0=0$, and three nontrivial solutions $E_1$, $E_2$ and $E_3$ as shown in Fig.~\ref{fig:KSE_FPs}. The repelling eigenvalues of these equilibrium solutions are listed in Table \ref{tab:KSE_eigenvals}. The corresponding eigenvectors of $E_0$, $E_1$, $E_2$ and $E_3$, are shown in Figs.~\ref{fig:eigenvectors_E0} to \ref{fig:eigenvectors_E3}, respectively.

\section{Parameters used in constructing connecting orbits of the KSE}
\label{sec:parameters}
In all calculations presented in Section \ref{sec:application_to_KSE} we have used $N=64$ Fourier modes in space, have set the center of the temporal distribution at the origin $s_0=0$, and have used time step size $\Delta\tau=0.01$. The temporal resolution $M$ and the scaling $S$ are listed in Table \ref{tab:parameters}. The temporal resolution is set high enough so that the convergence criterion $J_\mathrm{arc,min}<10^{-12}$ is achieved (see Section \ref{sec:results_and_discussion}.)
\begin{table}
    \centering
    \caption{Parameters used for numerically integrating the dynamics in the space of connecting curves between different equilibrium solutions of the KSE for $L=22$ to construct connecting orbits.}
    \begin{tabular}{>{\centering\arraybackslash}p{0.8cm} >{\centering\arraybackslash}p{0.8cm} >{\centering\arraybackslash}p{1.5cm} >{\centering\arraybackslash}p{1cm} >{\centering\arraybackslash}p{0.8cm} >{\centering\arraybackslash}p{1.5cm}}
    \toprule
         row & from & to & $M$ & $S$ & Figure\\
    \midrule
         1 & \multirow{3}{*}{$E_0$} & $E_1$ & 80 & 40 & \ref{fig:state_space_from_E0}a \& \ref{fig:contour_from_E0}a\\
         2 & & $E_2$ & 130 & 35 & \ref{fig:state_space_from_E0}b \& \ref{fig:contour_from_E0}b\\
         3 & & $E_3$ & 120 & 40 & \ref{fig:state_space_from_E0}c \& \ref{fig:contour_from_E0}c \vspace{2.75mm}\\
         4 & \multirow{2}{*}{$E_1$} & $E_2$ & 550 & 55 & \ref{fig:deformation} \& \ref{fig:initial_final_contours}\\
         5 & & $E_3$ & 500 & 60 & \ref{fig:state_space_from_E1} \& \ref{fig:contour_from_E1} \vspace{2.75mm}\\
         6 & \multirow{2}{*}{$E_2$} & $\tau(1/4)E_2$ & 400 & 35 & \ref{fig:state_space_from_E2}a \& \ref{fig:contour_from_E2}a\\
         7 & & $E_3$ & 450 & 50 & \ref{fig:state_space_from_E2}b \& \ref{fig:contour_from_E2}b \vspace{2.75mm}\\
         8 & \multirow{2}{*}{$E_3$} & \multirow{2}{*}{$E_2$} & 600 & 50 & \ref{fig:state_space_from_E3}a \& \ref{fig:contour_from_E3}a\\
         9 & & & 500 & 40 & \ref{fig:state_space_from_E3}b \& \ref{fig:contour_from_E3}b\\
    \bottomrule
    \end{tabular}
    \label{tab:parameters}
\end{table}

\section*{Acknowledgements}
This research has been supported by the European Research Council (ERC) under the European Union's Horizon 2020 research and innovation programme (grant agreement no. 865677). The authors would like to thank Sajjad Azimi and Jeremy P. Parker for helpful discussions.

\section*{Data Availability Statement}
The data that support the findings of this study are available from the corresponding author upon reasonable request.

\section*{References}
%

\end{document}